\documentclass[pra,aps,preprint,amsmath,amssymb]{revtex4}
\usepackage{graphicx}
\usepackage{epsfig}
\include{graphics}
\usepackage{epstopdf} 

\begin{document}

\title{Stabilizing solitons of the cubic-quintic nonlinear Schr\"odinger equation 
by frequency-dependent linear gain-loss and delayed Raman response}

\author{Avner Peleg$^{1}$ and Debananda Chakraborty$^{2}$}

\affiliation{$^{1}$ Department of Mathematics, Azrieli College of Engineering, 
Jerusalem 9371207, Israel}
\affiliation{$^{2}$ Department of Mathematics, New Jersey City University, 
Jersey City, NJ 07305, USA}

\date{\today}

\begin{abstract} 
We demonstrate transmission stabilization against radiation emission by frequency-dependent 
linear gain-loss and perturbation-induced frequency shifting for solitons of the cubic-quintic 
nonlinear Schr\"odinger (CQNLS) equation. We consider soliton propagation in a nonlinear optical 
waveguide with focusing cubic nonlinearity, defocusing quintic nonlinearity, 
and dissipative perturbations due to weak frequency-dependent linear gain-loss, 
cubic loss, and delayed Raman response. The frequency shifting 
is induced by delayed Raman response. Our perturbation analysis and numerical simulations with 
the perturbed CQNLS equation show that transmission stabilization with CQNLS solitons is indeed 
possible, and in this way provide the first demonstration of the stabilization method 
for solitons of a nonintegrable nonlinear wave model. Moreover, we find that 
transmission stabilization with energetic CQNLS solitons is realized with significantly 
smaller frequency shifts and pulse distortion compared with stabilization 
with energetic solitons of the cubic nonlinear Schr\"odinger equation. 
Therefore, our study also demonstrates that stabilization of energetic solitons 
by the method is significantly improved by the presence of defocusing quintic nonlinearity. 
\newline
\newline

Keywords: Soliton, nonlinear Schr\"odinger equation, soliton stabilization, 
quintic nonlinearity, dissipative perturbations, perturbation method. 
\end{abstract}

\maketitle

     

\section{Introduction}
\label{Introduction}

The cubic-quintic nonlinear Schr\"odinger (CQNLS) equation is a simple nonintegrable 
generalization of the integrable cubic nonlinear Schr\"odinger (CNLS) equation 
\cite{Yang2010,Fibich2015,Pushkarov96}. It describes the behavior of a variety of 
physical systems, including the propagation of optical pulses in nonlinear semiconductor-doped 
optical waveguides \cite{Pushkarov96,Agrawal2019,Gagnon89,Kartashov2004,Biswas2006,Wai2008}, the 
propagation of optical beams in certain nonlinear bulk optical media 
\cite{Fibich2015,Leblond2013,Sarkar2021} and thin films \cite{Mihalache88}, 
the dynamics of Bose-Einstein condensates \cite{Newman2000,Gammal2000,Kevrekidis2008,Luckins2018}, 
and a range of interaction phenomena in plasmas \cite{He92,He2008,Singh2019}. 
It is also a simple model of pattern formation 
\cite{Yang2010,Hohenberg92}, which describes, for example, the formation of 
fronts from pulses in the presence of deterministic or stochastic 
perturbations \cite{Hohenberg92,Kramer2002,PDC2005,PCDN2009}. Since the CQNLS equation 
is a nonintegrable model, its investigation allows one to uncover physical effects 
that do not exist in the integrable CNLS model, or dynamic behavior that is 
qualitatively different from the one observed in the integrable case. 
Examples for the latter behavior include radiation-induced decay of solitons 
due to the existence of internals modes \cite{Pelinovsky98,Yang2000}, 
radiation emission in fast two-soliton collisions \cite{Malomed86,SP2004}, 
and emergence of spatiotemporal chaos \cite{He92,He94}. Due to all these reasons, 
the study of the CQNLS equation and its soliton solutions received great attention 
in the last few decades, and it is still a very active area of research. Examples 
of works on this subject include investigation of stability of soliton solutions 
in dimension 1 \cite{Yang2010,Pelinovsky98,Yang2000,Ohta95,Martel21,Akhmediev99}  
and in higher dimensions \cite{Yang2010,Sparber21}, studies of soliton collisions 
and interaction \cite{Malomed86,SP2004,QMN2021,Wuster2022}, investigation of 
soliton dynamics in confining potentials \cite{Kevrekidis2008,Luckins2018,Tang2007},  
and studies of highly nonlinear dynamic behavior that is described 
by the equation and by related equations \cite{He92,He94,Singh2019}.

In the current paper, we are interested in stabilizing the long-distance 
transmission of CQNLS solitons in the presence of weak linear and nonlinear  
dissipative perturbations in dimension 1. Since we are interested in 
transmission stabilization, we consider soliton propagation in the presence 
of focusing cubic nonlinearity and defocusing quintic nonlinearity. In this 
case, it is known that the unperturbed CQNLS equation in dimension 1 possesses 
stable soliton solutions, which exist for values of the amplitude parameter 
$\eta$ satisfying $0<\eta<\eta_{m}$, where the constant $\eta_{m}$ depends on the quintic 
nonlinearity coefficient \cite{Yang2010,Pushkarov96,Gagnon89,PDC2005,Pelinovsky98,Ohta95}.   
Additionally, when $\eta$ is near $\eta_{m}$, the soliton solutions attain 
a flat-top shape \cite{PDC2005,PCDN2009}. Stability properties of CQNLS 
solitons in dimension 1 were studied in a number of works, see Refs. 
\cite{Yang2010,Pelinovsky98,Yang2000,Hohenberg92,Ohta95,Martel21,Akhmediev99}, 
and references therein. Importantly, orbital stability of solitons of the unperturbed 
CQNLS equation with focusing cubic nonlinearity and defocusing quintic nonlinearity 
was proved in Ref. \cite{Ohta95} for any $\eta$ satisfying $0<\eta<\eta_{m}$.    
Additionally, asymptotic stability of these solitons was proved in Ref. \cite{Martel21} 
for small-to-intermediate values of $\eta$. In Ref. \cite{Akhmediev99}, stability was 
established by considering the Hamiltonian-versus-energy diagram for the solitons. 
Furthermore, in Ref. \cite{Pelinovsky98} it was shown that there are no internal modes 
for CQNLS solitons in the case of focusing cubic nonlinearity and defocusing quintic 
nonlinearity, a result that leads to considerable simplification of the proof of their 
asymptotic stability \cite{Martel21}. Therefore, the studies in Refs. 
\cite{Pelinovsky98,Ohta95,Martel21,Akhmediev99} showed that solitons of the unperturbed  
CQNLS equation with focusing cubic nonlinearity and defocusing quintic nonlinearity in 
dimension 1 exhibit certain enhanced stability properties. From the physical point of view,  
the enhanced stability might be attributed to the fact that for these solitons, 
the effects of focusing cubic nonlinearity are balanced by both second-order dispersion 
and defocusing quintic nonlinearity \cite{Kartashov2004,Nath2017}.

In Ref. \cite{PC2018A}, we developed a method for transmission stabilization 
of solitons of the CNLS equation that propagate in the presence of weak 
linear and nonlinear dissipative perturbations. Due to the presence of 
perturbations the CNLS soliton emits small-amplitude waves (radiation). 
These small-amplitude waves can grow and interact with the soliton, and this 
can lead to distortion of the soliton's shape and to transmission 
destabilization \cite{PC2018A,CPN2016,PNT2016,PC2018B}. 
The stabilization method that we developed in Ref. \cite{PC2018A} was 
based on the interplay between perturbation-induced shifting of the soliton's 
frequency and frequency-dependent linear gain-loss. In this method, 
the perturbation-induced shifting of the soliton's frequency leads to 
the separation of the soliton's Fourier spectrum from the radiation's 
Fourier spectrum, and the frequency-dependent linear gain-loss leads to 
efficient suppression of radiation emission. We demonstrated the stabilization 
method for CNLS solitons with moderate power in two physically relevant 
nonlinear optical waveguide setups. In the first setup, the frequency shifting 
was caused by delayed Raman response, and in the second setup, the frequency 
shifting was caused by guiding filters with a varying central frequency \cite{PC2018A}. 
Our numerical simulations with the two perturbed CNLS equations validated the method, 
and showed stable long-distance transmission of solitons with moderate power 
with very weak pulse distortion \cite{PC2018A}. It should be pointed out that 
the method that we developed in Ref. \cite{PC2018A} generalized a method that 
was theoretically and experimentally used in the 1990s to stabilize soliton 
transmission in nonlinear optical fibers in the presence of weak linear loss, linear 
amplifier gain, and weak linear amplifier noise \cite{Mollenauer92,Mollenauer97,Mollenauer2006}. 
Therefore, our stabilization method is also very interesting from the experimental 
and the application points of view.

The study in Ref. \cite{PC2018A} was limited to solitons of the integrable CNLS 
equation. Therefore, it is unclear if the stabilization method works for solitons 
of nonintegrable nonlinear wave models, such as the CQNLS equation. Additionally, 
it is unclear whether soliton stability is enhanced or reduced in the nonintegrable 
model compared with the integrable one. Furthermore, in Ref. \cite{PC2018A} and in 
Refs. \cite{Mollenauer92,Mollenauer97,Mollenauer2006}, only CNLS solitons with 
moderate power values were considered. Therefore, it is also unclear if the 
stabilization method works for solitons with high power values. The moderate 
power assumption is valid for the optical fiber transmission application. 
However, this assumption is definitely insufficient for nonlinear optical 
waveguide setups, where intermediate and strongly nonlinear amplitude dynamics 
is observed, see Refs. \cite{PC2018B,P2009} for examples of such waveguide setups.

In the current paper, we address the important questions listed in the preceding 
paragraph. More specifically, we investigate the propagation of a highly energetic 
CQNLS soliton in a nonlinear optical waveguide with focusing cubic nonlinearity, 
defocusing quintic nonlinearity, and three weak perturbations due to weak 
linear gain-loss, cubic loss, and delayed Raman response. The shifting of the 
soliton's frequency is caused by the Raman perturbation. We assume a typical 
waveguide transmission system, in which the weak linear gain-loss counteracts 
the effects of weak cubic loss, such that the value of the soliton's amplitude 
tends to an equilibrium value. We consider two different types of linear gain-loss: 
(1) frequency-independent linear gain; (2) frequency-dependent linear gain-loss.  
For each type of gain-loss, we use perturbative calculations to derive 
equations for the dynamics of the CQNLS soliton's amplitude and frequency 
in the presence of the three perturbations. Furthermore, we perform numerical 
simulations with the perturbed CQNLS equation and characterize the distortion 
of the pulse shape and the Fourier spectrum. Additionally, we compare  
the simulations results with results of numerical simulations for 
propagation of a highly energetic CNLS soliton in similar waveguide 
setups in the absence of quintic nonlinearity.

In the case of waveguides with frequency-independent linear gain, we  
find that the transmission of an energetic CQNLS soliton and the dynamics of its 
amplitude and frequency parameters become unstable due to radiation emission. 
More specifically, our numerical simulations show that the soliton's and 
radiation's Fourier spectra become separated due to the perturbation-induced 
frequency shift experienced by the CQNLS soliton. However, due to the presence 
of frequency-independent linear gain and the lack of an efficient mechanism for 
radiation suppression, the emitted radiation forms a highly modulated hump 
that continues to grow and leads to transmission destabilization. A comparison 
with results of numerical simulations with the perturbed CNLS equation show that 
transmission instability is stronger for the CNLS soliton.

Significant enhancement of transmission stability of the energetic CQNLS soliton  
is demonstrated in waveguides with frequency-dependent linear gain-loss.   
In this case, our numerical simulations with the perturbed CQNLS equation 
show stable long-distance propagation of the soliton with almost no pulse 
distortion, and stable dynamics of its amplitude and frequency parameters in 
excellent agreement with the perturbation theory predictions. Suppression of 
pulse distortion is enabled by the separation of the soliton's and radiation's 
Fourier spectra due to the soliton's frequency shift, and by the efficient 
suppression of radiation emission due to the frequency-dependent linear gain-loss. 
Therefore, our numerical simulations and perturbative calculations provide the first 
demonstration of the transmission stabilization method for solitons of a nonintegrable 
nonlinear wave model, far from the integrable limit. Further numerical simulations with 
the perturbed CNLS equation indicate that the transmission of an energetic CNLS soliton 
in waveguides with frequency-independent linear gain-loss is stable as well. 
However, the pulse distortion and the perturbation-induced frequency shift 
of the CNLS soliton are significantly larger than the pulse distortion and 
the frequency shift of the CQNLS soliton. Therefore, our study also demonstrates 
that stabilization of energetic solitons by the method is significantly improved 
by the presence of defocusing quintic nonlinearity.

We choose to investigate soliton propagation in the presence of linear gain-loss 
and cubic loss, since these perturbations are very common in many optical waveguides, 
and are often the dominant dissipative processes in these systems. In this sense, 
linear gain-loss and cubic loss are central examples for dissipative perturbations 
in optical systems. Furthermore, linear gain or loss and cubic loss are also present 
in many pattern forming systems, which are described by the complex 
Ginzburg-Landau equation \cite{Hohenberg92,Kramer2002,Akhmediev2005,Kutz2006}. 
In optical waveguides, the cubic loss arises due to two-photon absorption or gain/loss 
saturation \cite{Agrawal2007,Dekker2007,Borghi2017}. Pulse propagation in the 
presence of cubic loss has been studied in many previous works, see Refs. 
\cite{Hohenberg92,Kramer2002,Agrawal2007,PC2018A,PCDN2009,PNC2010,PC2018B,Malomed89,Stegeman89,
Aceves92,Tsoy2001,Gaeta2012}, and references therein. The subject gained even more 
attention in recent years due to the importance of two-photon absorption in silicon 
nanowaveguides, which are expected to play a key role in many applications 
in optoelectronic devices \cite{Agrawal2007,Dekker2007,Borghi2017,Gaeta2008}.    
Delayed Raman response is another major nonlinear dissipative perturbation in optical 
waveguides. It is associated with the finite time of nonlinear response of the 
waveguide's medium to the propagation of light \cite{Agrawal2019,Gordon86,Kodama87}. 
Its main effects on the propagation of a single CNLS soliton in an optical waveguide 
are a continuous downshift of the soliton's frequency, which is known as the Raman 
self-frequency shift \cite{Agrawal2019,Gordon86,Kodama87,Mitschke86}, and the 
emission of radiation by the soliton \cite{Agrawal2019,Kaup95}. 
The Raman-induced self-frequency shift can be viewed as a nonlinear process, 
in which energy is transferred from higher frequency components of the pulse 
to its lower frequency components \cite{Agrawal2019,Gordon86}. More generally, 
the Raman perturbation can lead to destabilization of the transmission of 
CNLS solitons in optical waveguides \cite{Agrawal2019,PNT2016,Raman}. 
However, in Ref. \cite{PC2018A}, we showed that one can use the interplay 
between the Raman-induced frequency shift of a CNLS soliton and frequency-dependent 
linear gain-loss to stabilize the long-distance transmission of a CNLS soliton. 
In the current work, we show that this transmission stabilization method also works 
for solitons of the nonintegrable CQNLS equation with focusing cubic nonlinearity 
and defocusing quintic nonlinearity. Furthermore, we demonstrate that the introduction 
of defocusing quintic nonlinearity leads to remarkable enhancement of transmission 
stability of highly energetic solitons.

The rest of the paper is organized in the following manner. In Sec. \ref{dynamics1},  
we present the unperturbed CQNLS equation and discuss the properties of its soliton 
solutions. In Sec. \ref{dynamics2}, we study transmission stability and instability 
of energetic CQNLS and CNLS solitons in waveguides with weak frequency-independent 
linear gain, cubic loss, and delayed Raman response. In Sec. \ref{dynamics3}, 
we investigate transmission stabilization of energetic CQNLS solitons in waveguides 
with weak frequency-dependent linear gain-loss, cubic loss, and delayed Raman response. 
We also carry out a comparison with stabilization of energetic CNLS solitons 
in a similar waveguide setup without quintic nonlinearity. 
In Sec. \ref{conclusions}, we summarize our results. In Appendix \ref{appendA}, 
we derive the equations for dynamics of the CQNLS soliton's amplitude and frequency 
parameters for the different waveguide setups considered in the paper. 
Appendix \ref{appendB} is reserved for a description of the methods used to evaluate 
transmission stability from the numerical simulations results.

\section{Dynamics of CQNLS solitons in the presence of linear gain-loss, cubic loss, 
and delayed Raman response}
\label{dynamics}

\subsection{The unperturbed CQNLS equation with defocusing quintic nonlinearity}
\label{dynamics1}
Propagation of pulses of light in a nonlinear optical waveguide with 
second-order dispersion, focusing cubic nonlinearity, and defocusing quintic 
nonlinearity is described by the CQNLS equation 
\cite{Pushkarov96,Pelinovsky98,SP2004,Yang2010,Fibich2015}:   
\begin{eqnarray}&&
i\partial_z\psi+\partial_t^2\psi+2|\psi|^2\psi-\epsilon_q|\psi|^4\psi=0. 
\label{cqnls1}
\end{eqnarray}     
In Eq. (\ref{cqnls1}), $\psi$ is proportional to the envelope of the electric field, 
$z$ is distance, $t$ is time, and $\epsilon_{q}>0$ is the defocusing quintic 
nonlinearity coefficient \cite{dimensions1}. Equation (\ref{cqnls1}) is a nonintegrable 
nonlinear wave model \cite{Yang2010,Pelinovsky98}, and therefore its solutions can exhibit 
dynamical behavior that is different from the behavior of solutions of integrable models. 
The equation supports soliton solutions of the form 
\cite{Pushkarov96,Pelinovsky98,Yang2010,Gagnon89,PDC2005} 
\begin{equation}
\psi_{s}(t,z)=\Psi_{s}(x)\exp(i\chi), 
\label{cqnls2}
\end{equation}     
where  
\begin{eqnarray} &&
\Psi_{s}(x) =\sqrt{2}\eta
\left[(1-\eta^{2}/\eta_{m}^{2})^{1/2}\cosh(2x)+1\right]^{-1/2} , 
\label{cqnls3}
\end{eqnarray}   
$\chi = \alpha-\beta(t-y)+(\eta^2-\beta^2)z$, $x=\eta(t-y+2\beta z)$, 
and $\eta_{m}=(4\epsilon_{q}/3)^{-1/2}$. The parameters $\eta$, $\beta$, 
$y$, and $\alpha$ in these solutions are related to the soliton's amplitude, 
frequency, position, and phase, respectively. The soliton solutions (\ref{cqnls2}) 
exist for amplitude parameter values satisfying $0<\eta<\eta_{m}$. Additionally, 
for $\eta$ values near $\eta_{m}$ the solitons possess high power and attain a 
flat-top shape \cite{PDC2005,PCDN2009}.

Stability properties of solitons of unperturbed CQNLS equations were 
studied in a number of works, see Refs. 
\cite{Yang2010,Pelinovsky98,Yang2000,Hohenberg92,Ohta95,Martel21,Akhmediev99}, 
and references therein. In particular, the orbital stability of the solitons (\ref{cqnls2}) 
of Eq. (\ref{cqnls1}) was proved in Ref. \cite{Ohta95} for any $0<\eta<\eta_{m}$. 
Furthermore, asymptotic stability of these solitons was established 
in Ref. \cite{Martel21} for small-to-intermediate $\eta$ values. 
In Ref. \cite{Akhmediev99}, stability was shown by considering 
the Hamiltonian-versus-energy diagram for the solitons. 
Additionally, in Ref. \cite{Pelinovsky98} it was shown that there are 
no internal modes for the solitons (\ref{cqnls2}), and this leads to 
considerable simplification of the proof of their asymptotic 
stability \cite{Martel21}. Thus, the works in Refs. 
\cite{Ohta95,Martel21,Akhmediev99,Pelinovsky98} showed that the 
solitons (\ref{cqnls2}) of the CQNLS equation (\ref{cqnls1}) exhibit 
certain enhanced stability properties. 
In the current paper we address another important question concerning 
the stability properties of the solitons (\ref{cqnls2}). More specifically, 
we investigate whether propagation of these solitons in the presence of 
perturbations, which typically lead to radiative instability, can be 
stabilized by the interplay between frequency-dependent linear gain-loss 
and perturbation-induced shifting of the solitons' frequency. 
Furthermore, we compare transmission stabilization of the CQNLS solitons 
by this method with transmission stabilization for CNLS solitons 
in the presence of the same destabilizing perturbations.

The power ${\cal M}$, the momentum ${\cal P}$, and the Hamiltonian 
${\cal H}$ are conserved by the unperturbed CQNLS equation (\ref{cqnls1}). 
These physical quantities are expressed in terms of $\psi$ in the following 
manner \cite{Yang2010}: 
\begin{eqnarray}&&
\!\!\!\!\!\!\!\!\!\!
{\cal M}=
\int_{-\infty}^{\infty} \!\! dt \, |\psi|^{2}, 
\label{cqnls4}
\end{eqnarray}       
\begin{eqnarray}&&
\!\!\!\!\!\!\!\!\!\!
{\cal P}= i\int_{-\infty}^{\infty} \!\! dt \, 
\left(\psi \partial_{t}\psi^{*} - \psi^{*} \partial_{t}\psi\right), 
\label{cqnls5}
\end{eqnarray} 
and 
\begin{eqnarray}&&
\!\!\!\!\!\!\!\!\!\!
{\cal H}=
\int_{-\infty}^{\infty} \!\! dt \, 
\left( |\partial_{t}\psi|^{2}/2 - |\psi|^{2}/2 
+ \epsilon_{q}|\psi|^{4}/6 \right).  
\label{cqnls6}
\end{eqnarray} 
In general, ${\cal M}$, ${\cal P}$, and ${\cal H}$, are not conserved 
in the presence of {\it dissipative perturbations} to Eq. (\ref{cqnls1}). 
As a result, in this case, the values of the soliton's parameters $\eta$ 
and $\beta$ might change with distance $z$. One can then obtain the 
equations for $d\eta/dz$ and $d\beta/dz$ by employing energy and momentum 
balance calculations, which are based on the expressions for ${\cal M}$ and 
${\cal P}$ in Eqs. (\ref{cqnls4}) and (\ref{cqnls5}). In the next subsections, 
we study the dynamics of the CQNLS soliton and its parameters $\eta$ 
and $\beta$ in the presence of dissipative perturbations due to linear gain or 
loss, cubic loss, and delayed Raman response.

\subsection{Soliton dynamics in the presence of frequency-independent linear gain, cubic loss, and delayed Raman response}

\label{dynamics2}

We consider soliton propagation in a nonlinear optical waveguide  
with focusing cubic nonlinearity, defocusing quintic nonlinearity, weak 
frequency-independent linear gain, cubic loss, and delayed Raman response.   
Quintic nonlinearity and delayed Raman response are both first-order 
corrections to cubic nonlinearity, and therefore, including both effects 
in the same propagation equation is consistent in terms of the mathematical 
modeling. The propagation is described by the following perturbed 
CQNLS equation \cite{quintic_model}: 
\begin{eqnarray}&&
i\partial_z\psi+\partial_t^2\psi+2|\psi|^2\psi-\epsilon_q|\psi|^4\psi=
ig_{0}\psi/2 - i\epsilon_{3}|\psi|^2\psi
+\epsilon_{R}\psi\partial_{t}|\psi|^2 ,  
\label{cqnls11}
\end{eqnarray} 
where the third term on the right hand side is due to delayed Raman response. 
The linear gain, cubic loss, and Raman coefficients, $g_{0}$, $\epsilon_{3}$, 
and $\epsilon_{R}$, satisfy $g_{0}>0$,  $0 < \epsilon_{3} \ll 1$, and 
$0 < \epsilon_{R} \ll 1$ \cite{dimensions2,Chi89}.

The equations for the dynamics of the amplitude and frequency parameters 
of the CQNLS soliton are obtained in Appendix \ref{appendA} by employing 
energy balance and momentum balance calculations. These calculations yield 
the following equations: 
\begin{eqnarray} &&
\!\!\!\!\!\!\!\!
\frac{d \eta}{dz}=
\frac{\left(\eta_{m}^{2}-\eta^{2}\right)}{\eta_{m}}
\left\{g_{0}\mbox{arctanh}\left(\frac{\eta}{\eta_{m}}\right)
-4\epsilon_{3}\eta_{m}^{2}\left[\mbox{arctanh}\left(\frac{\eta}{\eta_{m}}\right)- 
\frac{\eta}{\eta_{m}} \right]\right\},     
 \label{cqnls12}
\end{eqnarray}   
and  
\begin{eqnarray} &&
\!\!\!\!\!\!\!\!
\frac{d \beta}{dz}=
\frac{-4\epsilon_{R}\eta_{m}^{4}}{3\mbox{arctanh}\left(\eta/\eta_{m}\right)}
\left[\frac{3\eta}{\eta_{m}} - \frac{2\eta^{3}}{\eta_{m}^{3}}
-3\left(1-\frac{\eta^{2}}{\eta_{m}^{2}}\right)
\mbox{arctanh}\left(\frac{\eta}{\eta_{m}}\right) \right].     
\label{cqnls13}
\end{eqnarray}    
Note that the amplitude shift in Eq. (\ref{cqnls12}) is induced only 
by the effects of linear gain and cubic loss. Additionally, 
the frequency shift in Eq. (\ref{cqnls13}) is induced solely by the 
Raman effect. Thus, Eq. (\ref{cqnls13}) describes the soliton's Raman 
self-frequency shift in optical waveguides with focusing cubic nonlinearity 
and defocusing quintic nonlinearity.

In optical waveguide transmission systems it is usually desired to realize 
stable steady-state transmission with constant pulse amplitudes 
\cite{Agrawal2019,NP2010A}. We therefore require that $\eta=\eta_{0}$ with 
$0 < \eta_{0} < \eta_{m}$ is a stable equilibrium point of Eq. (\ref{cqnls12}), 
and obtain the following expression for $g_{0}$: 
\begin{equation}
g_{0} = 4\epsilon_{3}\eta_{m}^{2}
\left[1-\frac{\eta_{0}/\eta_{m}}
{\mbox{arctanh}\left(\eta_{0}/\eta_{m}\right)}\right].  
\label{cqnls14}
\end{equation}   
Substitution of  Eq. (\ref{cqnls14}) into  Eq. (\ref{cqnls12}) yields 
\begin{eqnarray} &&
\!\!\!\!\!\!\!\!
\frac{d \eta}{dz}=
4\epsilon_{3}\eta_{m}^{3}
\left(1-\frac{\eta^{2}}{\eta_{m}^{2}}\right)
\mbox{arctanh}\left(\frac{\eta}{\eta_{m}}\right)
\left[\frac{\eta/\eta_{m}}{\mbox{arctanh}\left(\eta/\eta_{m}\right)}
-\frac{\eta_{0}/\eta_{m}}{\mbox{arctanh}\left(\eta_{0}/\eta_{m}\right)}\right]. 
\label{cqnls15}
\end{eqnarray}   
One can show that $\eta=\eta_{0}$ is the only equilibrium point of 
Eq. (\ref{cqnls15}) with a nonnegative amplitude value and that it is a stable 
equilibrium point. In the current work, we are particularly interested in 
transmission stabilization of energetic CQNLS solitons, which occurs for 
$\eta_{0}$ values close to $\eta_{m}$. This means that we are studying 
transmission stabilization near the flat-top soliton limit of Eq. (\ref{cqnls1}),   
which is also the front formation limit \cite{PDC2005,PCDN2009}.

It is useful to consider the form of Eqs. (\ref{cqnls15}) and  (\ref{cqnls13}) 
in the following two cases. (1) The limit $\epsilon_{q} \rightarrow 0^{+}$ 
$(\eta_{m} \rightarrow \infty)$ that corresponds to the CNLS limit. 
(2) The case where $\eta$ is close to $\eta_{0}$, and $\eta_{0}$ is 
close to $\eta_{m}$, i.e., for highly energetic CQNLS solitons 
near the flat-top soliton limit. In the limit 
$\epsilon_{q} \rightarrow 0^{+}$, we retrieve the equations for the 
amplitude shift and the Raman-induced self-frequency shift of the 
CNLS soliton (see Refs. \cite{PC2018A,Mitschke86,Gordon86,Kodama87}): 
\begin{equation}
\frac{d\eta}{dz}\simeq
\frac{4}{3}\epsilon_{3}\left(\eta_{0}^{2} - \eta^{2}\right)\eta, 
\label{cqnls16}
\end{equation} 
and 
\begin{equation}
\frac{d\beta}{dz} \simeq -\frac{8}{15}\epsilon_{R}\eta^{4}.
\label{cqnls17}
\end{equation}    
In case (2), we denote $\eta=\eta_{0}+\delta\eta$, where $|\delta\eta| \ll 1$, 
$\delta\eta < \eta_{m} - \eta_{0}$, and $\eta_{0} \lesssim \eta_{m}$. We then 
expand the right hand sides of Eqs. (\ref{cqnls15}) and  (\ref{cqnls13}) in 
Taylor series about $\eta_{0}$, keeping only the leading term in each 
expansion. This approximation amounts to a linearization of Eq. (\ref{cqnls15}) 
about the equilibrium point $\eta=\eta_{0}$, and to a calculation of the 
constant rate of change of $\beta$ for $\eta=\eta_{0}$. We obtain: 
\begin{eqnarray} &&
\!\!\!\!\!\!\!\!
\frac{d \eta}{dz} \simeq  
4\epsilon_{3}
\left[\left(\eta_{m}^{2}-\eta_{0}^{2}\right)
-\frac{\eta_{0}\eta_{m}}{\mbox{arctanh}\left(\eta_{0}/\eta_{m}\right)}\right]
\delta\eta,
\label{cqnls18}
\end{eqnarray}               
and  
\begin{equation}
\frac{d\beta}{dz} \simeq 
\frac{-4\epsilon_{R}\eta_{m}^{4}}{3\mbox{arctanh}\left(\eta_{0}/\eta_{m}\right)}
\left[\frac{3\eta_{0}}{\eta_{m}} - \frac{2\eta_{0}^{3}}{\eta_{m}^{3}}
-3\left(1-\frac{\eta_{0}^{2}}{\eta_{m}^{2}}\right)
\mbox{arctanh}\left(\frac{\eta_{0}}{\eta_{m}}\right) \right]. 
\label{cqnls19}
\end{equation}  
Note that for $\eta_{0}$ values near $\eta_{m}$, the value of the 
expression inside the square brackets on the right hand side of 
Eq. (\ref{cqnls19}) is close to 1, while the value of the denominator 
is much larger than 1. Therefore, according to the perturbation theory, 
in case (2), the rate of change of the CQNLS soliton's frequency due to the 
Raman effect tends to a {\it small} negative constant value at large distances.  
This behavior is strikingly different from the behavior expected for the 
CNLS soliton. Indeed, it follows from Eq. (\ref{cqnls17}) that the rate 
of change of the CNLS soliton's frequency is not necessarily small for 
large $\eta$ and $\eta_{0}$ values.

{\it Numerical simulations}.      
According to Eq. (\ref{cqnls15}), the dynamics of the CQNLS soliton's 
amplitude is expected to be stable. However, this prediction is based 
on an energy-balance calculation, which neglects radiation emission 
effects. The latter effects can become important at intermediate and 
large distances, and they can lead to distortion of the soliton's shape 
and to the breakdown of the energy-balance approximation used in the 
derivation of Eq. (\ref{cqnls15}). This is particularly true for the 
optical waveguide setup considered here, since the presence of linear 
gain leads to unstable growth of small amplitude waves that are associated 
with radiation. For these reasons, it is important to check the validity 
of the predictions obtained with Eq. (\ref{cqnls13}) and Eq. (\ref{cqnls15}), 
and the stability of the CQNLS soliton in the presence of perturbations by 
numerical simulations with Eq. (\ref{cqnls11}).

The setup of the numerical simulations with the perturbed CQNLS equation 
(\ref{cqnls11}) is somewhat similar to the setup used in Ref. \cite{PC2018A} 
for simulations with the perturbed CNLS equation. More 
specifically, we solve Eq. (\ref{cqnls11}) numerically on a time domain 
$[t_{\mbox{min}},t_{\mbox{max}}]=[-400,400]$ using the split-step method 
with periodic boundary conditions \cite{Agrawal2019,Yang2010}.     
The initial input is in the form of the CQNLS soliton of Eq. (\ref{cqnls2}) 
with amplitude parameter $\eta(0)$, frequency $\beta(0)=0$, position $y(0)=0$, 
and phase $\alpha(0)=0$. Since we are interested in transmission stabilization 
near the flat-top soliton limit, we choose an $\eta_{0}$ value that is close 
to $\eta_{m}$. Therefore, the soliton's shape at equilibrium is not far from a 
flat-top shape. Additionally, the soliton's power at equilibrium is 
significantly larger than the equilibrium value of the soliton's power 
that was used in Ref. \cite{PC2018A} for CNLS solitons.     
As a typical example, we consider here the results of the simulations 
with $\epsilon_{q}=0.5$, $\epsilon_{3}=0.01$, $\epsilon_{R}=0.04$, 
$\eta_{0}=1.2$, and $\eta(0)=0.9$. We emphasize, however, that similar 
results are obtained with other values of the physical parameters.      
Note that the value $\eta_{0}=1.2$ is indeed close to the $\eta_{m}$ value, 
$\eta_{m}=1.2247...$. Additionally, the CQNLS soliton's power at equilibrium 
is 5.6172..., which is significantly larger than the value 2 that was used 
in Ref. \cite{PC2018A} for CNLS solitons. The application of periodic 
boundary conditions means that the simulations describe propagation 
in a closed optical waveguide loop. This setup is very relevant for applications, 
since most long-distance optical waveguide transmission experiments 
are carried out in closed loops, see, e.g., 
Refs. \cite{Mollenauer2006,Mollenauer97,Nakazawa2000,Nakazawa91,Mollenauer2003}.

Transmission quality and pulse-shape distortion are measured from the 
simulations results by the transmission quality integral $I(z)$, which 
is defined in Eq. (\ref{Iz4}) in Appendix \ref{appendB}. $I(z)$ measures 
the deviation of the numerically obtained pulse shape $|\psi^{(num)}(t,z)|$ 
from the perturbation theory prediction $|\psi^{(th)}(t,z)|$ 
that is given by Eq. (\ref{Iz1}). Therefore, $I(z)$ measures both distortion 
in the pulse shape due to radiation emission and deviations in the numerically 
obtained values of the soliton's parameters from the values predicted by 
Eqs. (\ref{cqnls13}) and Eq. (\ref{cqnls15}). Further insight into transmission 
quality and pulse-shape distortion is obtained by measuring the distance $z_{q}$ 
at which the value of $I(z)$ first exceeds 0.075. We refer to $z_{q}$ as the 
transmission quality distance. Additionally, we characterize soliton stability 
and pulse-shape distortion at larger distances, by running the simulations up 
to a final distance $z_{f}$ at which the value of $I(z)$ first 
exceeds 0.655 \cite{thresholds}. For the physical parameter values specified 
in the preceding paragraph, we find $z_{q}=140$ and $z_{f}=300$.

\begin{figure}[ptb]
\begin{tabular}{cc}
\epsfxsize=8.1cm  \epsffile{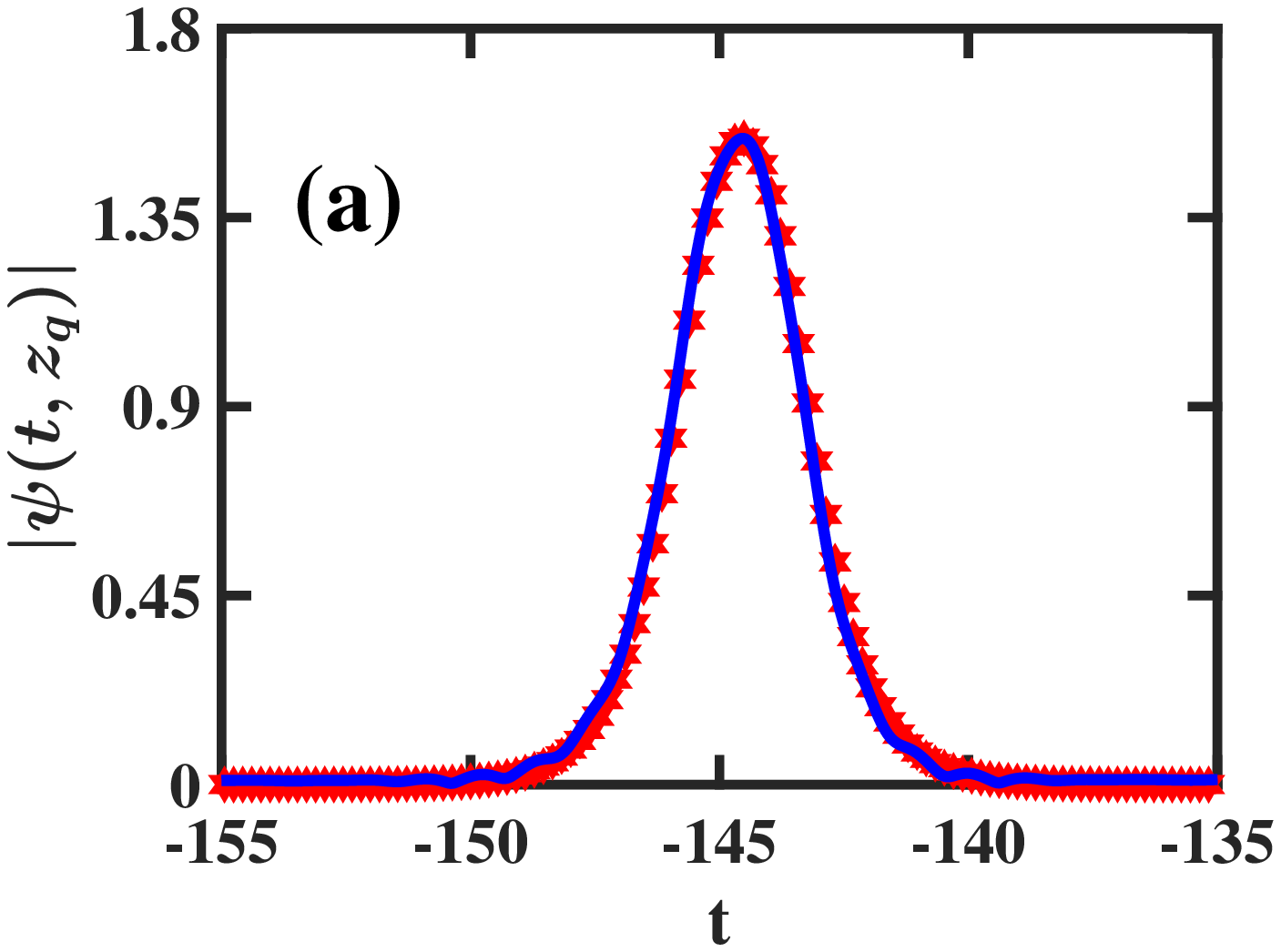} &
\epsfxsize=8.1cm  \epsffile{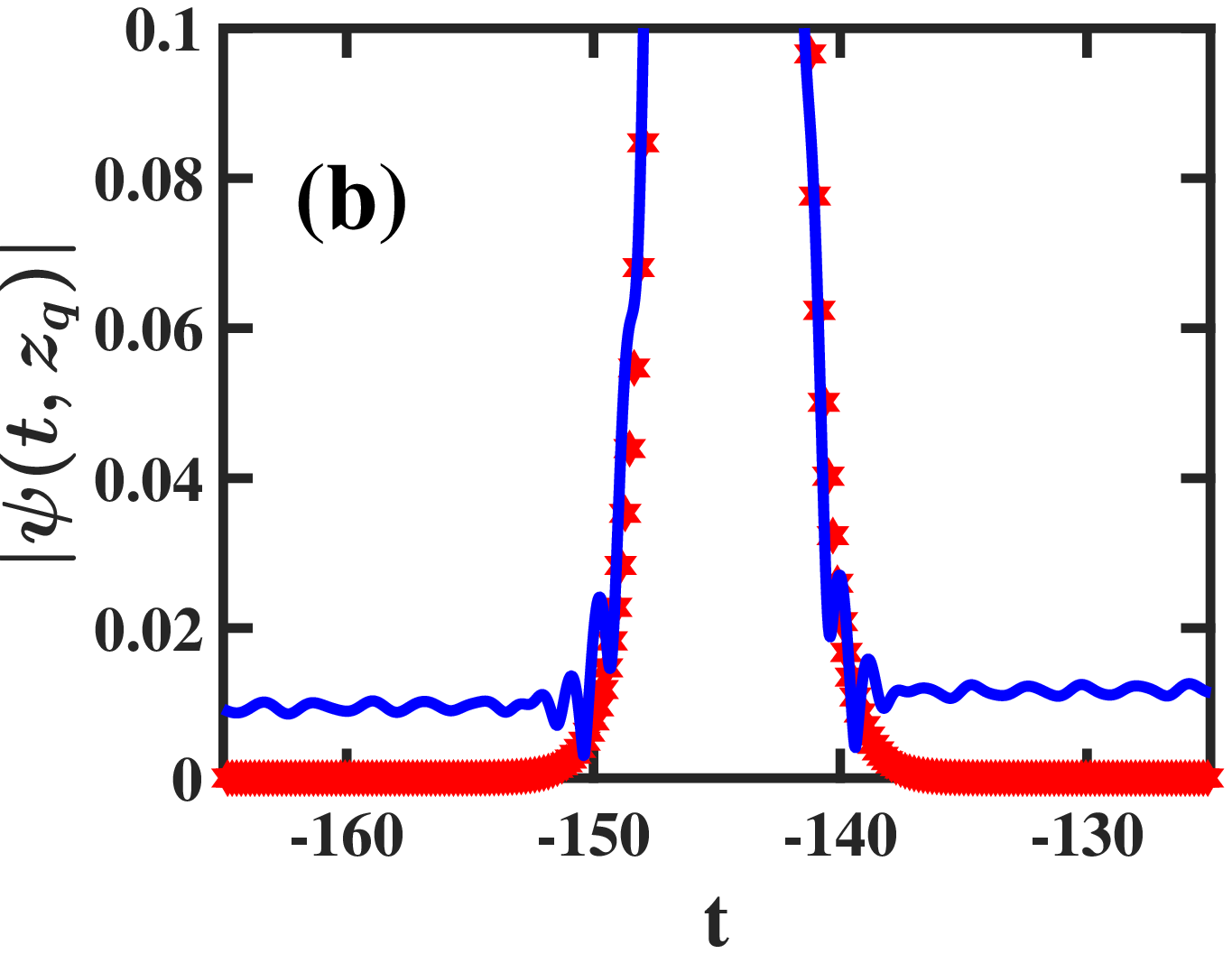} \\
\epsfxsize=8.1cm  \epsffile{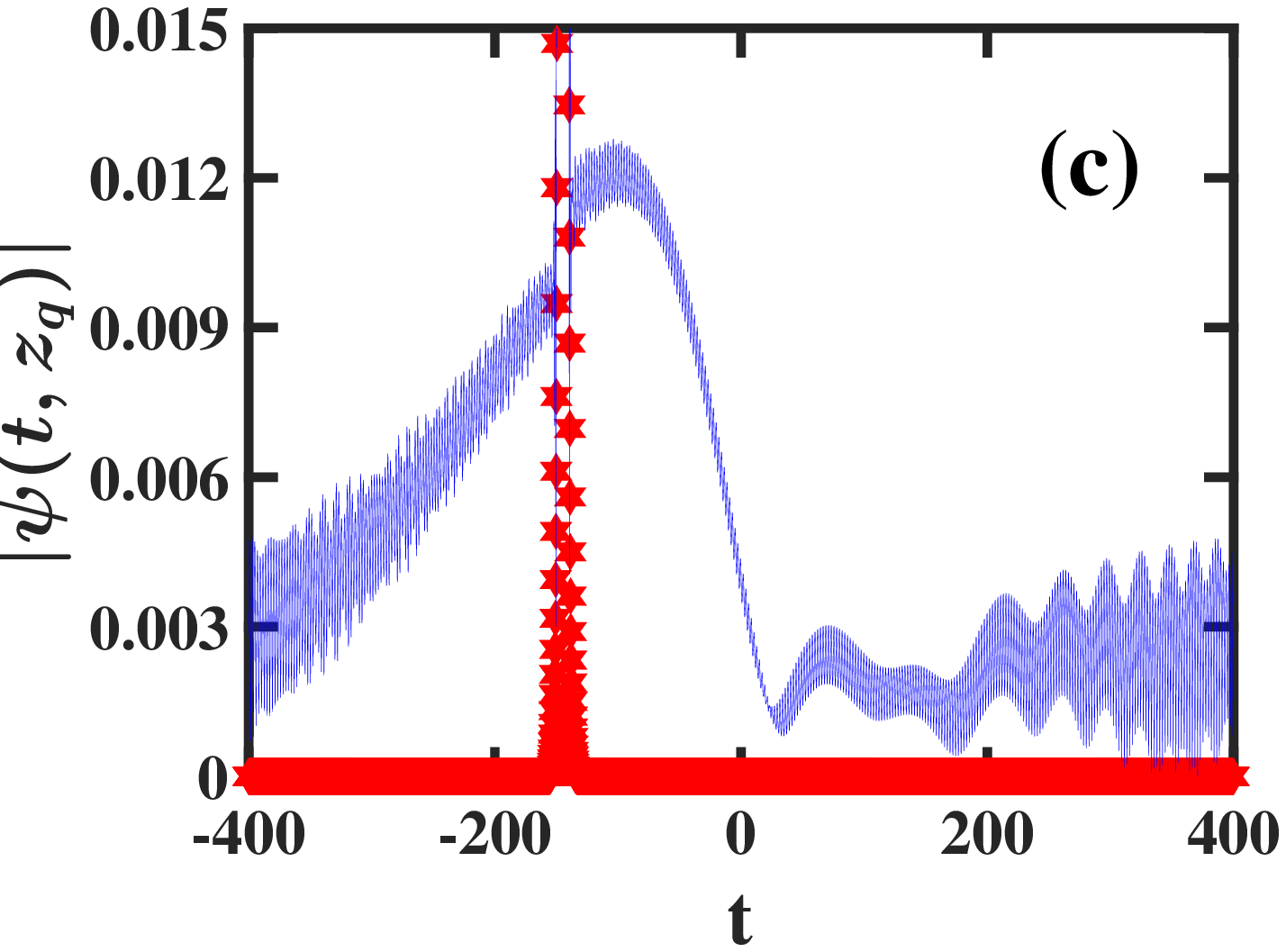} &
\epsfxsize=8.1cm  \epsffile{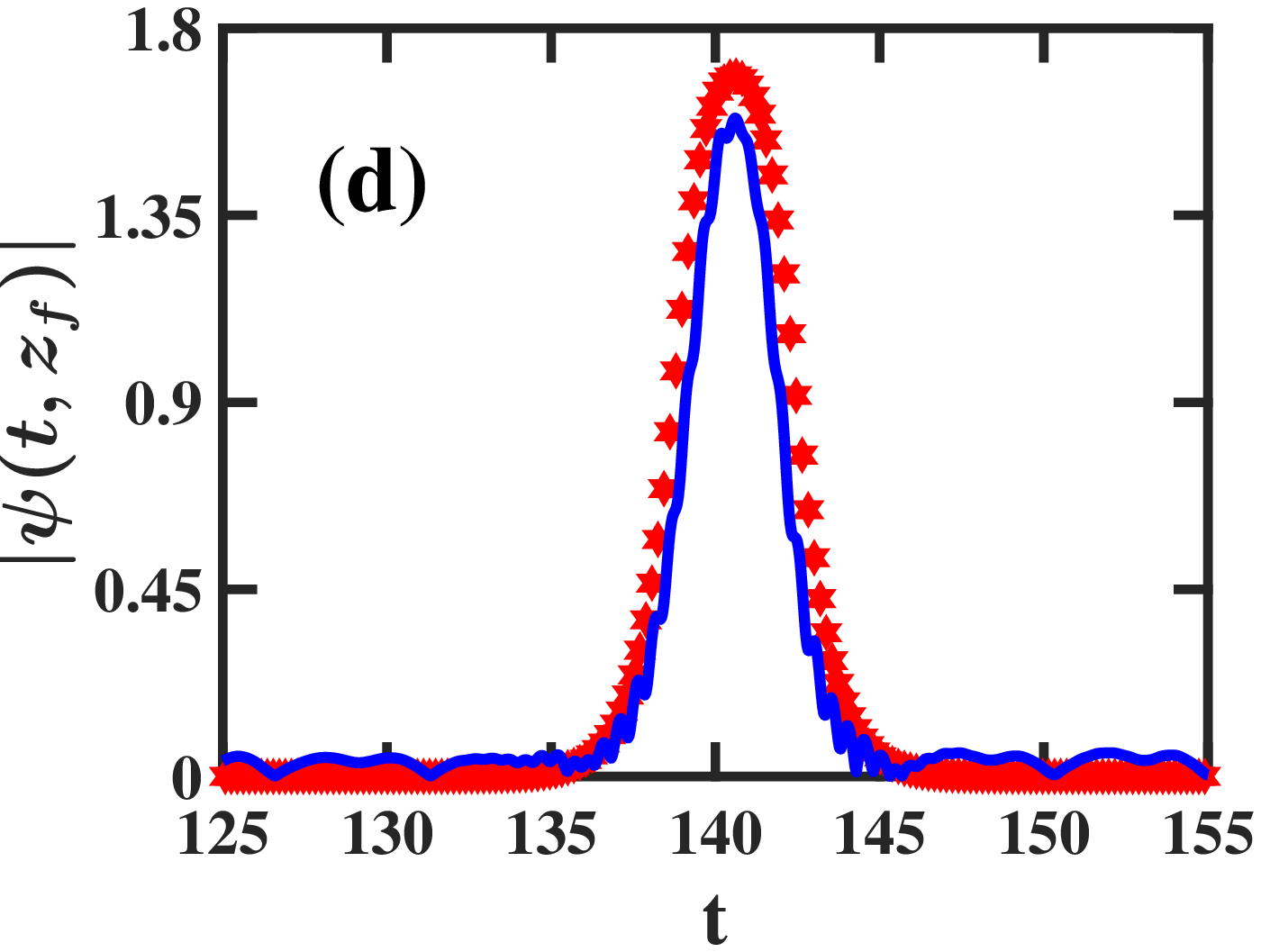} \\ 
\epsfxsize=8.1cm  \epsffile{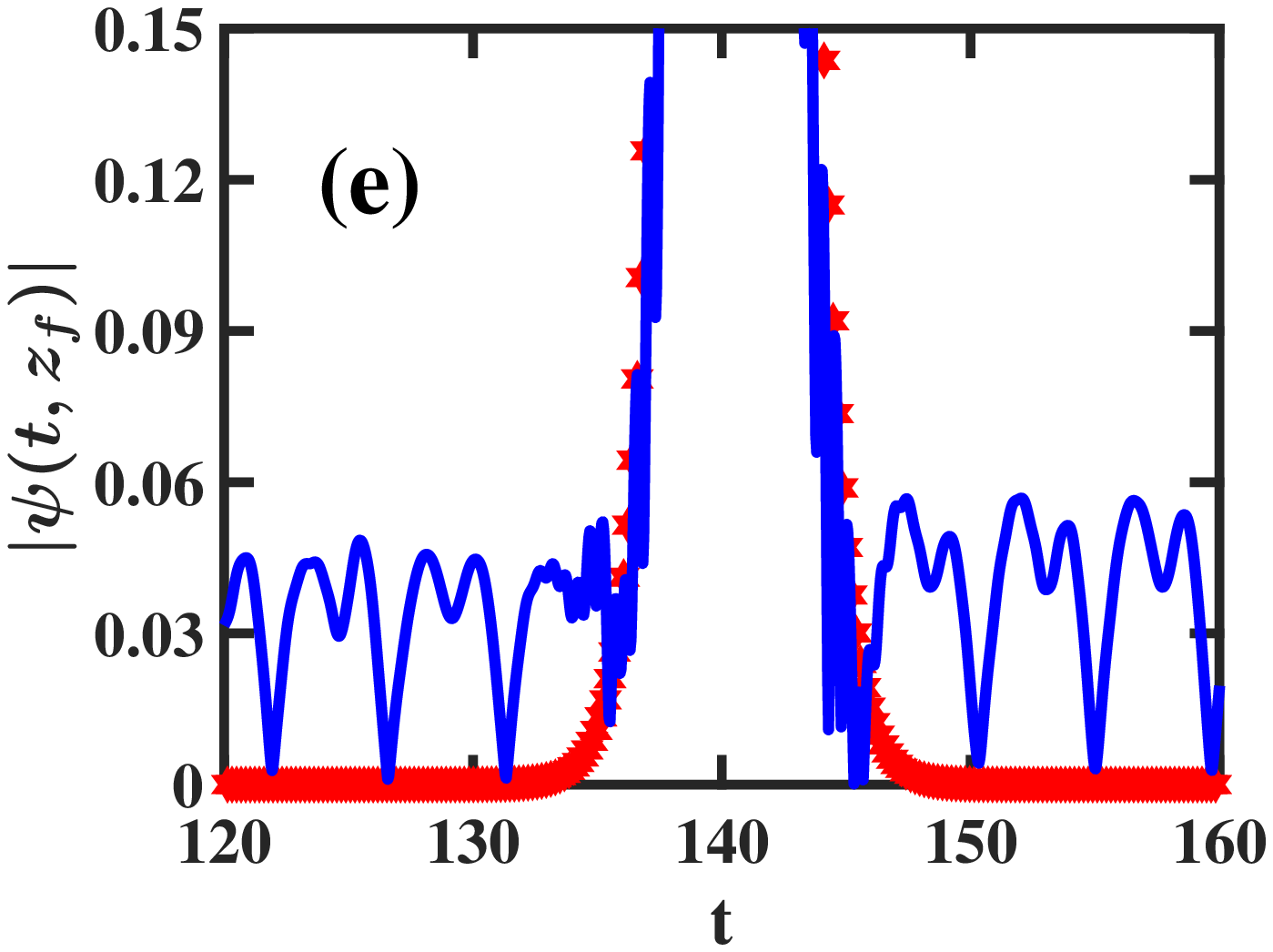} &
\epsfxsize=8.1cm  \epsffile{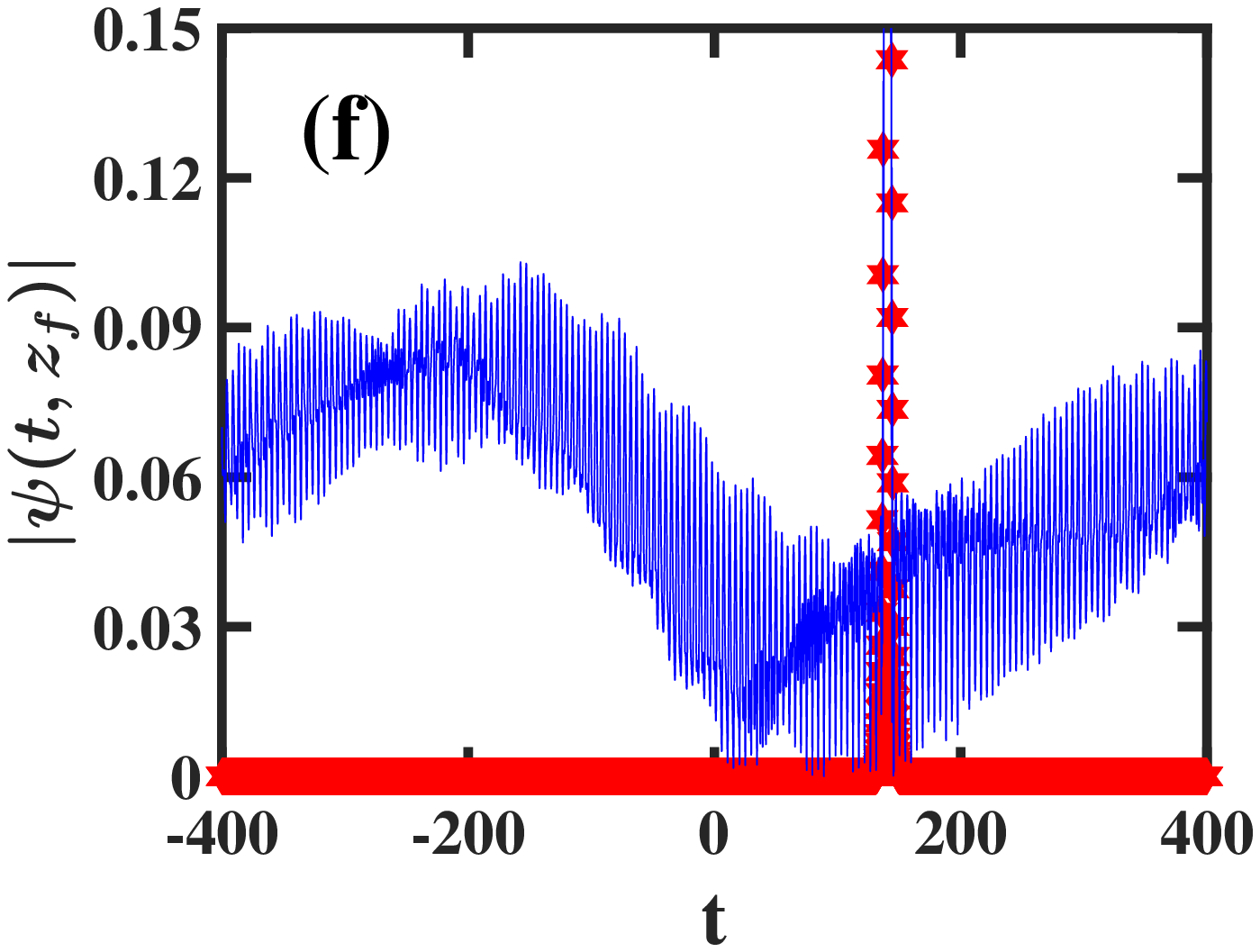}
\end{tabular}
\caption{The pulse shape $|\psi(t,z)|$ at $z_{q}=140$ [(a), (b), (c)] and at   
$z_{f}=300$ [(d), (e), (f)] for propagation of a CQNLS soliton in a waveguide loop 
with weak frequency-independent linear gain, cubic loss, and delayed Raman response. 
The physical parameter values are $\epsilon_{q}=0.5$, $\epsilon_{3}=0.01$, 
$\epsilon_{R}=0.04$, $\eta_{0}=1.2$, and $\eta(0)=0.9$. 
The solid blue curve represents the result obtained by the numerical 
simulation with Eq.  (\ref{cqnls11}). The red stars correspond to 
the perturbation theory prediction, obtained with Eqs. (\ref{Iz1}) 
and (\ref{cqnls15}).}
\label{fig1}
\end{figure}

\begin{figure}[ptb]
\begin{tabular}{cc}
\epsfxsize=8.5cm  \epsffile{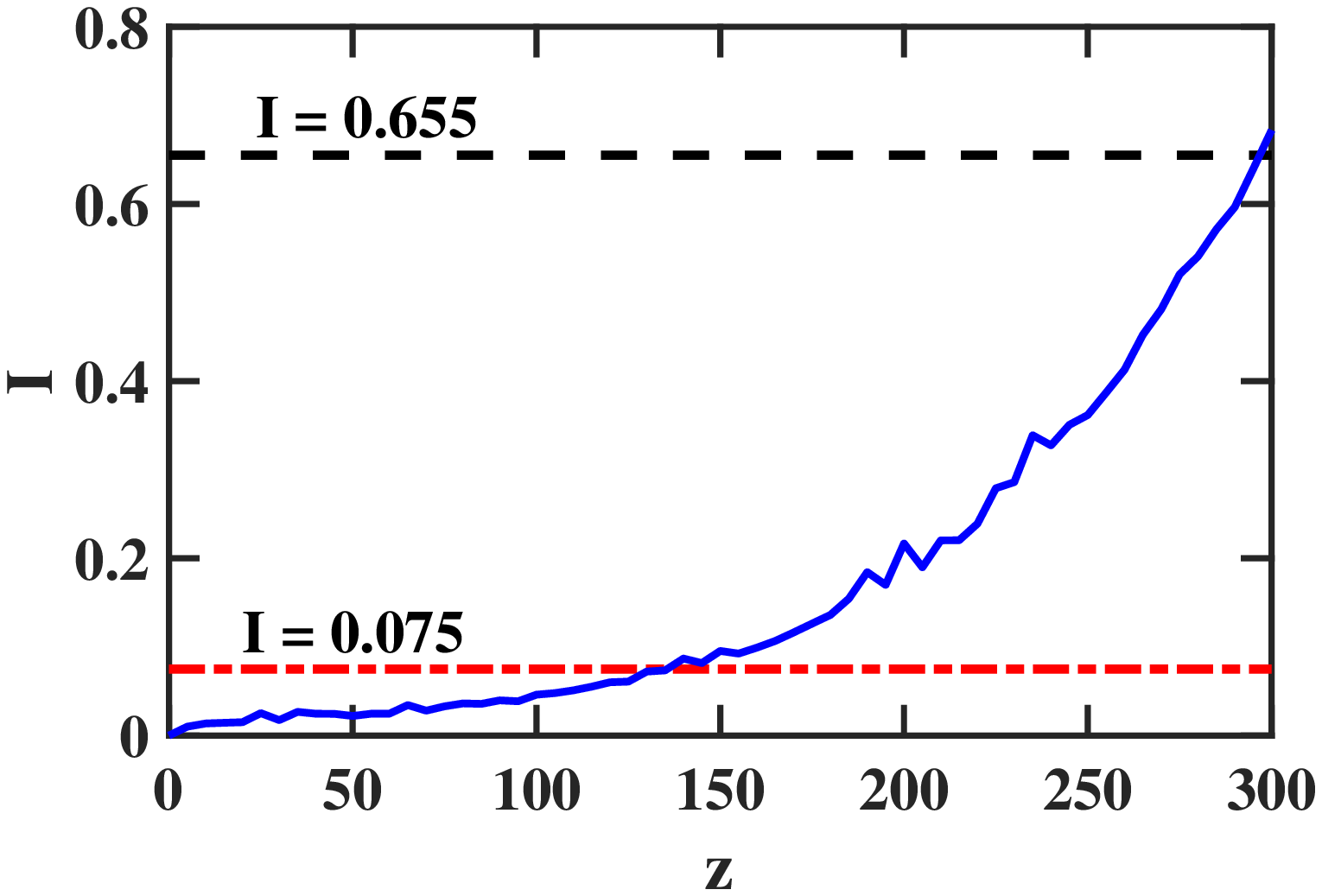} 
\end{tabular}
\caption{The $z$ dependence of the transmission quality integral $I(z)$ obtained 
by the numerical simulation with Eq. (\ref{cqnls11}) for the same waveguide 
setup as in Fig. \ref{fig1}. The solid blue curve stands for the simulation's result. 
The dashed black and dashed-dotted red horizontal lines correspond to $I=0.655$ 
and $I=0.075$, respectively.}
\label{fig2}
\end{figure}

The pulse shape $|\psi(t,z)|$ obtained in the numerical simulation at 
$z=z_{q}$ and at $z=z_{f}$ is shown in Fig. \ref{fig1}. A comparison 
with the perturbation theory prediction of Eqs. (\ref{Iz1}) 
and (\ref{cqnls15}) is also shown. As seen in Fig. \ref{fig1}(a), 
the pulse shape obtained in the simulation at $z=z_{q}$ is close to 
the perturbation theory prediction. However, the comparison in Figs. \ref{fig1}(b) 
and \ref{fig1}(c) for small $|\psi(t,z_{q})|$ values shows 
that a noticeable radiative tail exists already at this distance. 
This radiative tail, which is induced by the three perturbation terms 
in Eq. (\ref{cqnls11}), is highly oscillatory and is spread over the 
entire computational domain. It attains the form of a highly 
modulated hump. As the soliton continues to propagate, 
the radiative tail continues to grow [see Figs. \ref{fig1}(d)-\ref{fig1}(f)]. 
Furthermore, as seen in Fig. \ref{fig2}, the growth of the radiative tail 
leads to the increase of the value of the transmission quality 
integral $I(z)$ with increasing $z$.

\begin{figure}[ptb]
\begin{center}
\begin{tabular}{cc}
\epsfxsize=8.2cm  \epsffile{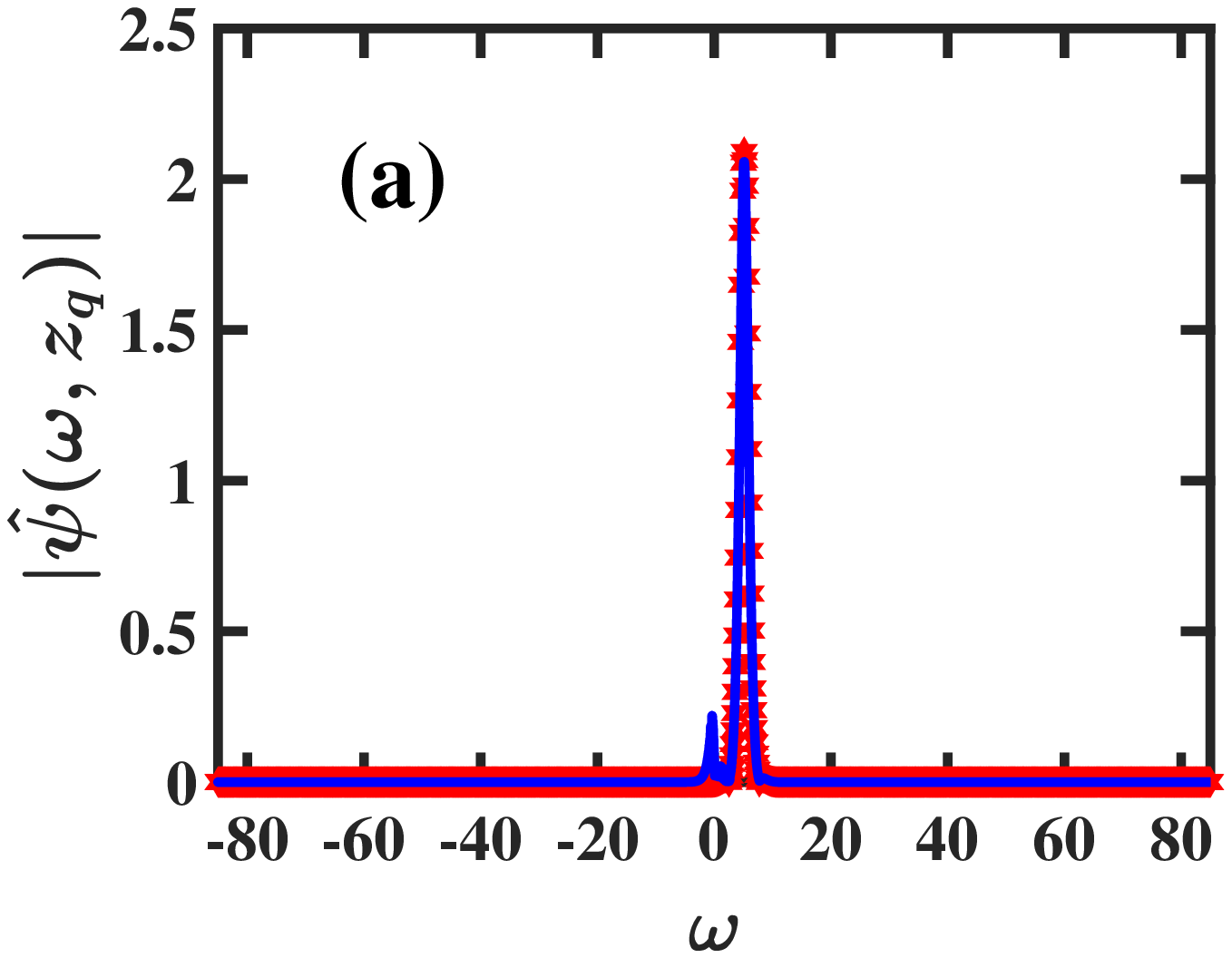} &
\epsfxsize=8.2cm  \epsffile{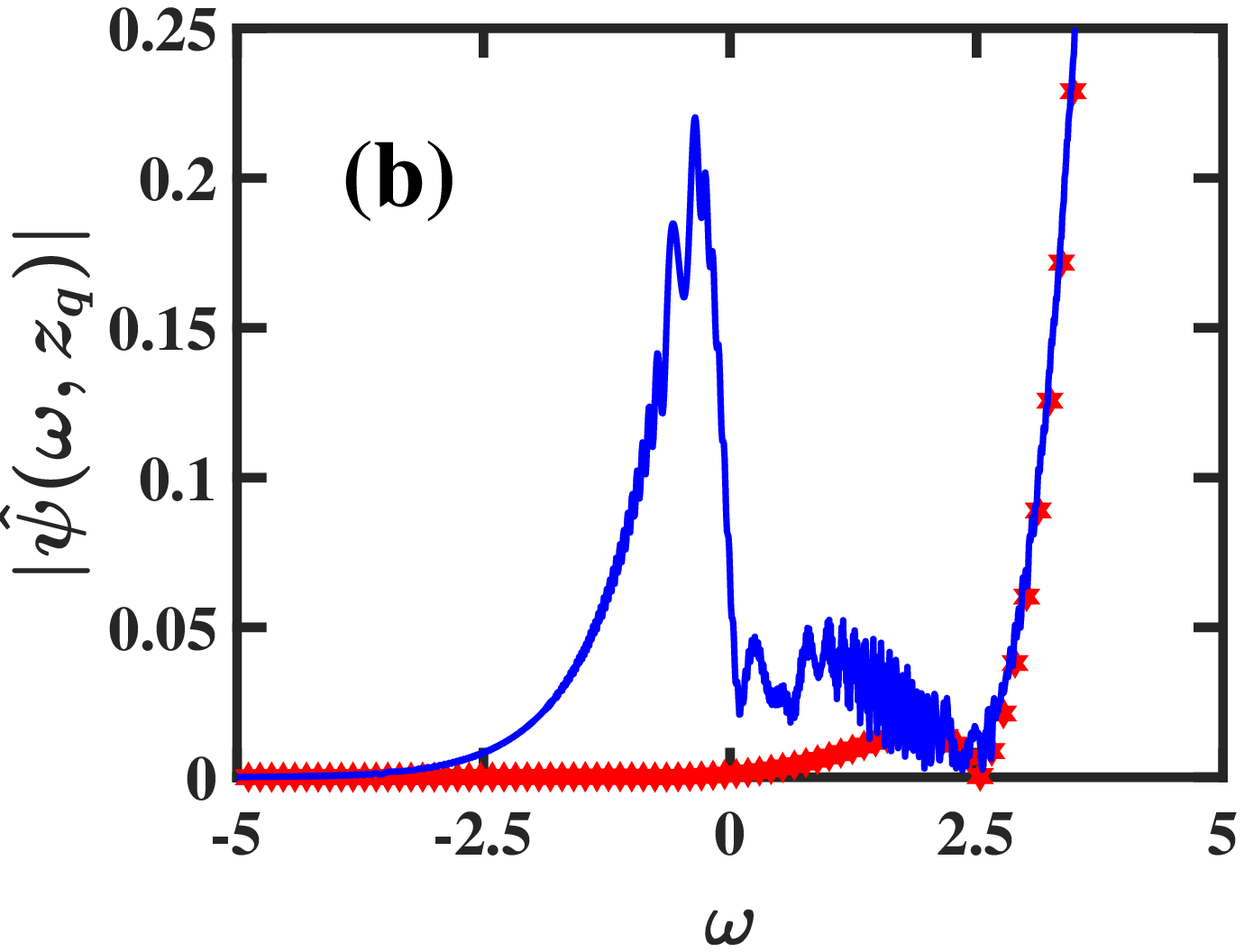} \\
\epsfxsize=8.2cm  \epsffile{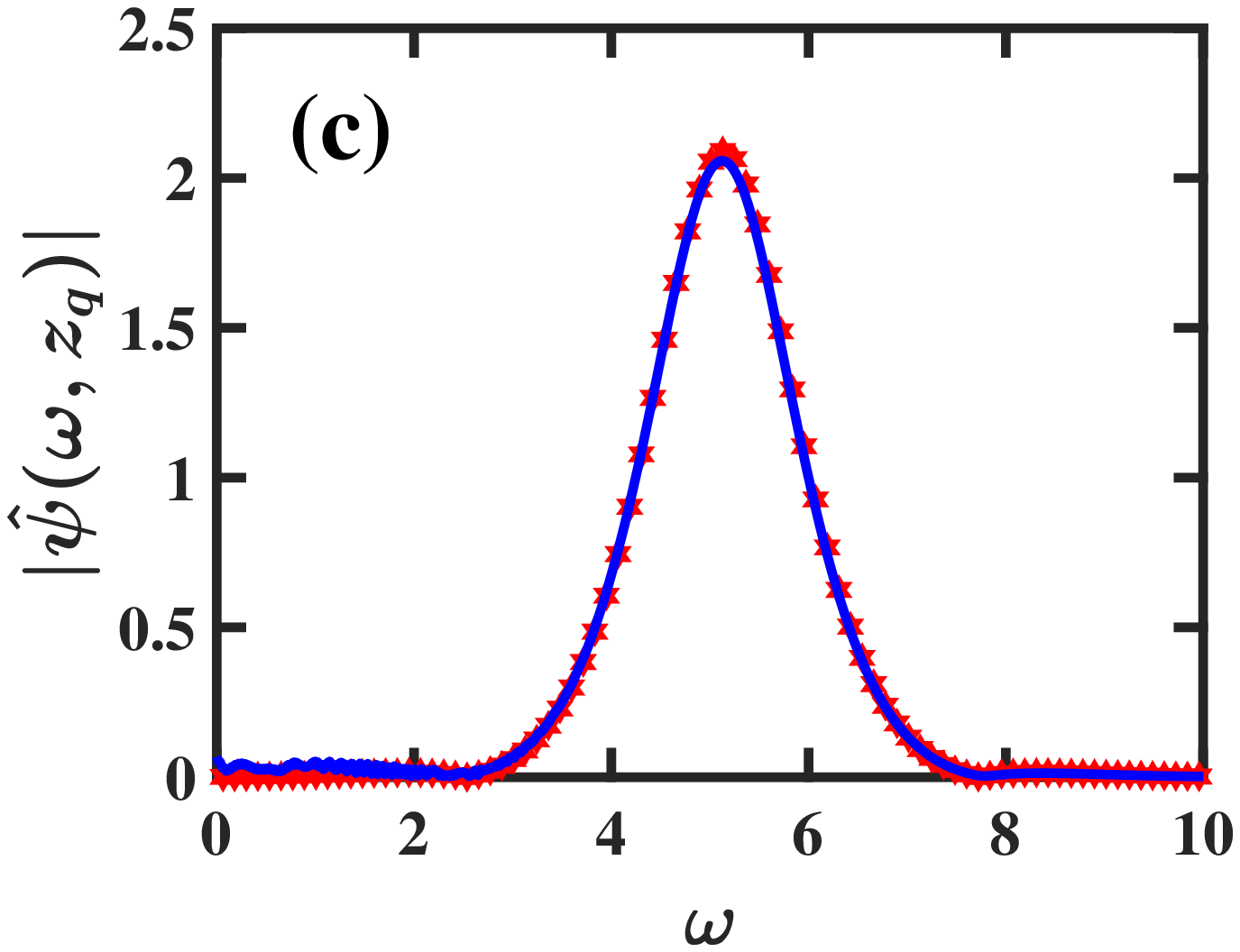} &
\epsfxsize=8.2cm  \epsffile{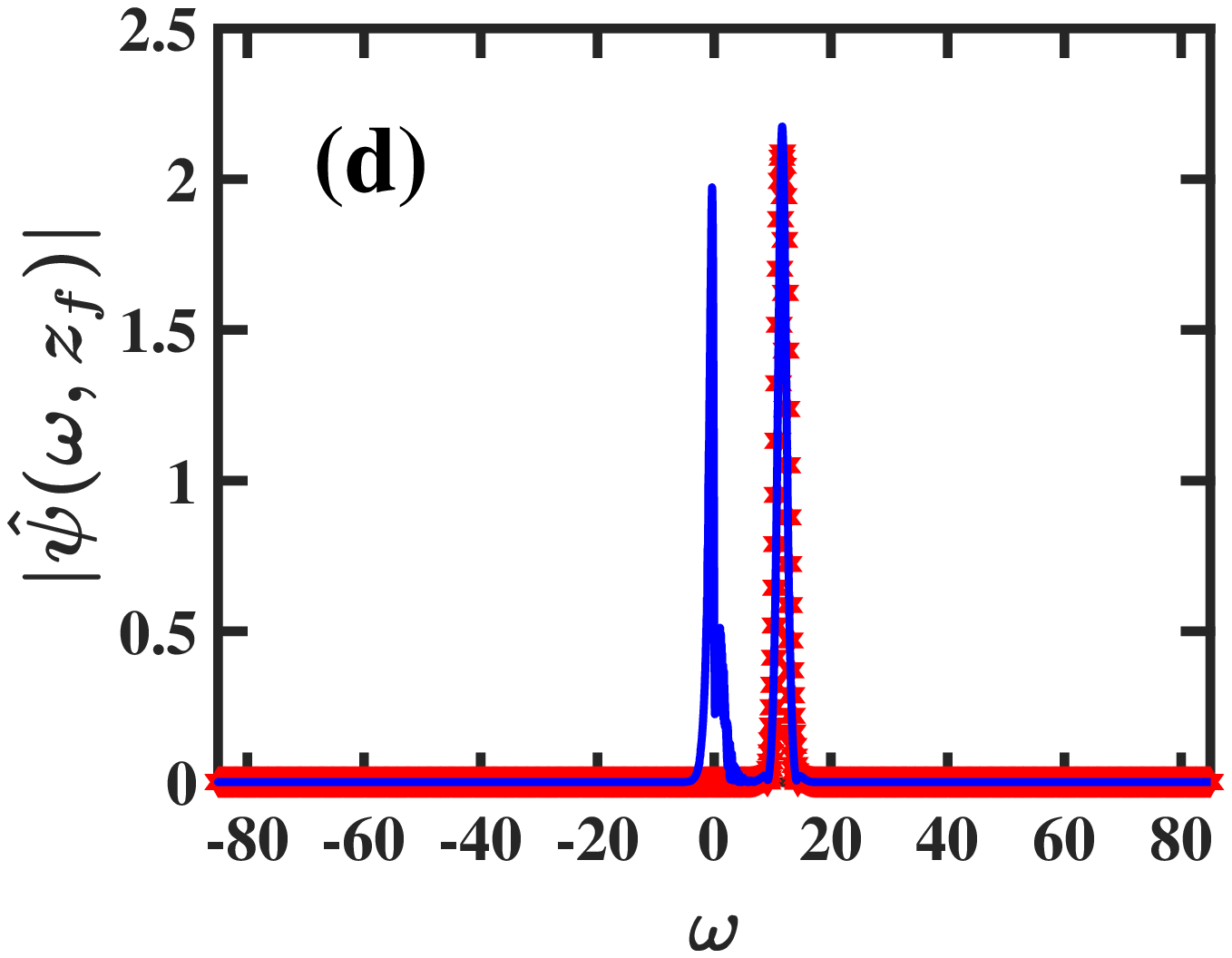} \\
\epsfxsize=8.2cm  \epsffile{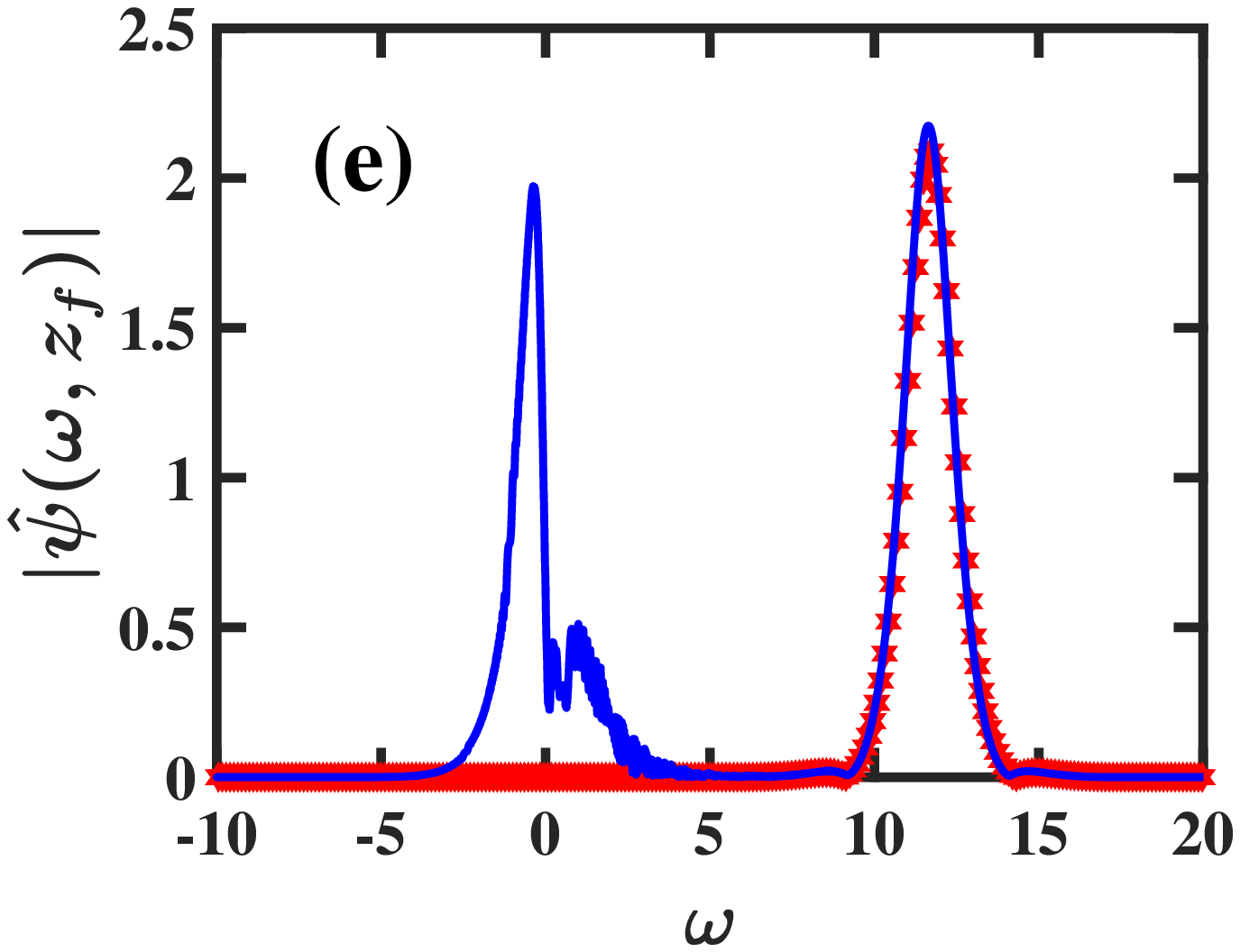}   
\end{tabular}
\end{center}
\caption{The shape of the Fourier spectrum $|\hat\psi(\omega,z)|$ 
at $z_{q}=140$ [(a), (b), (c)] and at $z_{f}=300$ [(d), (e)]  
for propagation of a CQNLS soliton in a waveguide loop with weak 
frequency-independent linear gain, cubic loss, and delayed Raman response. 
The physical parameter values are the same as in Fig. \ref{fig1}. 
The solid blue curve represents the result obtained by numerical 
solution of Eq. (\ref{cqnls11}), and the red stars correspond 
to the perturbation theory prediction of Eqs. (\ref{Iz3}), 
(\ref{cqnls13}), and (\ref{cqnls15}).}                        
 \label{fig3}
\end{figure}

Further insight into pulse-shape distortion and soliton instability 
is gained by considering the shape of the Fourier 
spectrum $|\hat\psi(\omega,z)|$. The numerically obtained spectrum 
$|\hat\psi(\omega,z)|$ at $z=z_{q}$ and at $z=z_{f}$ is shown in 
Fig. \ref{fig3} along with the perturbation theory prediction
of Eqs. (\ref{Iz3}), (\ref{cqnls13}), and (\ref{cqnls15}). It is seen 
that the soliton's Fourier spectrum is centered around the $z$ dependent 
soliton's frequency $\beta(z)$, and is shifted relative to the radiation's 
spectrum, which is centered near $\omega=0$ and is growing 
with increasing $z$. The separation of the 
soliton and radiation spectra, which is clear already at $z=z_{q}$ 
[see Figs. \ref{fig3}(a) - \ref{fig3}(c)], is a result of the 
Raman-induced self-frequency shift experienced by the soliton. 
It grows with increasing $z$ due to the increase in $|\beta(z)|$ 
[compare Figs. \ref{fig3}(a) and \ref{fig3}(d)]. 
Due to the separation between the two spectra, the soliton part 
of the numerically obtained graph of $|\hat\psi(\omega,z)|$ is very 
close to the perturbation theory prediction [see Figs. \ref{fig3}(c) 
and \ref{fig3}(e)]. However, the growth of the radiation spectrum 
with increasing $z$ leads to the strong oscillatory distortion in 
the pulse shape and to the soliton destabilization seen 
in Figs. \ref{fig1} and \ref{fig2}.

\begin{figure}[ptb]
\begin{tabular}{cc}
\epsfxsize=8.2cm  \epsffile{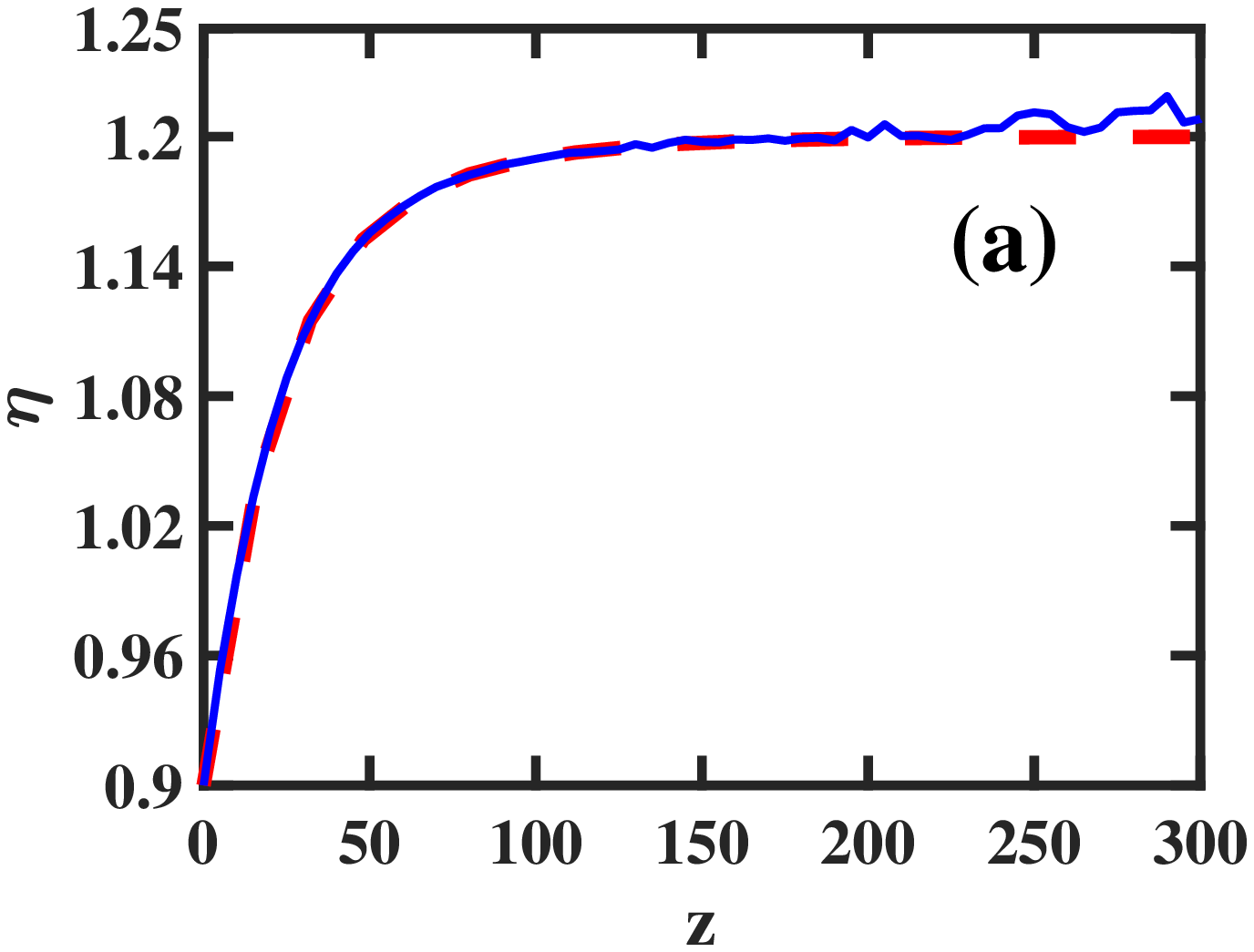} & 
\epsfxsize=8.2cm  \epsffile{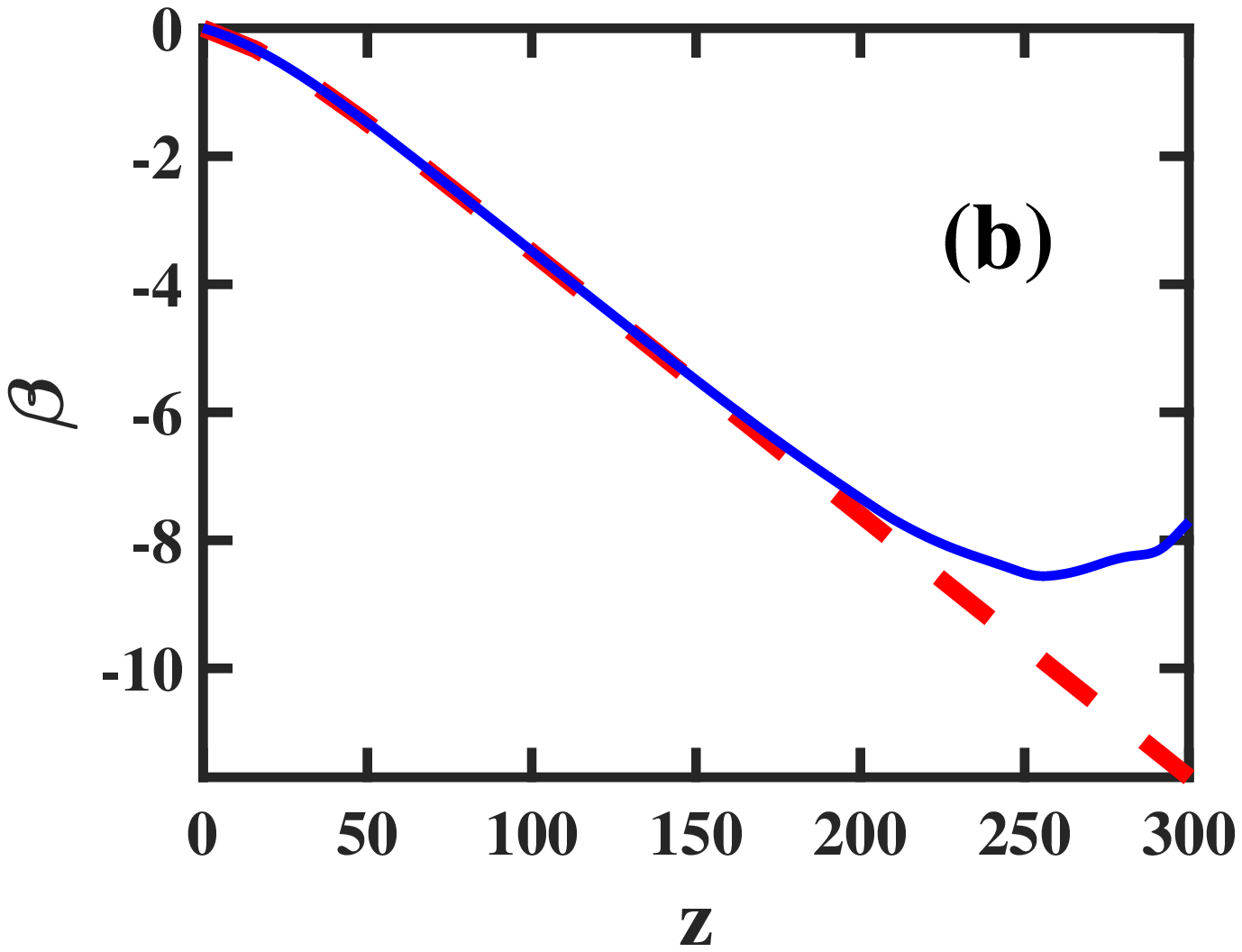}
\end{tabular}
\caption{The $z$ dependence of the CQNLS soliton's amplitude $\eta(z)$ (a) 
and frequency $\beta(z)$ (b) for the same waveguide setup as in 
Figs. \ref{fig1}-\ref{fig3}. The solid blue curves represent the results 
obtained by numerical solution of Eq. (\ref{cqnls11}). The dashed red curves 
correspond to the perturbation theory predictions of Eq. (\ref{cqnls15}) 
in (a) and of Eqs. (\ref{cqnls13}) and (\ref{cqnls15}) in (b).}
\label{fig4}
\end{figure}

The destabilization of the CQNLS soliton in the current waveguide setup 
is also evident in the dynamics of the soliton's amplitude and frequency. 
Figure \ref{fig4} shows the $z$ dependence of the soliton's amplitude and 
frequency obtained in the simulation together with the perturbation 
theory predictions of Eqs. (\ref{cqnls15}) and (\ref{cqnls13}). 
In both graphs we observe good agreement between the simulation's 
results and the predictions of Eqs. (\ref{cqnls15}) and (\ref{cqnls13}) 
for $0 \le z \le 200$. In particular, we see that in this interval, 
the numerically obtained amplitude value tends to the equilibrium 
value $\eta_{0}=1.2$. However, for $200 < z \le 300$, the numerical  
curve of $\eta(z)$ deviates from the prediction of Eq. (\ref{cqnls15})  
and the numerical curve of $\beta(z)$ deviates significantly from 
the prediction of Eq. (\ref{cqnls13}). 
These increasing deviations coincide with the increase in pulse-shape 
distortion and in the value of $I(z)$ observed in Figs. \ref{fig1} 
and \ref{fig2}. Thus, based on the results presented in Figs. 
\ref{fig1}-\ref{fig4} and on similar results obtained with other 
values of the physical parameters we conclude that transmission of 
the CQNLS soliton in waveguide loops with weak frequency-independent 
linear gain, cubic loss, and delayed Raman response is unstable.

{\it Comparison with transmission stability of the CNLS soliton}.   
It is useful to compare the results of the simulation for transmission 
stability of the CQNLS soliton with the results of numerical simulations 
for transmission stability of the CNLS soliton in the same optical 
waveguide setup. For this purpose, we carry out numerical simulations 
with the following perturbed CNLS model \cite{PC2018A}:
\begin{eqnarray}&&
i\partial_z\psi+\partial_t^2\psi+2|\psi|^2\psi=
ig_{0}\psi/2 - i\epsilon_{3}|\psi|^2\psi
+\epsilon_{R}\psi\partial_{t}|\psi|^2 ,  
\label{cqnls25}
\end{eqnarray} 
which is similar to Eq. (\ref{cqnls11}), except for the absence of the 
quintic nonlinearity term. We consider an initial condition in the form of 
the fundamental soliton of the CNLS equation 
\begin{eqnarray} 
\psi^{(c)}_{s}(t,z)\!=\!
\eta^{(c)}\exp(i\chi)/\cosh(x),
\label{cqnls26}
\end{eqnarray}
where $x=\eta^{(c)}\left(t-y^{(c)}+2\beta^{(c)} z\right)$, 
$\chi=\alpha^{(c)}-\beta^{(c)}(t-y^{(c)})+\left(\eta^{(c)2}-\beta^{(c)2}\right)z$, 
and $\eta^{(c)}$, $\beta^{(c)}$, $y^{(c)}$, and $\alpha^{(c)}$ are 
the CNLS soliton's amplitude, frequency, position, and phase.

By the adiabatic perturbation theory for the CNLS soliton, the dynamics 
of the soliton's amplitude and frequency is described by \cite{PC2018A}:  
\begin{equation}
\frac{d\eta^{(c)}}{dz} = 
\frac{4}{3}\epsilon_{3}\left(\eta_{0}^{(c)2} - \eta^{(c)2}\right)\eta^{(c)}, 
\label{cqnls27}
\end{equation} 
and 
\begin{equation}
\frac{d\beta^{(c)}}{dz} = -\frac{8}{15}\epsilon_{R}\eta^{(c)4}, 
\label{cqnls28}
\end{equation}    
where $\eta_{0}^{(c)}$ is the amplitude value at stable equilibrium, 
and $g_{0}=4\epsilon_{3}\eta_{0}^{(c)2}/3$ is used.  
Therefore, $\eta^{(c)}(z)$ and $\beta^{(c)}(z)$ are given by: 
\begin{equation}
\eta^{(c)}(z) = \eta^{(c)}_{0}\left[1 + \left(\frac{\eta_{0}^{(c)2}}{\eta^{(c)2}(0)}-1\right)
\exp\left(-8\epsilon_{3}\eta_{0}^{(c)2}z/3\right)\right]^{-\frac{1}{2}} ,  
\label{cqnls29} 
\end{equation} 
and 
\begin{eqnarray}&&
\!\!\!\!\!\!\!\!\!
\beta^{(c)}(z) = \beta^{(c)}(0) 
-\frac{\epsilon_{R}\eta_{0}^{(c)2}}{5\epsilon_{3}}
\left\{\ln\left[
\frac{\eta_{0}^{(c)2} - \eta^{(c)2}(0) 
+ \eta^{(c)2}(0)\exp\left(8\epsilon_{3}\eta_{0}^{(c)2}z/3\right)}
{\eta_{0}^{(c)2}}\right]
\right.
\nonumber \\&&
\!\!\!\!\!\!\!\!\!
\left.
+\frac{\eta^{(c)2}(0)}{\eta_{0}^{(c)2}}
-\frac{\eta^{(c)2}(0)}{\eta^{(c)2}(0)+\left[\eta_{0}^{(c)2} - \eta^{(c)2}(0)\right]
\exp\left(-8\epsilon_{3}\eta_{0}^{(c)2}z/3\right)}
\right\} .
\label{cqnls30}
\end{eqnarray}

Equation (\ref{cqnls25}) is solved numerically on a time domain 
$[t_{\mbox{min}},t_{\mbox{max}}]=[-400,400]$ with periodic boundary conditions. 
The initial condition is in the form of the CNLS soliton in Eq. (\ref{cqnls26}) 
with parameter values $\eta^{(c)}(0)$, $\beta^{(c)}(0)=0$, $y^{(c)}(0)=0$, 
and $\alpha^{(c)}(0)=0$. To make a meaningful comparison with the results 
presented in Figs. \ref{fig1}-\ref{fig4}, we use the parameter values 
$\epsilon_{3}=0.01$ and $\epsilon_{R}=0.04$. 
Additionally, we require that the initial power and the equilibrium power 
of the CNLS soliton would be equal to the initial power and the equilibrium power   
of the CQNLS soliton, respectively. That is, we determine the values of $\eta^{(c)}(0)$ 
and $\eta_{0}^{(c)}$ that should be used in the CNLS simulation by requiring:     
$2\eta^{(c)}(0) = 2\eta_{m} \mbox{arctanh}\left[\eta(0)/\eta_{m}\right]$  
and $2\eta^{(c)}_{0} = 2\eta_{m} \mbox{arctanh}\left(\eta_{0}/\eta_{m}\right)$, 
where $\eta(0)$ and $\eta_{0}$ are the values of the initial amplitude and 
the equilibrium amplitude used in the CQNLS simulation. 
For the values $\eta_{m}=1.2247...$, $\eta_{0}=1.2$, and $\eta(0)=0.9$
used in the CQNLS simulation, we obtain: $\eta_{0}^{(c)}=2.8086...$ 
and $\eta^{(c)}(0)=1.1503...$.

The values of the transmission quality distance and the final propagation 
distance obtained in the simulation with Eq. (\ref{cqnls25}) are $z_{q}=20$ 
and $z_{f}=76$. These values are significantly smaller compared with the 
values $z_{q}=140$ and $z_{f}=300$, obtained for the CQNLS soliton. As we 
will see in the next paragraphs, the reduction in the values of $z_{q}$ and $z_{f}$ 
for the CNLS soliton is due to the fact that the radiation-induced pulse distortion 
effects in this case are significantly stronger than the pulse distortion effects 
observed for the CQNLS soliton in Figs. \ref{fig1}-\ref{fig4}.

\begin{figure}[ptb]
\begin{tabular}{cc}
\epsfxsize=8.1cm  \epsffile{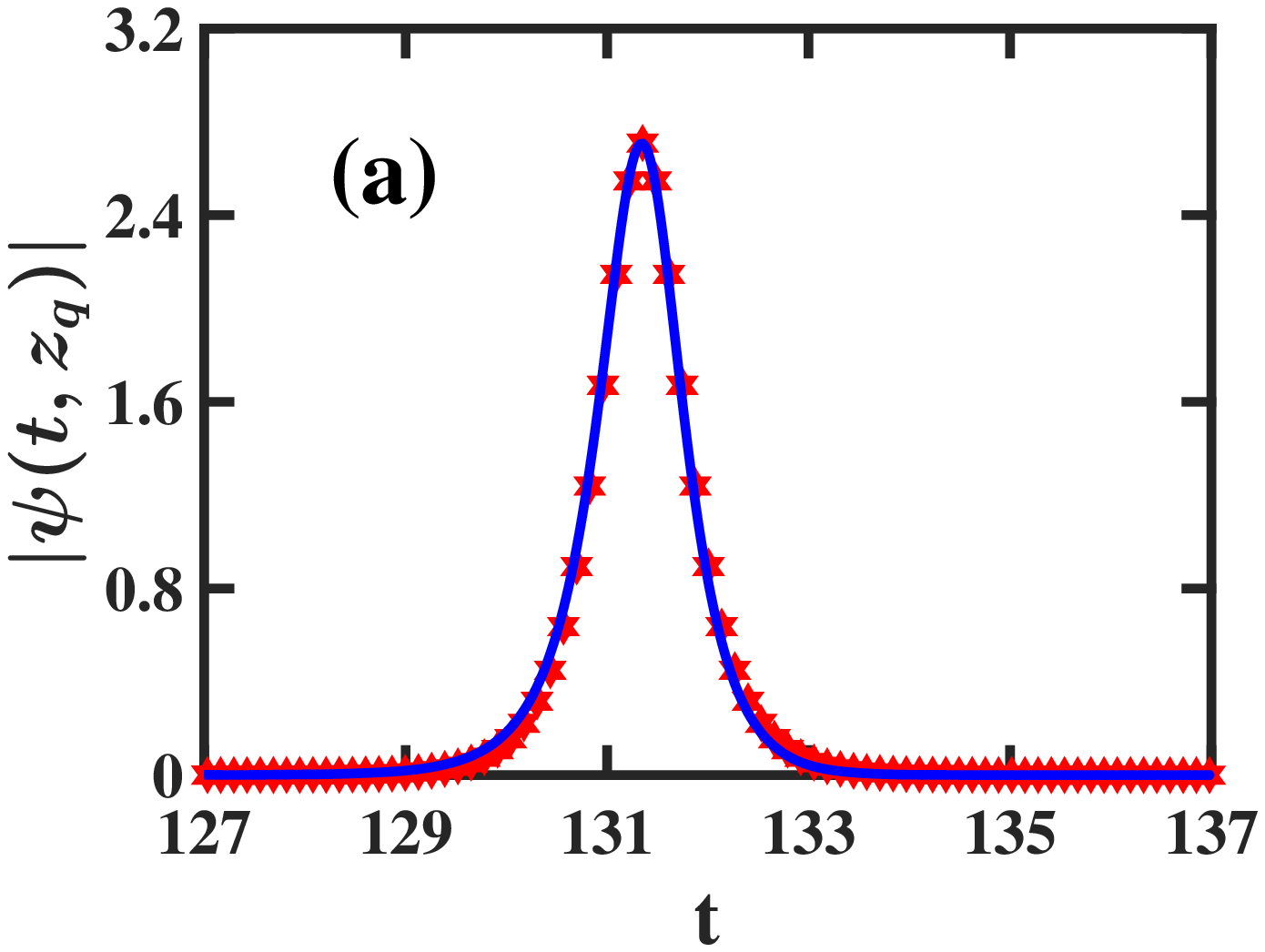} &
\epsfxsize=8.1cm  \epsffile{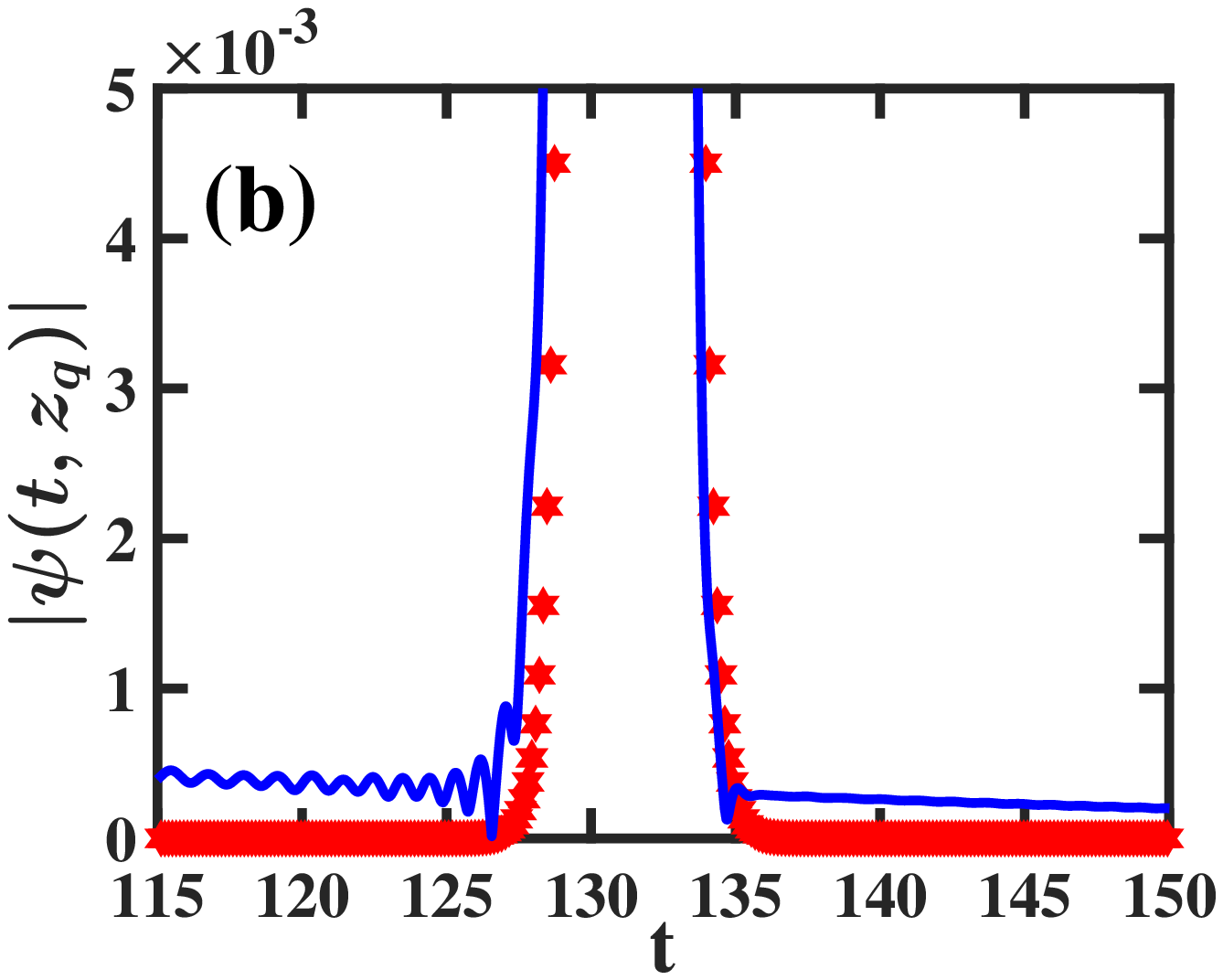} \\
\epsfxsize=8.1cm  \epsffile{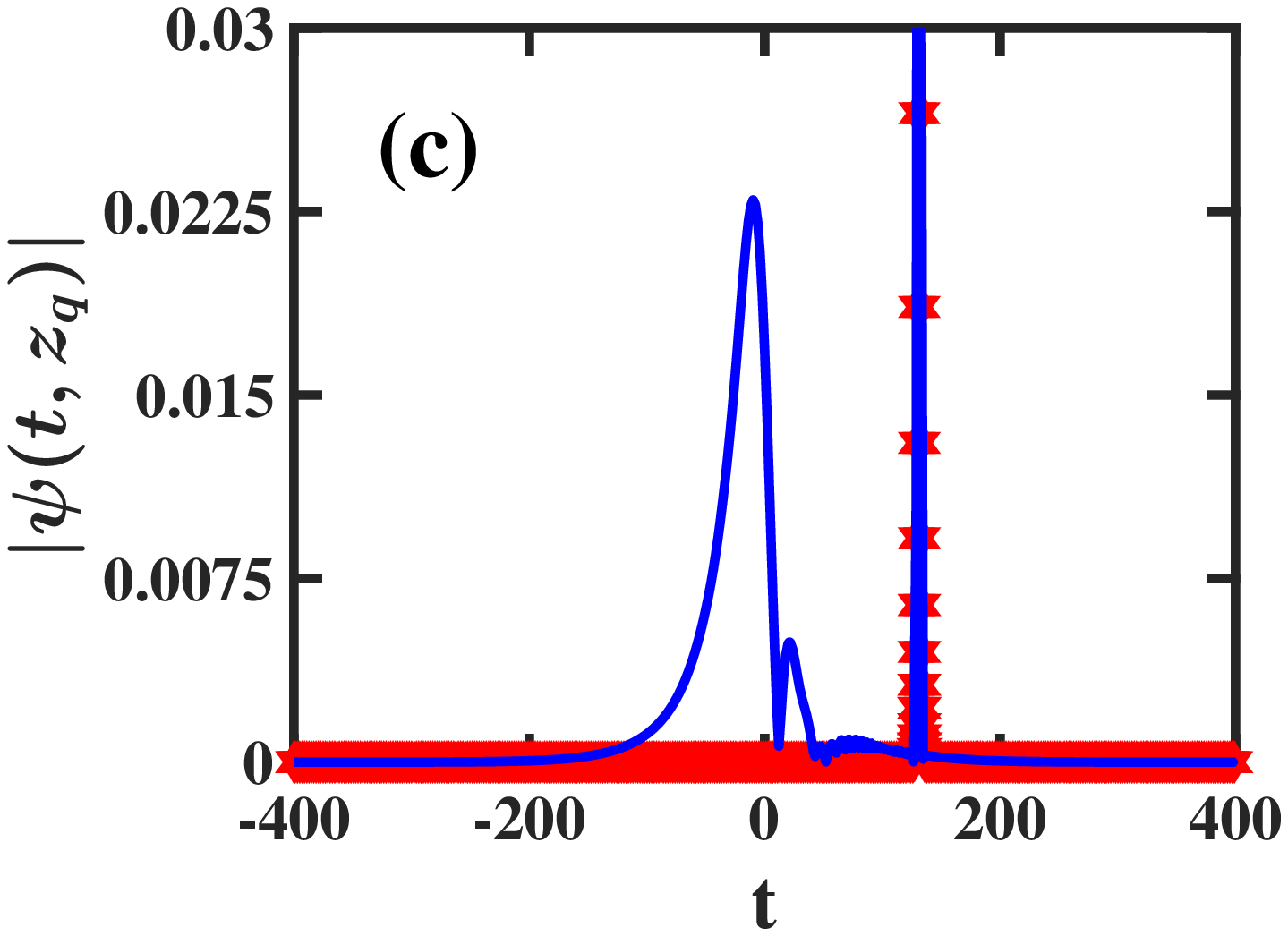} &
\epsfxsize=8.1cm  \epsffile{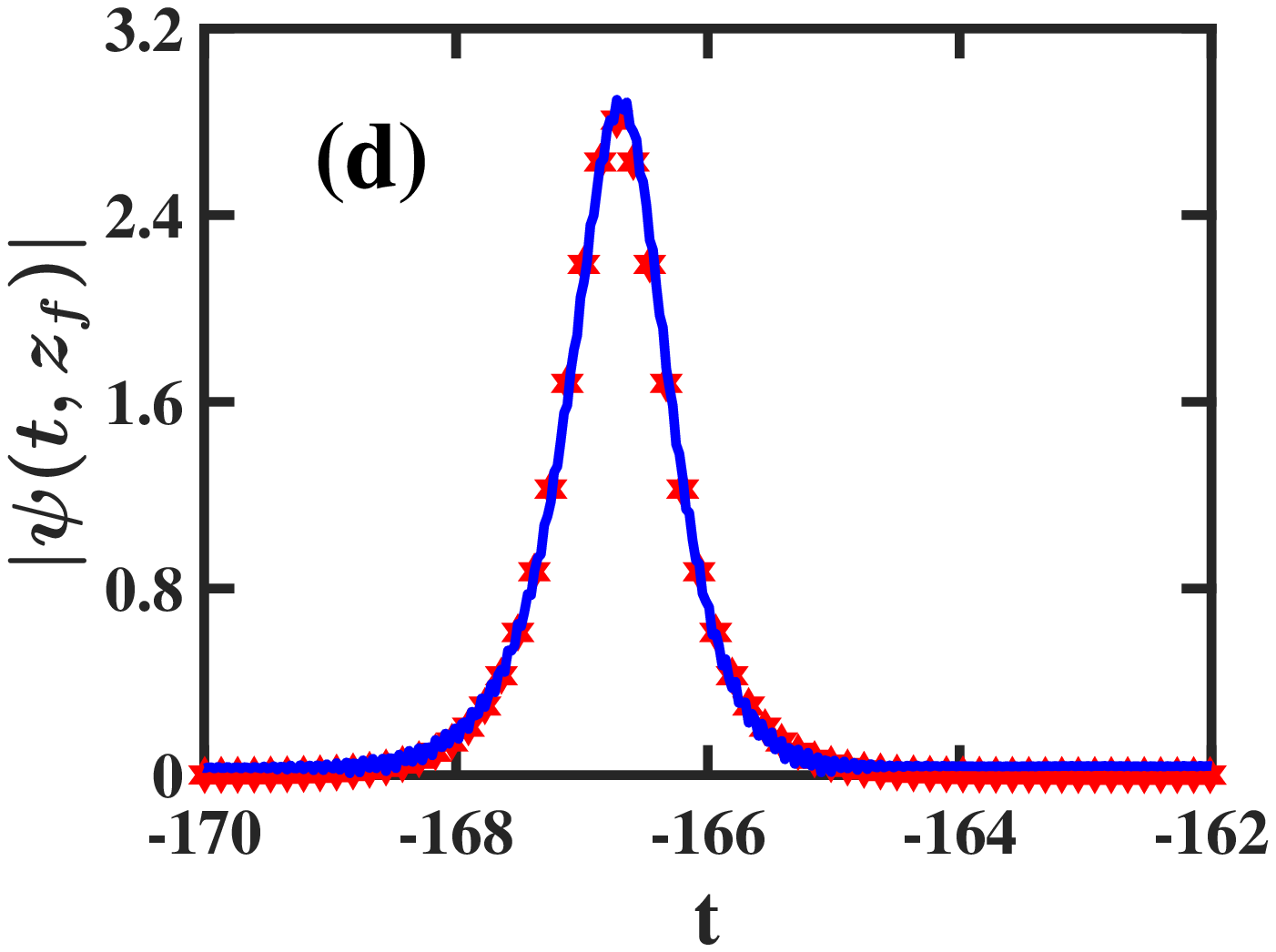} \\ 
\epsfxsize=8.1cm  \epsffile{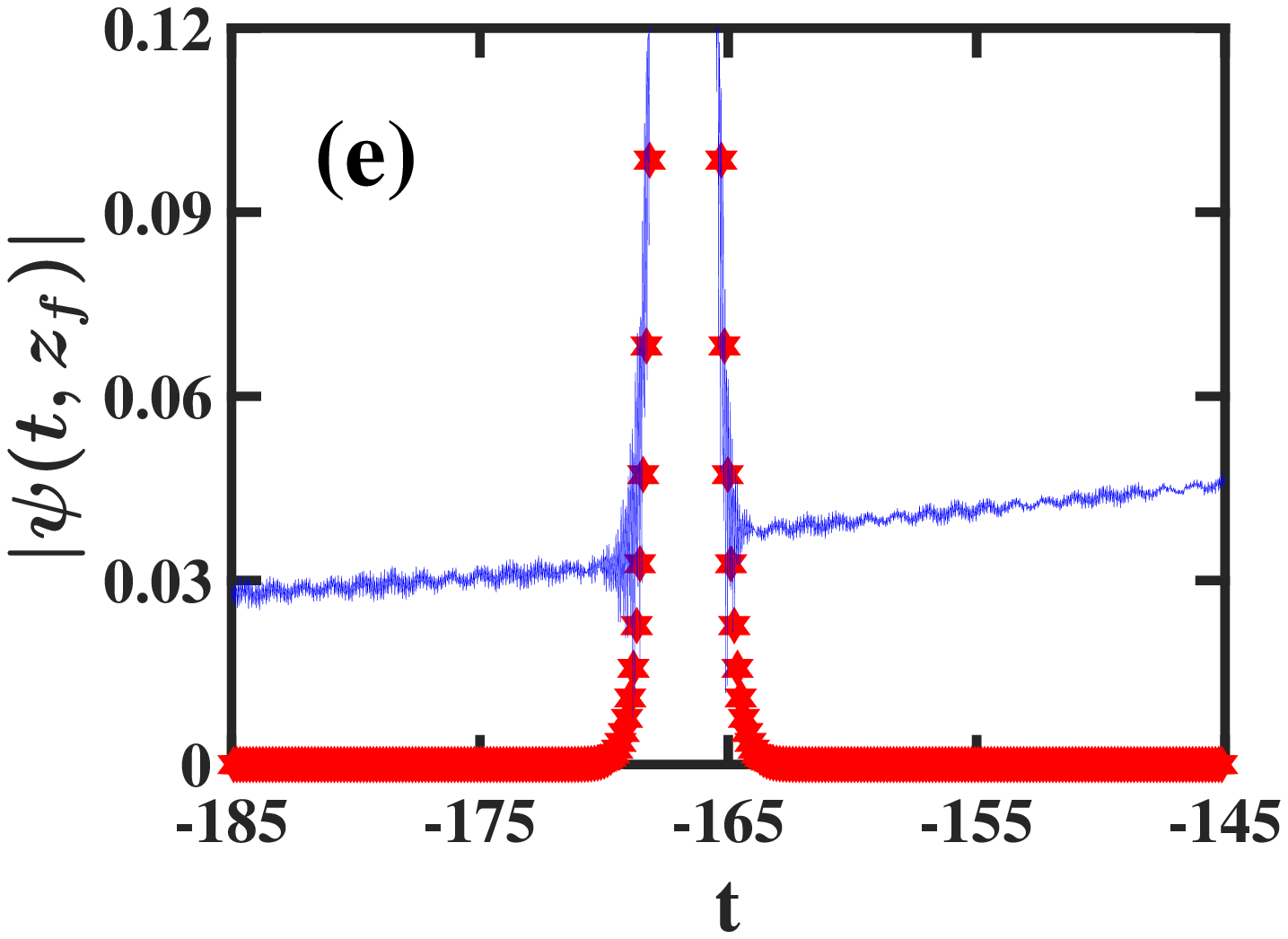} &
\epsfxsize=8.1cm  \epsffile{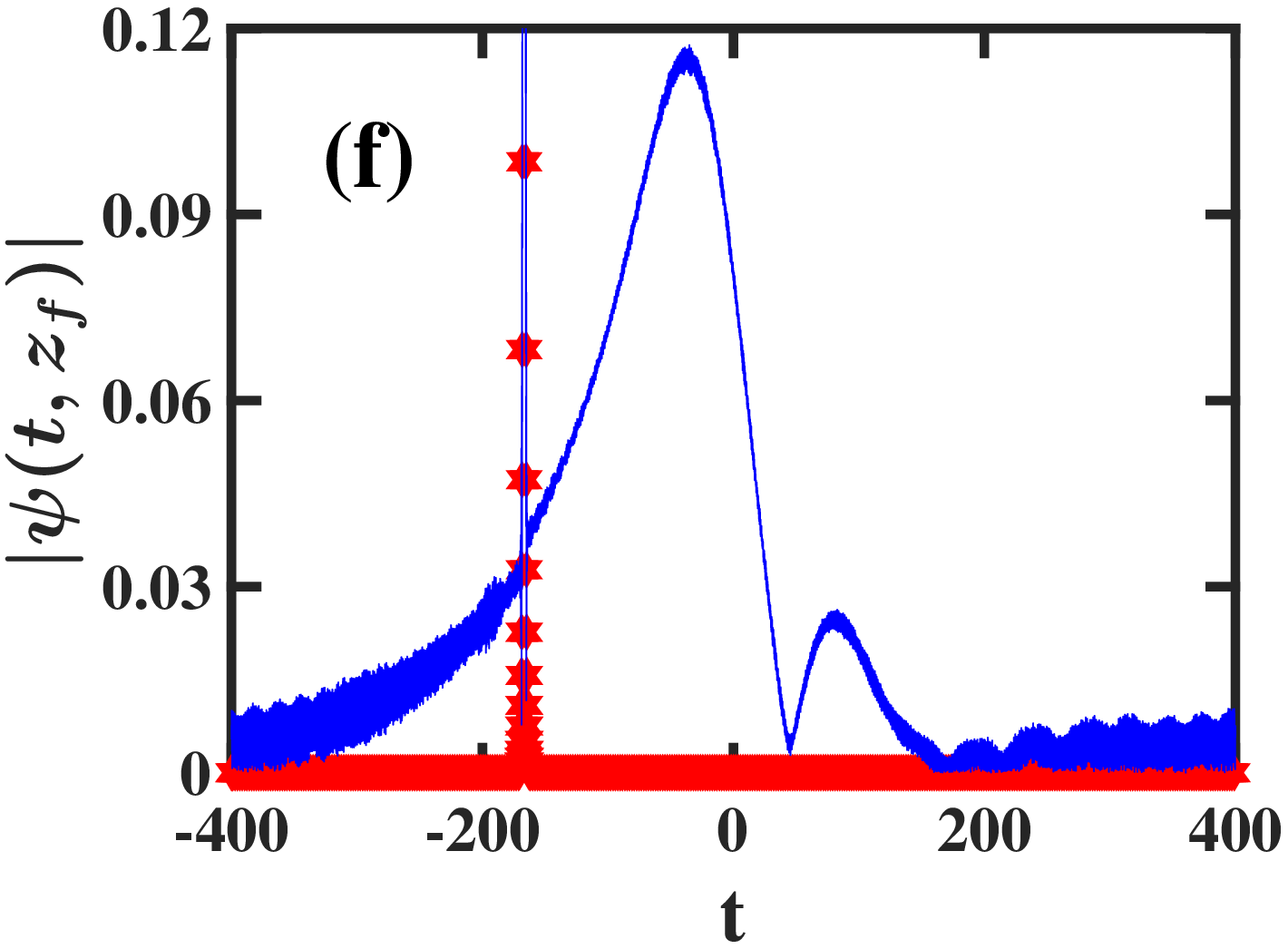}
\end{tabular}
\caption{The pulse shape $|\psi(t,z)|$ at $z_{q}=20$ [(a), (b), (c)] and at   
$z_{f}=76$ [(d), (e), (f)] for propagation of a CNLS soliton in a waveguide loop 
with weak frequency-independent linear gain, cubic loss, and delayed Raman response. 
The parameter values are $\epsilon_{3}=0.01$, $\epsilon_{R}=0.04$, 
$\eta_{0}^{(c)}=2.8086...$ and $\eta^{(c)}(0)=1.1503...$. 
The solid blue curve represents the result obtained by numerical 
solution of Eq.  (\ref{cqnls25}). The red stars correspond to 
the perturbation theory prediction, obtained with Eqs. (\ref{Iz5}) 
and (\ref{cqnls29}).}
\label{fig5}
\end{figure}

\begin{figure}[ptb]
\begin{tabular}{cc}
\epsfxsize=8.5cm  \epsffile{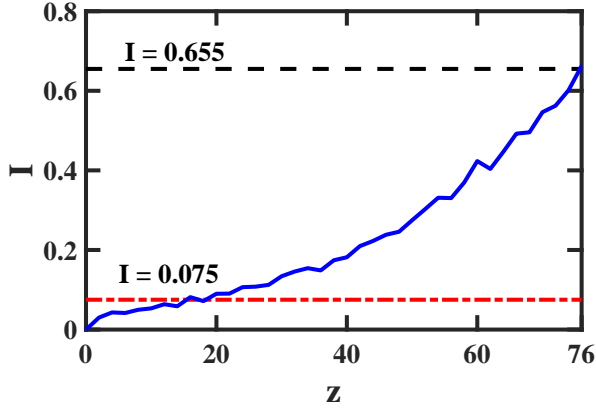} 
\end{tabular}
\caption{The $z$ dependence of the transmission quality integral $I(z)$ obtained 
by the numerical simulation with Eq. (\ref{cqnls25}) for the waveguide setup  
considered in Fig. \ref{fig5}. The solid blue curve represents the simulation's result. 
The dashed black and dashed-dotted red horizontal lines correspond to $I=0.655$ 
and $I=0.075$, respectively.}
\label{fig6}
\end{figure}

We consider first the dynamics of the radiation-induced pulse distortion 
for the CNLS soliton. Figure \ref{fig5} shows the pulse shape $|\psi(t,z)|$ 
obtained in the simulation with Eq. (\ref{cqnls25}) at $z=z_{q}$ and $z=z_{f}$
together with the prediction of the perturbation theory, which is given by 
Eqs. (\ref{Iz5}) and (\ref{cqnls29}). As seen in Figs. \ref{fig5}(a)-\ref{fig5}(c), 
the numerical result for the pulse shape at $z=z_{q}$ is close to the 
perturbation theory prediction, but an appreciable radiative tail is present 
already at this distance. We note that the main part of the radiative tail forms 
a double hump, which is more pronounced than the radiative hump that is formed  
for the CQNLS soliton. We also note that the part of the radiative tail near 
the CNLS soliton is weaker (smaller) than the corresponding part of the radiative 
tail for the CQNLS soliton. Moreover, as seen in \ref{fig5}(d)-\ref{fig5}(f), 
the radiative tail of the CNLS soliton grows significantly with increasing distance 
\cite{oscillations}. As a result, the transmission quality integral increases 
significantly and exceeds the value 0.655 at $z_{f}=76$ (see Fig. \ref{fig6}).

\begin{figure}[ptb]
\begin{center}
\begin{tabular}{cc}
\epsfxsize=8.2cm  \epsffile{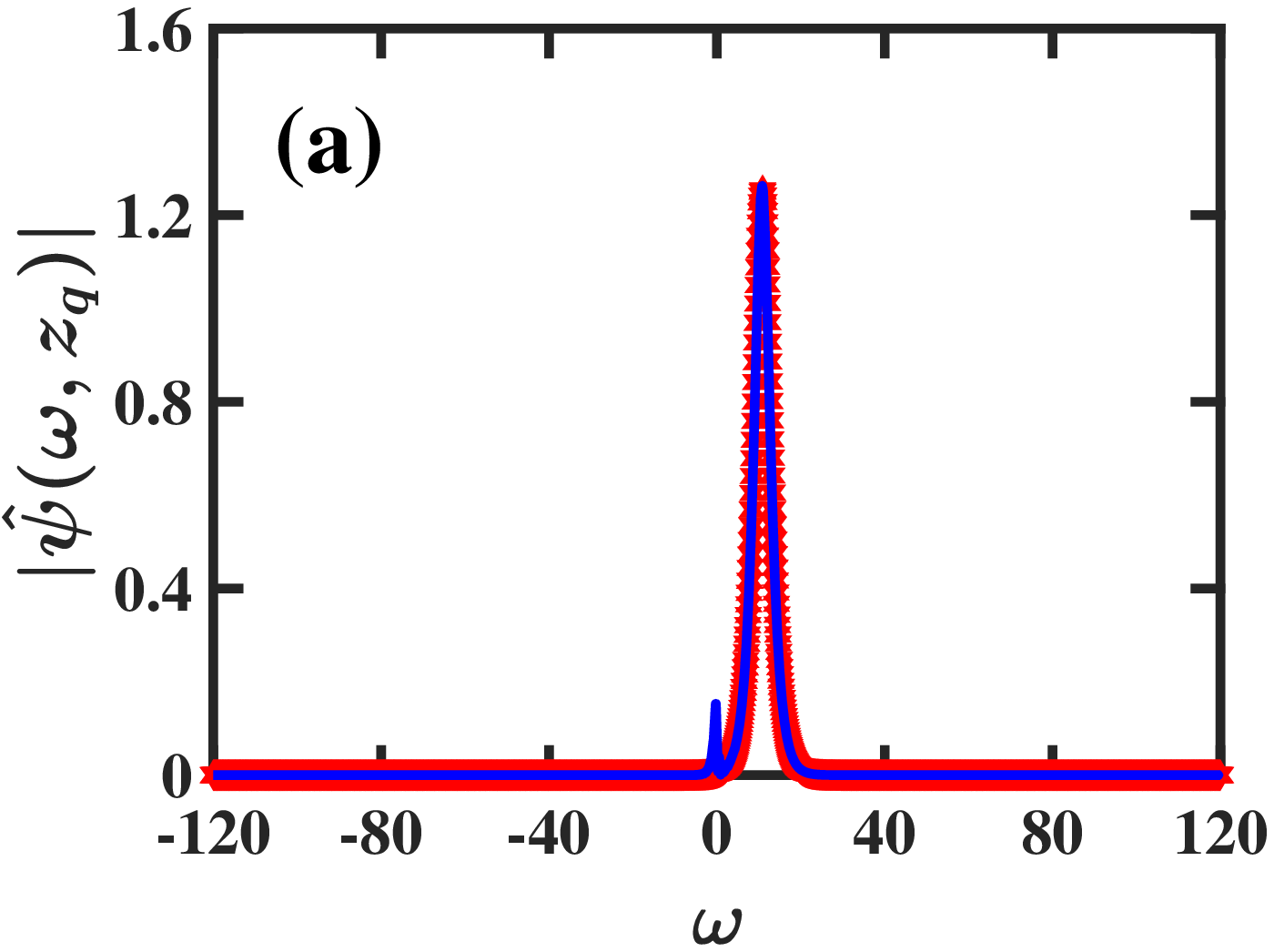} &
\epsfxsize=8.2cm  \epsffile{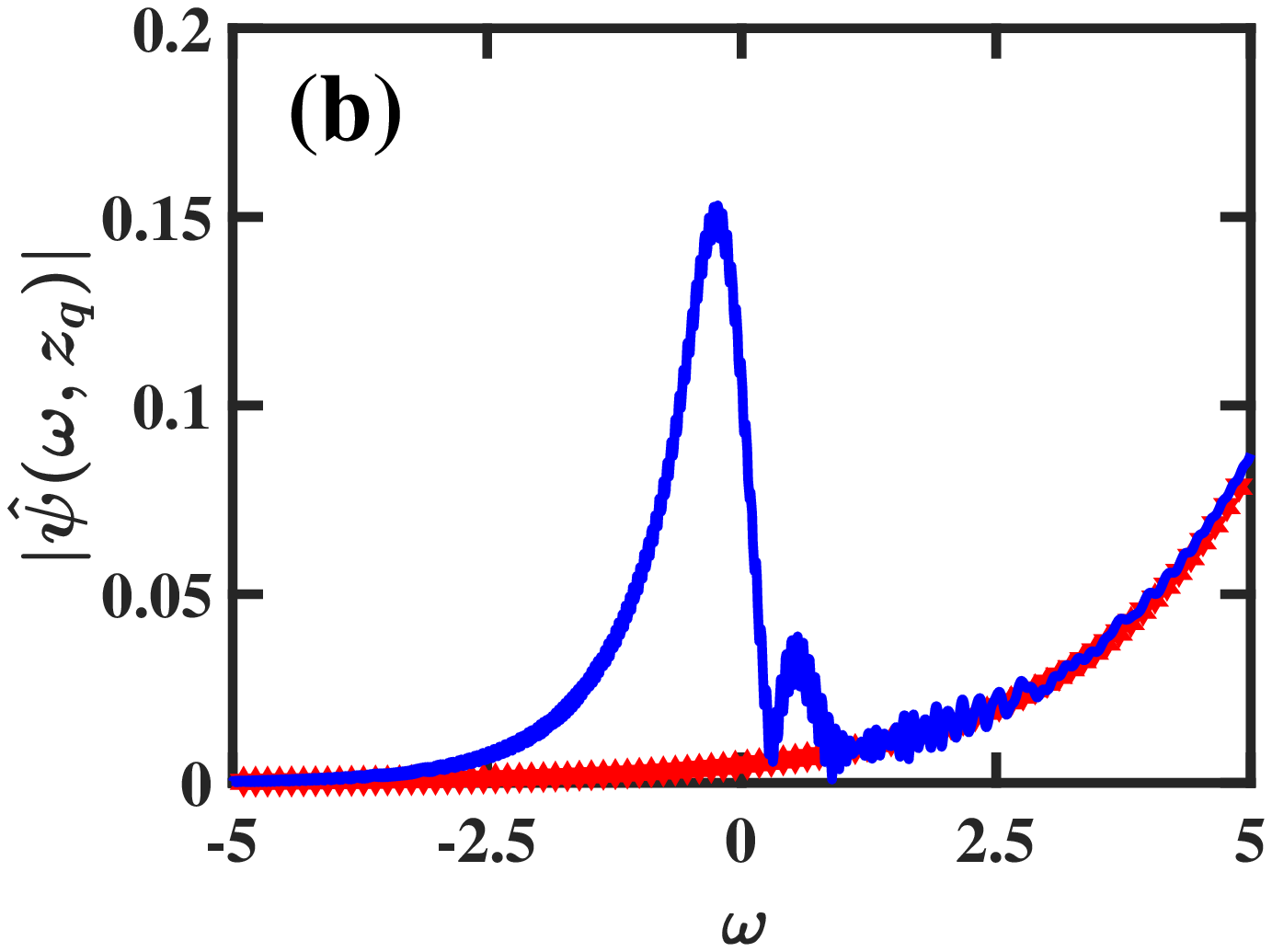} \\
\epsfxsize=8.2cm  \epsffile{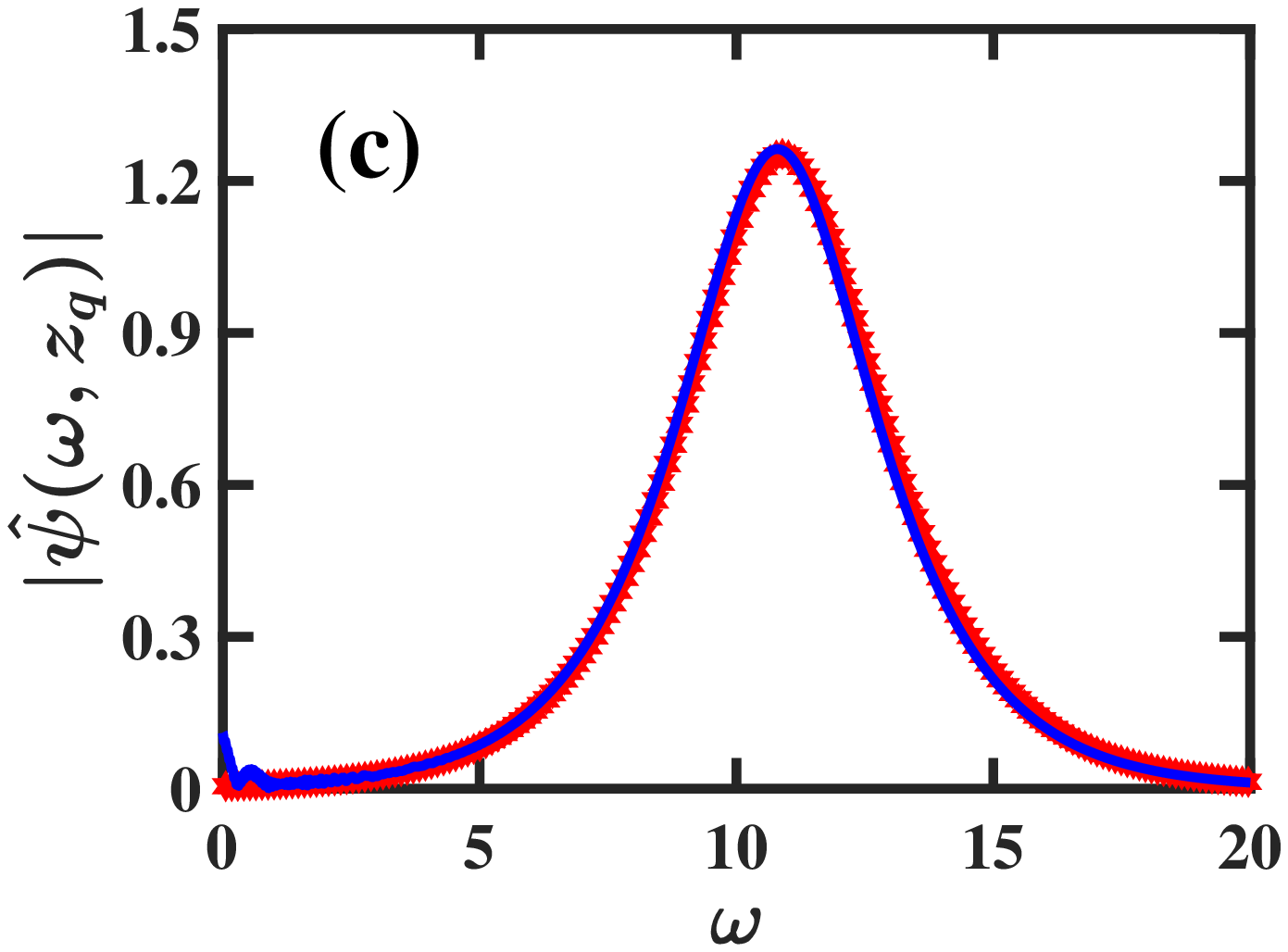} &
\epsfxsize=8.2cm  \epsffile{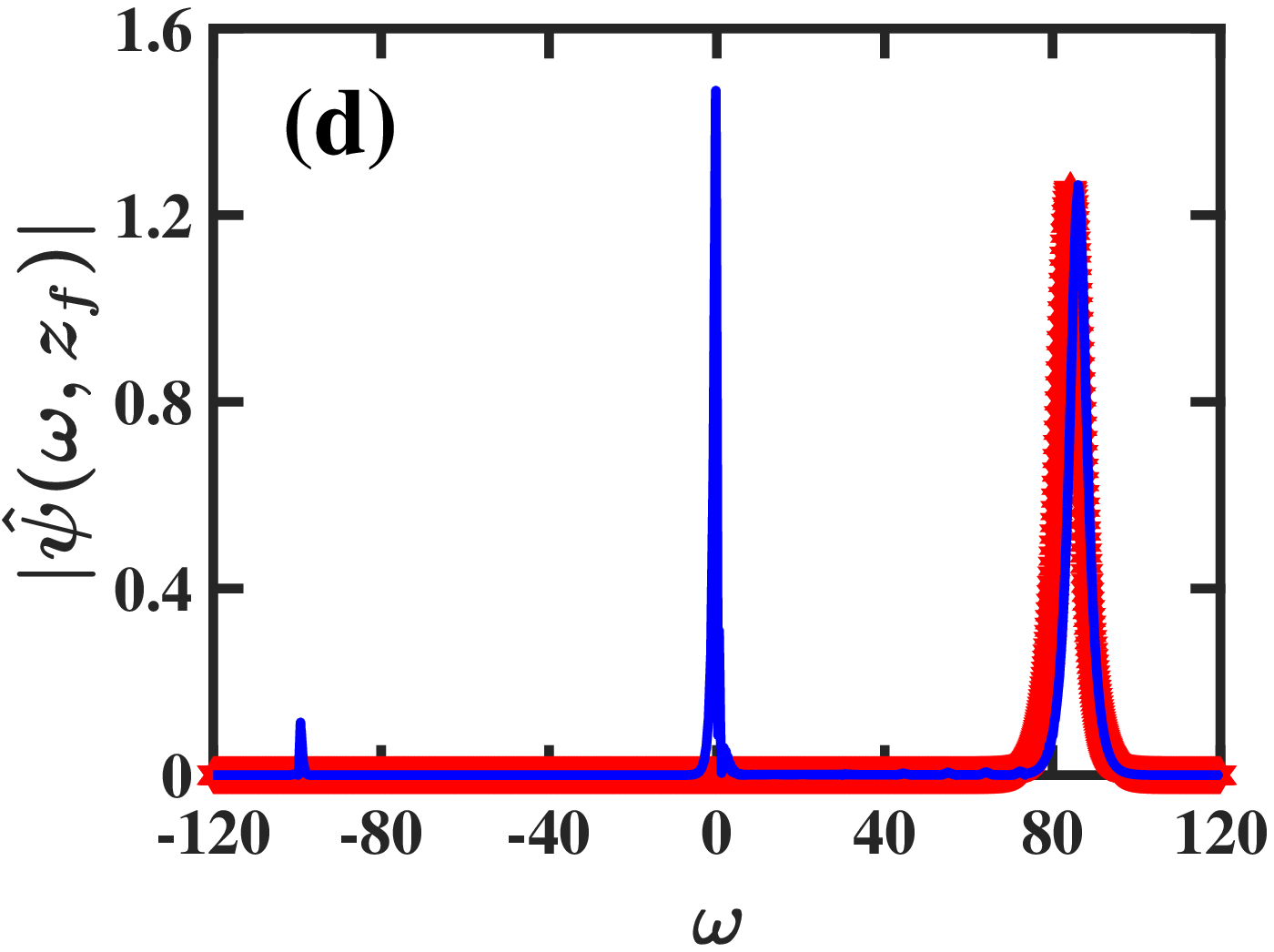} \\
\epsfxsize=8.2cm  \epsffile{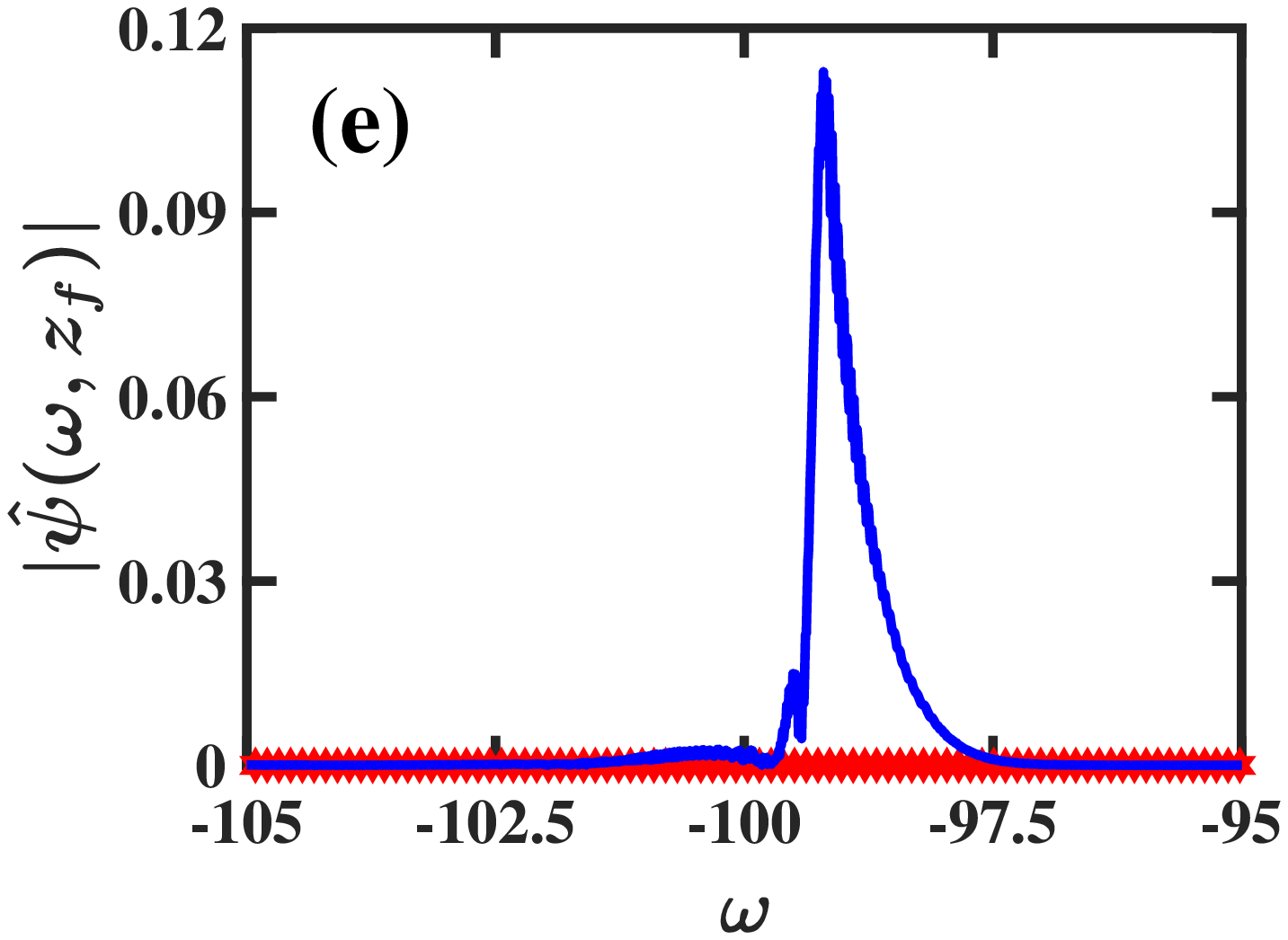} &
\epsfxsize=8.2cm  \epsffile{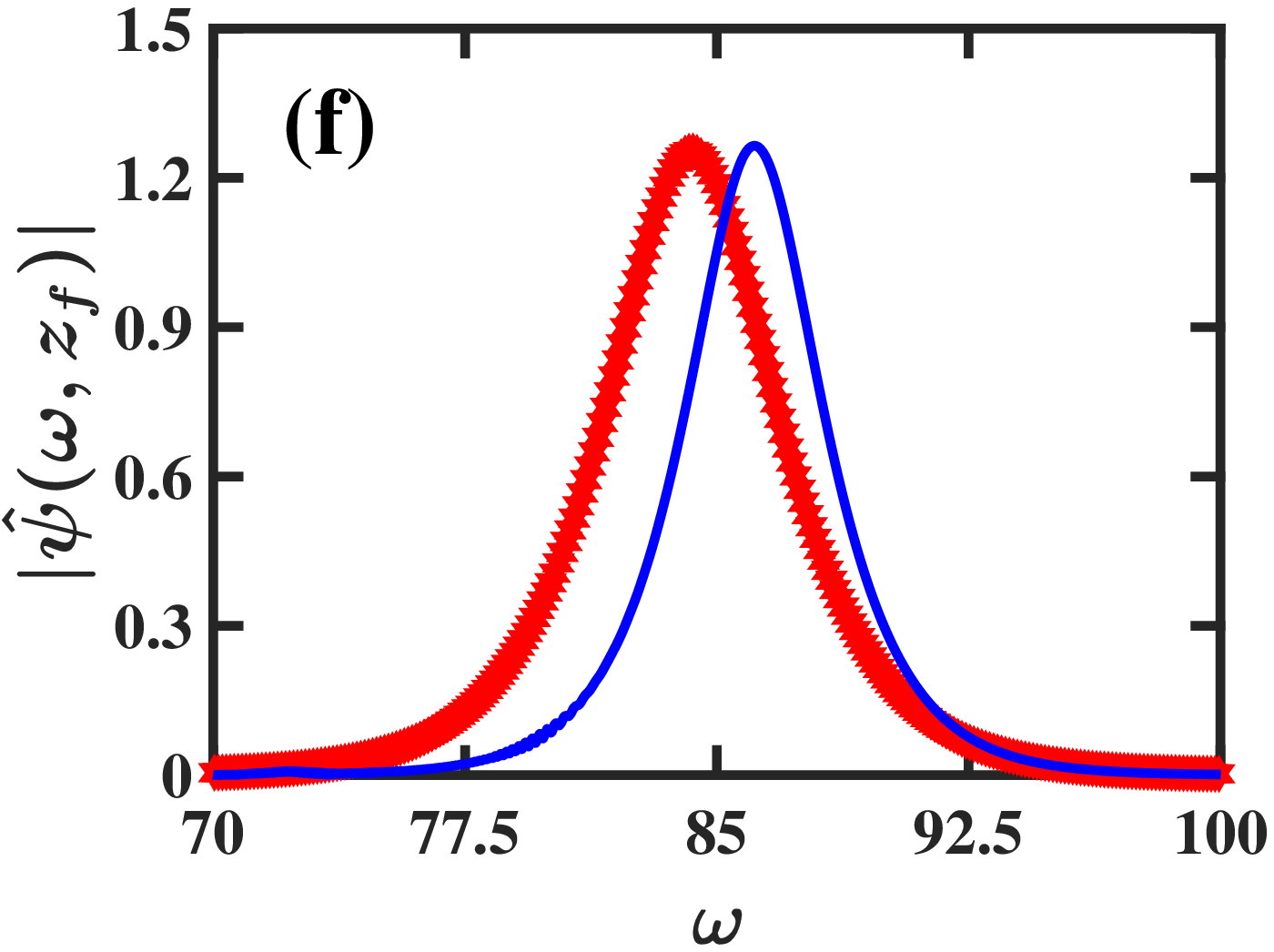}    
\end{tabular}
\end{center}
\caption{The shape of the Fourier spectrum $|\hat\psi(\omega,z)|$ 
at $z_{q}=20$ [(a), (b), (c)] and at $z_{f}=76$ [(d), (e), (f)]  
for propagation of a CNLS soliton in a waveguide loop with weak 
frequency-independent linear gain, cubic loss, and delayed Raman response. 
The parameter values are the same as in Fig. \ref{fig5}. 
The solid blue curve represents the result obtained by the numerical 
simulation with Eq. (\ref{cqnls25}), and the red stars correspond 
to the perturbation theory prediction of Eqs. (\ref{Iz6}), 
(\ref{cqnls29}), and (\ref{cqnls30}).}                        
 \label{fig7}
\end{figure}

The shape of the Fourier spectrum $|\hat\psi(\omega,z)|$ provides 
valuable insight into pulse-shape distortion and soliton instability.    
Figure \ref{fig7} shows the simulation's result for the Fourier 
spectrum at $z=z_{q}$ and at $z=z_{f}$ together with the perturbation 
theory prediction of Eqs. (\ref{Iz6}), (\ref{cqnls29}), 
and (\ref{cqnls30}). Similar to the CQNLS case, the soliton's and 
radiation's spectra are separated already at $z=z_{q}$ due to the 
Raman-induced self-frequency shift experienced by the CNLS soliton. 
Accordingly, the soliton's spectrum is centered around $\beta^{(c)}(z)$, 
while the main part of the radiation's spectrum is centered near 
$\omega=0$ [see Figs. \ref{fig7}(a)-\ref{fig7}(c)]. As the soliton 
continues to propagate along the waveguide, the radiation's spectrum 
grows significantly, and a new radiative peak, which was not observed 
in the CQNLS case, appears near $\omega=-100$ [see Figs. \ref{fig7}(d)-\ref{fig7}(f)]. 
Additionally, the separation between the soliton's and radiation's 
spectra increases significantly, and is much larger at $z=z_{f}$ 
than the spectral separation that was observed for the CQNLS soliton. 
Due to this spectral separation, the soliton part of the numerical 
result for $|\hat\psi(\omega,z)|$ is not distorted, i.e., its 
shape is close to the one predicted by the perturbation theory 
[see Figs. \ref{fig7}(c) and \ref{fig7}(f)]. 
However, the growth of the radiative spectrum leads to the strong 
pulse distortion and to the transmission destabilization that are 
observed in Figs. \ref{fig5} and \ref{fig6}. 
We also note that the numerical curve of $|\hat\psi(\omega,z_{f})|$ 
is shifted considerably to the right compared with the theoretical curve. 
This shift can be explained by noting that since 
$\eta_{0}^{(c)}=2.8086...$, the CNLS soliton attains large 
amplitude values at large distances. As a result, the Raman perturbation 
term in Eq. (\ref{cqnls25}) is not weak at these distances, 
and its effects are larger than the effects predicted by 
the perturbation theory.

\begin{figure}[ptb]
\begin{tabular}{cc}
\epsfxsize=8.2cm  \epsffile{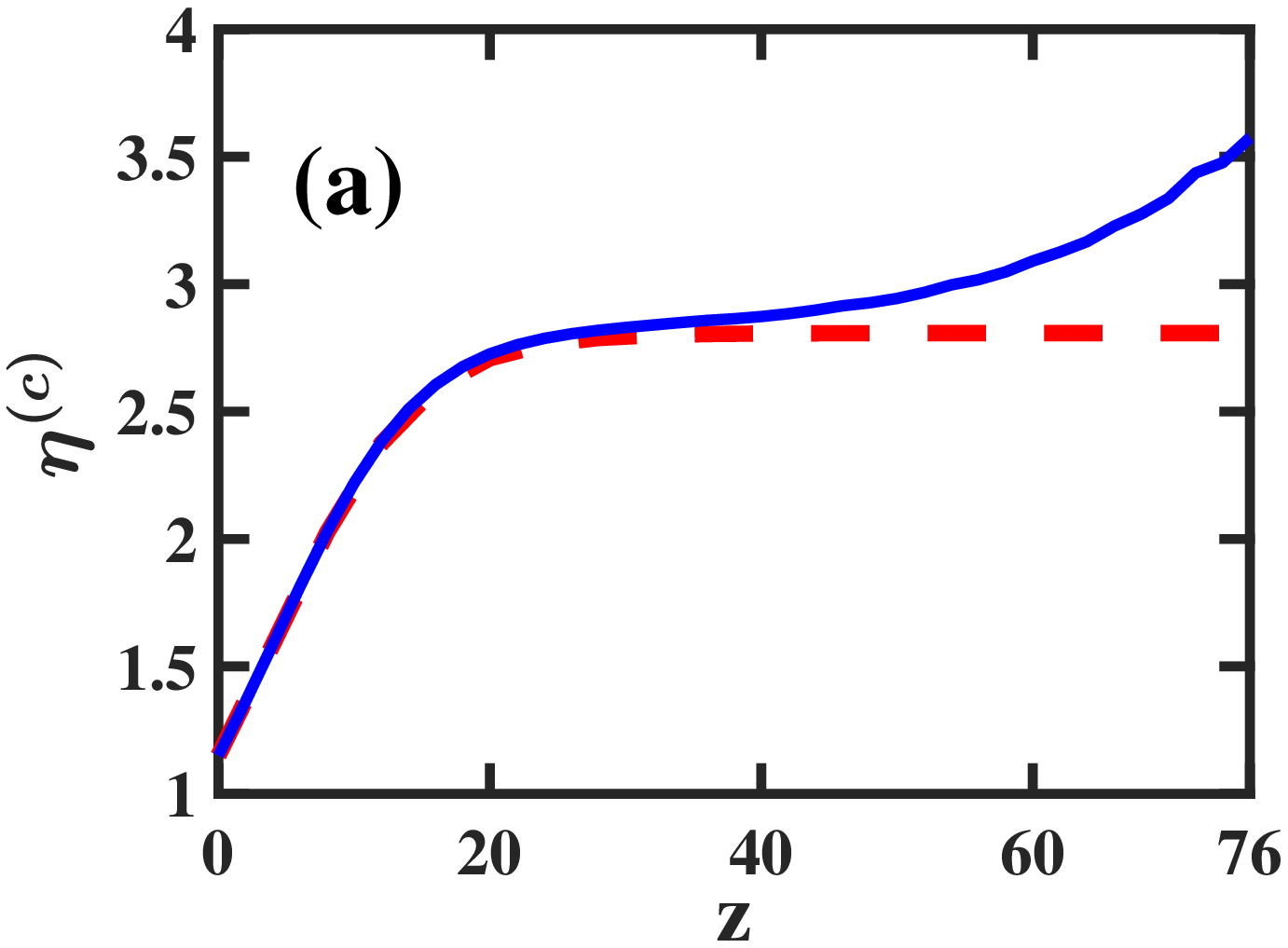} &
\epsfxsize=8.2cm  \epsffile{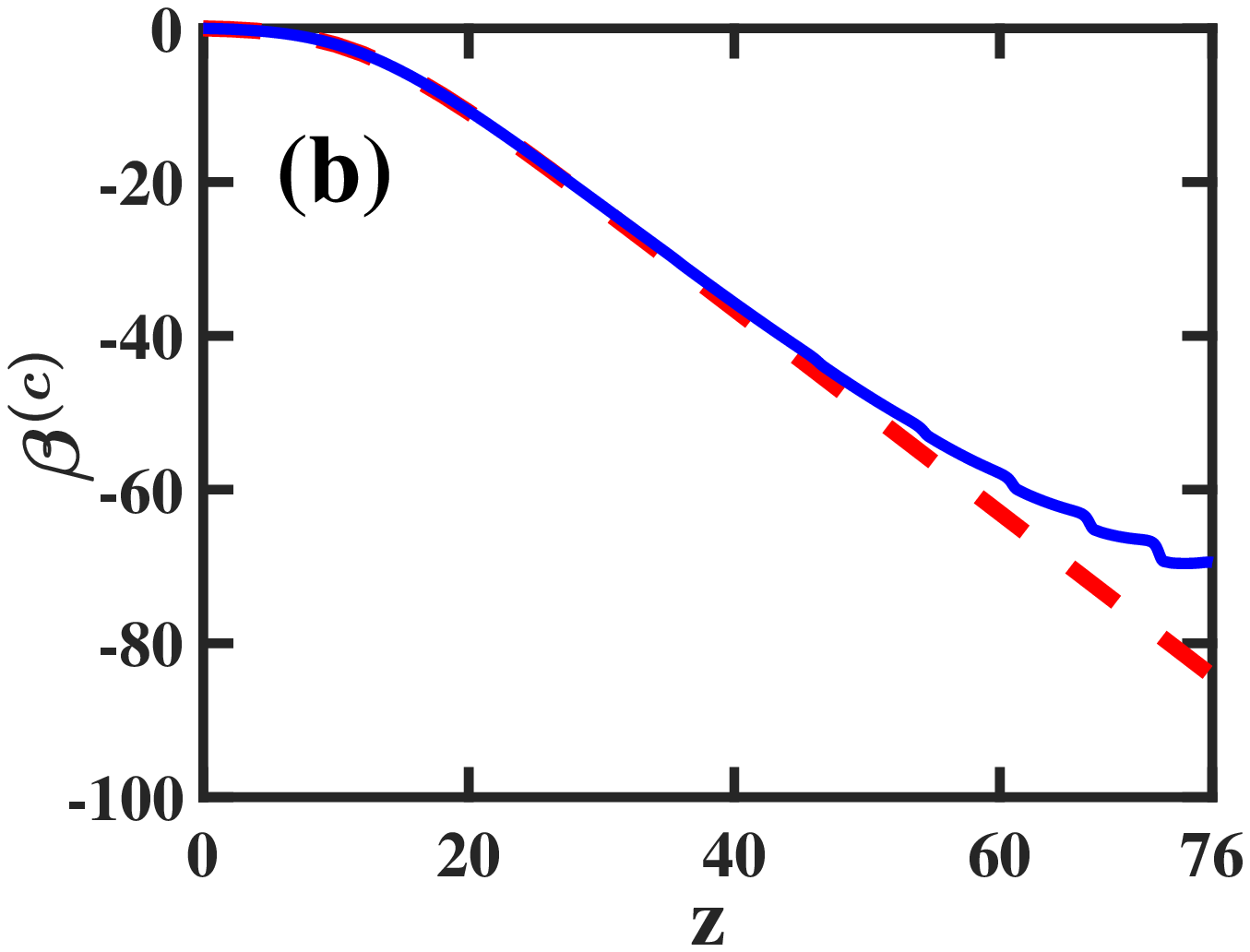}
\end{tabular}
\caption{The $z$ dependence of the CNLS soliton's amplitude $\eta^{(c)}(z)$ (a) 
and frequency $\beta^{(c)}(z)$ (b) for the waveguide setups considered in 
Figs. \ref{fig5}-\ref{fig7}. The solid blue curves represent the results 
obtained by numerical solution of Eq. (\ref{cqnls25}). The dashed red curves 
correspond to the perturbation theory predictions of Eq. (\ref{cqnls29}) 
in (a) and of Eq. (\ref{cqnls30}) in (b).}
\label{fig8}
\end{figure}

The effects of pulse distortion and transmission destabilization for 
the CNLS soliton can also be detected by following the dynamics of 
its amplitude and frequency. Figures \ref{fig8}(a) and \ref{fig8}(b) 
show the $z$ dependence of the soliton's amplitude and frequency 
obtained in the simulation together with the perturbation theory 
predictions of Eqs. (\ref{cqnls29}) and (\ref{cqnls30}). 
We observe that for $0 \le z \le 50$, the simulation's results 
for $\eta^{(c)}$ and $\beta^{(c)}$ are in good agreement with the 
perturbation theory predictions. However, for $50 < z \le 76$, 
the numerical results deviate significantly from the perturbation 
theory predictions. In particular, the absolute differences between 
the theoretical and numerical results at $z=z_{f}$ are   
$|\eta^{(c)(th)}(z_{f}) - \eta^{(c)(num)}(z_{f})| = 0.7670...$ 
and $|\beta^{(c)(th)}(z_{f}) - \beta^{(c)(num)}(z_{f})| = 14.79...$. 
These values are significantly larger than the corresponding 
differences that were obtained for the CQNLS soliton: 
$|\eta^{(th)}(z_{f}) - \eta^{(num)}(z_{f})| = 0.0083...$ 
and $|\beta^{(th)}(z_{f}) - \beta^{(num)}(z_{f})| = 3.99...$.         
The increasing deviations between the numerical and theoretical 
results for $\eta^{(c)}$ and $\beta^{(c)}$ occur together with the 
increase in pulse-shape distortion and in the value of $I(z)$, 
which are seen in Figs. \ref{fig5} and \ref{fig6}. Therefore, the 
results for amplitude and frequency dynamics together with the 
results shown in Figs. \ref{fig5}-\ref{fig7} and similar results 
obtained with other physical parameter values show that transmission 
of the CNLS soliton in waveguide loops with weak frequency-independent 
linear gain, cubic loss, and delayed Raman response is unstable.      
We also note that both the numerical and the theoretical values 
of the CNLS soliton's frequency $\beta^{(c)}(z)$ are much larger than 
the frequency values $\beta(z)$ observed for the CQNLS soliton. 
This finding is in agreement with the prediction of Eq. (\ref{cqnls19}) 
for the decrease in the rate of change of the Raman self-frequency 
shift of the CQNLS soliton for $\eta$ and $\eta_{0}$ values 
near $\eta_{m}$.

{\it Summary.} It is worthwhile to summarize the features of pulse dynamics, 
which show that pulse distortion and transmission destabilization are stronger 
for the CNLS soliton than for the CQNLS soliton. 
\begin{enumerate}         
\item {\it Transmission quality distances.} The distances $z_{q}$ and 
$z_{f}$, which characterize transmission quality and pulse distortion,  
are considerably smaller for the CNLS soliton.   

\item {\it Pulse-shape dynamics.} The radiative hump for the CNLS soliton 
is much more pronounced than the radiative hump for the CQNLS soliton. 


\item {\it The Fourier spectrum.} The Fourier spectrum for the CNLS soliton 
at $z \le z_{f}$ contains pronounced radiative features at high $|\omega|$ 
values, which are not present in the spectrum for the CQNLS soliton at 
$z \le z_{f}$. In addition, the numerically obtained spectrum of the CNLS soliton 
at $z=z_{f}$  is considerably shifted compared with the theoretical prediction, 
whereas the numerically obtained spectrum of the CQNLS soliton is not.

\item {\it Amplitude and frequency dynamics.} The deviations of the numerically 
obtained amplitude and frequency of the soliton from the perturbation theory predictions 
at distances close or equal to $z_{f}$ are significantly larger for the CNLS soliton. 

\end{enumerate} 

Based on these observations, we conclude that transmission instability in waveguide loops 
with weak frequency-independent linear gain, cubic loss, and delayed Raman response 
is weaker for the CQNLS soliton than for the CNLS soliton. We also point out that 
the Raman-induced frequency shift for the CQNLS soliton is much smaller than the 
frequency shift experienced by the CNLS soliton. This fact combined with the weaker pulse 
distortion of the CQNLS soliton indicate that waveguides with focusing cubic 
nonlinearity and defocusing quintic nonlinearity are advantageous compared with 
waveguides with focusing cubic nonlinearity for transmission of highly energetic 
solitons in the presence of weak nonlinear dissipative perturbations.

\subsection{Soliton dynamics in the presence of  
frequency-dependent linear gain-loss, cubic loss, and delayed Raman response}

\label{dynamics3}

We now turn to study soliton propagation in nonlinear waveguides 
with focusing cubic nonlinearity, defocusing quintic nonlinearity, 
weak frequency-dependent linear gain-loss, cubic loss, and delayed Raman 
response. Our main goal is to investigate whether the transmission 
stabilization method that was developed in Ref. \cite{PC2018A} for 
CNLS solitons also works in the nonintegrable case of the CQNLS soliton. 
Furthermore, we compare the transmission stabilization of the CQNLS 
soliton with that of the CNLS soliton in the same nonlinear waveguide 
setup. The propagation of the CQNLS soliton is described by 
the following perturbed CQNLS equation \cite{quintic_model}:    
\begin{eqnarray} &&
i\partial_z\psi+\partial_t^2\psi+2|\psi|^2\psi-\epsilon_q|\psi|^4\psi=
i{\cal F}^{-1}(\hat g(\omega,z) \hat\psi)/2 - i\epsilon_{3}|\psi|^2\psi
+\epsilon_{R}\psi\partial_{t}|\psi|^2,  
\label{cqnls31}
\end{eqnarray} 
where $\omega$ is angular frequency, $\hat\psi$ is the Fourier transform of $\psi$ 
with respect to time, $\hat g(\omega,z)$ is the frequency-dependent linear gain-loss,   
${\cal F}^{-1}$ is the inverse Fourier transform with respect to time, 
and the other notations are the same as in Eq. (\ref{cqnls11}).

The form of the frequency and distance-dependent linear gain-loss $\hat g(\omega,z)$ 
is chosen such that radiation emission effects are mitigated, soliton shape is retained, 
and the value of the amplitude parameter $\eta$ approaches the desired equilibrium value 
$\eta_{0}$. More specifically, we choose the form 
\begin{eqnarray} &&
\hat g(\omega,z) = -g_{L} + \frac{1}{2}\left(g_{0} + g_{L}\right)
\left[\tanh \left\lbrace \rho \left[\omega + \beta(z)+W/2\right] 
\right\rbrace 
\right.
\nonumber \\&&
\left.
- \tanh \left\lbrace \rho \left[\omega + \beta(z)- W/2\right] 
\right\rbrace\right],
\!\!\!\!\!\!\!
\label{cqnls32}
\end{eqnarray} 
where $g_{0}$ is the linear gain coefficient, $\beta(z)$ is the soliton 
frequency at distance $z$, $g_{L}$ is an $O(1)$ positive constant, 
and $g_{0}$, $W$, and $\rho$ satisfy $0<g_{0}\ll 1$, $W \gg 1$, and $\rho\gg 1$. 
The form (\ref{cqnls32}) is similar to the one used in Ref. \cite{PC2018A} for 
stabilization of the CNLS soliton, and also to the form used in 
Refs. \cite{CPN2016,PNT2016,PC2018B} in studies of multisequence 
transmission of CNLS solitons in the presence of various weak dissipative 
nonlinear perturbations. Since $\rho\gg 1$, $\hat g(\omega,z)$ 
can be approximated by the following step function: 
\begin{eqnarray} &&
\hat g(\omega,z) \simeq 
\left\{ \begin{array}{l l}
g_{0} &  \mbox{ if $-\beta(z)-W/2 < \omega < -\beta(z)+W/2$,}\\
(g_{0}-g_{L})/2 & \mbox{ if $\omega=-\beta(z) - W/2 \;\;$
or $\;\;\omega=-\beta(z) + W/2$,} \\ 
-g_{L} &  \mbox{elsewhere.}\\
\end{array} \right. 
\label{cqnls33}
\end{eqnarray}       
We point out that the constant weak linear gain $g_{0}$ in the frequency interval 
$(-\beta(z)-W/2, -\beta(z)+W/2)$ balances the effects of the weak cubic loss, 
such that the soliton amplitude approaches the equilibrium value $\eta_{0}$ with 
increasing distance. Additionally, the strong linear loss $-g_{L}$ suppresses 
emission of radiation with frequencies outside of the interval 
$(-\beta(z)-W/2, -\beta(z)+W/2)$. Furthermore, the Raman perturbation 
is expected to generate a relatively large separation between the soliton's 
spectrum and the radiation's spectrum. As a result, the introduction of the 
frequency-dependent linear gain-loss $\hat g(\omega,z)$ is expected to lead 
to efficient suppression of pulse-shape distortion and to significant enhancement 
of transmission stability. We also note that the flat gain in the interval 
$(-\beta(z)-W/2, -\beta(z)+W/2)$ can be realized by flat-gain amplifiers \cite{Becker99}, 
and the strong loss outside of this interval can be achieved by filters \cite{Becker99} 
or by waveguide impurities \cite{Agrawal2019}.

Using an energy balance calculation, we show in Appendix \ref{appendA} that 
the dynamics of the soliton's amplitude is described by 
\begin{eqnarray} &&
\!\!\!\!\!\!\!\!
\frac{d \eta}{dz}=
\frac{\left(\eta_{m}^{2}-\eta^{2}\right)}{\eta_{m}}
\left\{-g_{L}\mbox{arctanh}\left(\frac{\eta}{\eta_{m}}\right) 
+ \frac{(g_{0}+g_{L})\eta}{\pi\eta_{m}} J_{1}(\eta) 
\right.
\nonumber \\&&
\!\!\!\!\!\!\!\!\!
\left.
-4\epsilon_{3}\eta_{m}^{2}\left[\mbox{arctanh}
\left(\frac{\eta}{\eta_{m}}\right)- 
\frac{\eta}{\eta_{m}} \right]\right\},     
\label{cqnls35}
\end{eqnarray}   
where 
\begin{eqnarray} &&
\!\!\!\!\!\!\!\!
J_{1}(\eta)=
\int_{-\infty}^{\infty} \!\!\! ds \frac{\sin(Ws/2)}{s}  
\nonumber \\&&
\!\!\!\!\!\!\!\!
\int_{-\infty}^{\infty} \frac{dx}{
\left[(1-\eta^{2}/\eta_{m}^{2})^{1/2}\cosh(2x)+1\right]^{1/2}
\left[(1-\eta^{2}/\eta_{m}^{2})^{1/2}\cosh(2x-2\eta s)+1\right]^{1/2}} .  
\label{cqnls36}
\end{eqnarray}  
Additionally, the modification of the frequency-independent linear gain 
term of Eq. (\ref{cqnls11}) to the frequency-dependent linear gain-loss term 
of Eq. (\ref{cqnls31}) does not affect the dynamics of the soliton's 
frequency in first order in $\epsilon_{R}$ and $\epsilon_{3}$.     
Therefore, the dynamics of $\beta$ is still given by Eq. (\ref{cqnls13}).  
As in subsection \ref{dynamics2}, stable steady-state transmission with 
constant amplitudes can be realized by requiring that $\eta=\eta_{0}$ 
with $0 < \eta_{0} < \eta_{m}$ is a stable equilibrium point of 
Eq. (\ref{cqnls35}). This requirement yields the following expression 
for $g_{0}$: 
\begin{equation}
g_{0} = -g_{L}
+ \frac{\pi\eta_{m}}{\eta_{0}J_{1}(\eta_{0})}
\left\{g_{L}\mbox{arctanh}\left(\frac{\eta_{0}}{\eta_{m}}\right) 
+4\epsilon_{3}\eta_{m}^{2}
\left[\mbox{arctanh}\left(\frac{\eta_{0}}{\eta_{m}}\right) 
-\frac{\eta_{0}}{\eta_{m}}\right]
\right\}.  
\label{cqnls37}
\end{equation}      
Thus, the complete description of amplitude dynamics subject to 
the requirement on $\eta_{0}$ is given by Eqs. (\ref{cqnls35}), 
(\ref{cqnls36}), and (\ref{cqnls37}). Furthermore, using 
Eq. (\ref{cqnls37}), we can write the equation for the 
dynamics of $\eta$ in the form 
\begin{eqnarray} &&
\!\!\!\!\!\!\!\!
\frac{d \eta}{dz}=
\left(1 - \frac{\eta^{2}}{\eta_{m}^{2}}\right)\eta
\tilde J_{1}\left(\frac{\eta}{\eta_{m}}\right)  
\left[G\left(\frac{\eta}{\eta_{m}}\right)  
-G\left(\frac{\eta_{0}}{\eta_{m}}\right)\right],        
\label{cqnls38}
\end{eqnarray}   
where $\tilde J_{1}(\eta/\eta_{m}) = J_{1}(\eta)$, and      
\begin{equation}
G(x) = \frac{1}{x \tilde J_{1}(x)}
\left\{4\epsilon_{3}\eta_{m}^{2}
\left[x - \mbox{arctanh}(x)\right] - g_{L}\mbox{arctanh}(x) 
\right\}.  
\label{cqnls39}
\end{equation}      
It is possible to show that $\eta=\eta_{0}$ is the only equilibrium 
point of Eq. (\ref{cqnls38}) with a nonnegative amplitude value and 
that it is a stable equilibrium point for a range of physical parameter 
values, including the values considered in the current paper.

Let us discuss the approximate form of the equations for 
$d\eta/dz$ and $d\beta/dz$ in the following two cases that were 
also considered in subsection \ref{dynamics2}. 
(1) The limit $\epsilon_{q} \rightarrow 0^{+}$ 
$(\eta_{m} \rightarrow \infty)$ that corresponds to the CNLS limit. 
(2) The case where $\eta$ is close to $\eta_{0}$, and $\eta_{0}$ is 
close to $\eta_{m}$ (that is, for highly energetic CQNLS solitons  
near the flat-top soliton limit). 
In the limit $\epsilon_{q} \rightarrow 0^{+}$, we retrieve the 
equations for the dynamics of the CNLS soliton's amplitude 
and frequency that were obtained in Ref. \cite{PC2018A}. 
More specifically, we obtain 
\begin{eqnarray} &&
\frac{d\eta}{dz} \simeq \left\lbrace g_{L}\left[\frac{\tanh(V)}{\tanh\left(V_{0}\right)} - 1 \right]
+\frac{4}{3}\epsilon_{3}\left[\eta_{0}^{2}\frac{\tanh(V)}{\tanh\left(V_{0}\right)} - \eta^{2} \right]
\right\rbrace \eta, 
\label{cqnls40}
\end{eqnarray}  
where $V=\pi W/(4\eta)$ and $V_{0}=\pi W/(4\eta_{0})$.  
Additionally, since the dynamics of $\beta$ is described 
by Eq. (\ref{cqnls13}), the approximate form of this equation 
in the limit $\epsilon_{q} \rightarrow 0^{+}$ is still given by 
Eq. (\ref{cqnls17}). In case (2), we linearize Eq. (\ref{cqnls38}) 
about $\eta=\eta_{0}$, and obtain 
\begin{eqnarray} &&
\!\!\!\!\!\!\!\!
\frac{d \eta}{dz} \simeq  
\left\{-(12\epsilon_{3}\eta_{0}^{2} + g_{L}) 
+2\left(4\epsilon_{3}\eta_{m} + \frac{g_{L}}{\eta_{m}}\right)
\mbox{arctanh}\left(\eta_{0}/\eta_{m}\right)
\right. 
\nonumber \\&&
\left.
-\frac{G\left(\eta_{0}/\eta_{m}\right)}{\eta_{m}^{2}}
\left[\left(\eta_{m}^{2}-3\eta_{0}^{2}\right)J_{1}(\eta_{0}) 
+\left(\eta_{m}^{2}-\eta_{0}^{2}\right)\eta_{0}J'_{1}(\eta_{0})  
\right]\right\}
\delta\eta,
\label{cqnls41}
\end{eqnarray}       
where $\delta\eta=\eta - \eta_{0}$. In addition, in this case, the 
approximate form of the equation for $\beta$ is given by Eq. (\ref{cqnls19}).

{\it Numerical simulations}. The prediction for enhanced stability of 
the CQNLS soliton due to the interplay between frequency-dependent linear 
gain-loss and the Raman-induced self-frequency shift is based on a heuristic 
argument. It is therefore important to check the validity of this 
prediction by numerical simulations with Eqs. (\ref{cqnls31}) and 
(\ref{cqnls32}). To enable comparison with the simulations results 
of Sec. \ref{dynamics2}, we use a simulation setup that is similar 
to the one used in Sec. \ref{dynamics2}. More specifically, we solve 
Eqs. (\ref{cqnls31}) and (\ref{cqnls32}) numerically on a time domain 
$[t_{\mbox{min}},t_{\mbox{max}}]=[-400,400]$ with the split-step method 
and with periodic boundary conditions. Since we use periodic boundary 
conditions, the simulation describes propagation in a closed waveguide 
loop. Therefore, the simulation and the analysis are highly relevant 
for most long-distance optical waveguide transmission experiments, 
which are carried out in closed loops \cite{Mollenauer2006,Mollenauer97,Nakazawa2000,Nakazawa91,Mollenauer2003}. 
The initial condition is in the form of the CQNLS soliton of Eq. (\ref{cqnls2}) 
with parameter values $\eta(0)$, $\beta(0)=0$, $y(0)=0$, and 
$\alpha(0)=0$. The comparison with the results of Sec. \ref{dynamics2} 
is facilitated by choosing similar parameter values to the ones used 
in that section. That is, we consider here the results of the simulation  
with $\epsilon_{q}=0.5$, $\epsilon_{3}=0.01$, $\epsilon_{R}=0.04$, 
$\eta_{0}=1.2$, and $\eta(0)=0.9$. Since $\eta_{0}=1.2$ is close to 
$\eta_{m}=1.2247...$, we are in fact considering transmission 
stabilization of highly energetic solitons near the flat-top soliton limit. 
The values of the parameters $W$, $\rho$, and $g_{L}$ in Eq. (\ref{cqnls32}) 
are the same as the values used in Refs. \cite{PC2018A,CPN2016,PNT2016,PC2018B}
in studies of stabilization of CNLS solitons with different perturbed CNLS and 
coupled-CNLS models: $W=10$, $\rho=10$, and $g_{L}=0.5$. We point out that numerical 
simulations with other values of the physical parameters yield similar 
results. We run the simulation up to a final distance $z_{f}=2000$, 
and check if the value of the transmission quality integral $I(z)$ 
remains smaller than 0.075 throughout the simulation.

\begin{figure}[ptb]
\begin{tabular}{cc}
\epsfxsize=8.1cm  \epsffile{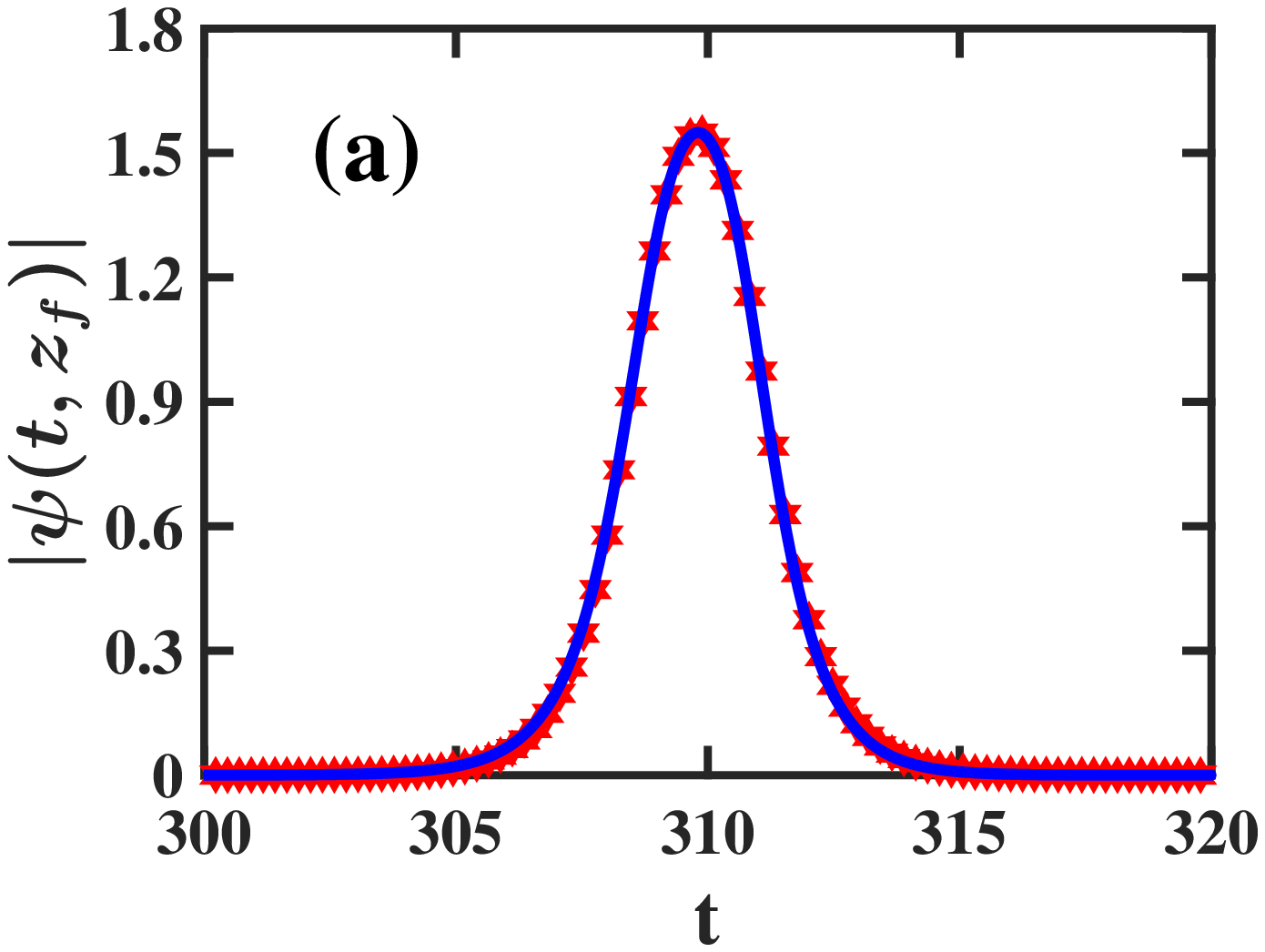} &
\epsfxsize=8.1cm  \epsffile{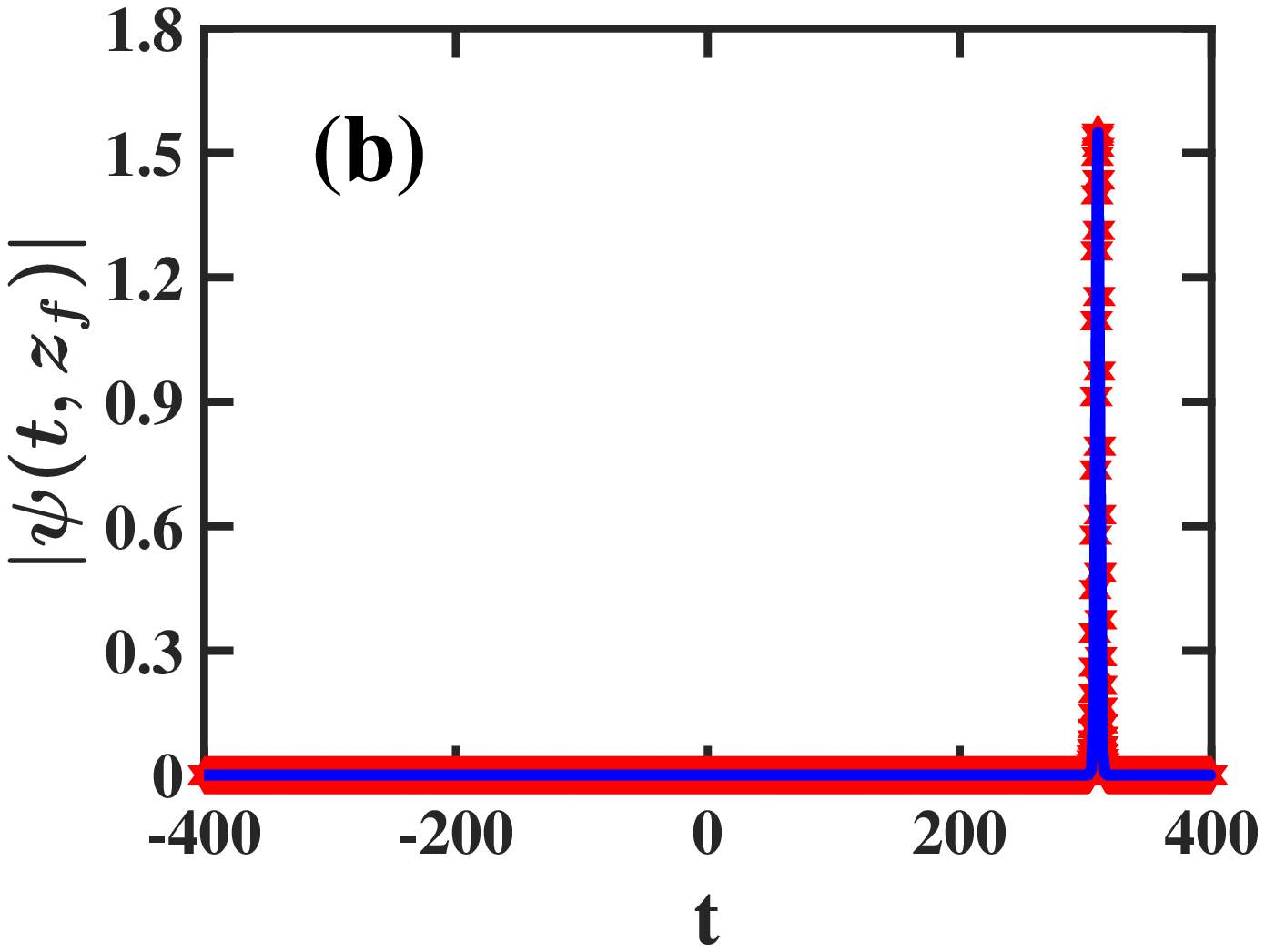} \\
\epsfxsize=8.1cm  \epsffile{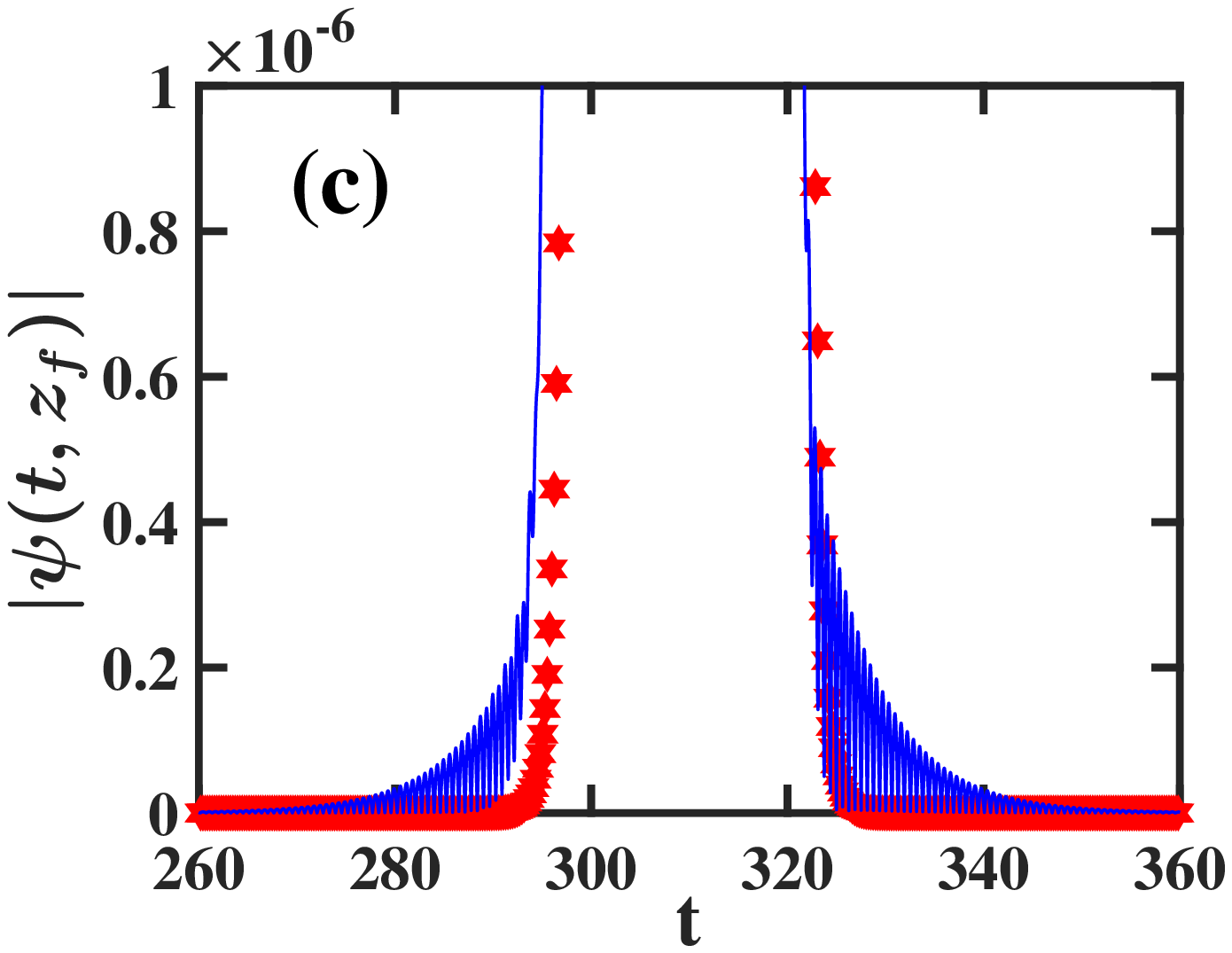} 
\end{tabular}
\caption{The pulse shape $|\psi(t,z_{f})|$, where $z_{f}=2000$, for 
propagation of a CQNLS soliton in a waveguide loop with weak 
frequency-dependent linear gain-loss, cubic loss, and delayed Raman response. 
The physical parameter values are $\epsilon_{q}=0.5$, $\epsilon_{3}=0.01$, 
$\epsilon_{R}=0.04$, $\eta_{0}=1.2$, $\eta(0)=0.9$, $W=10$, $\rho=10$, 
and $g_{L}=0.5$. The solid blue curve corresponds to the result obtained 
by the numerical simulation with Eqs. (\ref{cqnls31}) and (\ref{cqnls32}).  
The red stars represent the perturbation theory prediction of 
Eqs. (\ref{Iz1}) and (\ref{cqnls38}).}
\label{fig9}
\end{figure}

\begin{figure}[ptb]
\begin{tabular}{cc}
\epsfxsize=8.5cm  \epsffile{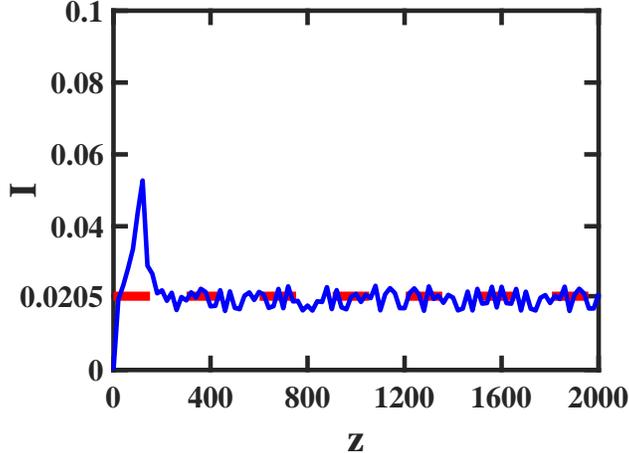} 
\end{tabular}
\caption{The $z$ dependence of the transmission quality integral $I(z)$ obtained 
by the numerical simulation with Eqs. (\ref{cqnls31}) and (\ref{cqnls32}) for the 
same waveguide setup as in Fig. \ref{fig9}. The solid blue curve represents $I(z)$ 
and the dashed red horizontal line corresponds to $\langle I(z) \rangle$.}
\label{fig10}
\end{figure}

Figure \ref{fig9} shows the numerical simulation's result for the 
pulse shape $|\psi(t,z)|$ at $z=z_{f}$ together with the perturbation 
theory prediction of Eqs. (\ref{Iz1}) and (\ref{cqnls38}). 
We observe that the numerically obtained pulse shape at $z=z_{f}$ is 
very close to the theoretical prediction and no appreciable pulse distortion 
due to radiation emission is present [see Figs. \ref{fig9}(a) and \ref{fig9}(b)]. 
Furthermore, as seen in Fig. \ref{fig9}(c), the deviation of the simulation's 
result from the theoretical prediction at $z=z_{f}$ is smaller than $10^{-6}$ 
for all $t$ values. Surprisingly, this deviation is of the same order of 
magnitude as the deviation observed in Ref. \cite{PC2018A} for a CNLS 
soliton with an equilibrium power value that is smaller than the CQNLS 
soliton's equilibrium power in the current simulation by a factor of $2.8086$. 
The enhanced transmission stability of the CQNLS soliton is also clearly 
visible in Fig. \ref{fig10}, which shows the numerically obtained $I(z)$ curve. 
As seen in this figure, the value of $I(z)$ is smaller than 0.053 throughout 
the propagation. Additionally, the average $\langle I(z) \rangle$, 
which is defined by $\langle I(z) \rangle \equiv \int_{0}^{z_{f}} dz' I(z') /z_{f}$, 
is $\langle I(z) \rangle=0.02053$, in agreement with the very weak pulse 
distortion observed in Fig. \ref{fig9}.

\begin{figure}[ptb]
\begin{center}
\begin{tabular}{cc}
\epsfxsize=8.2cm  \epsffile{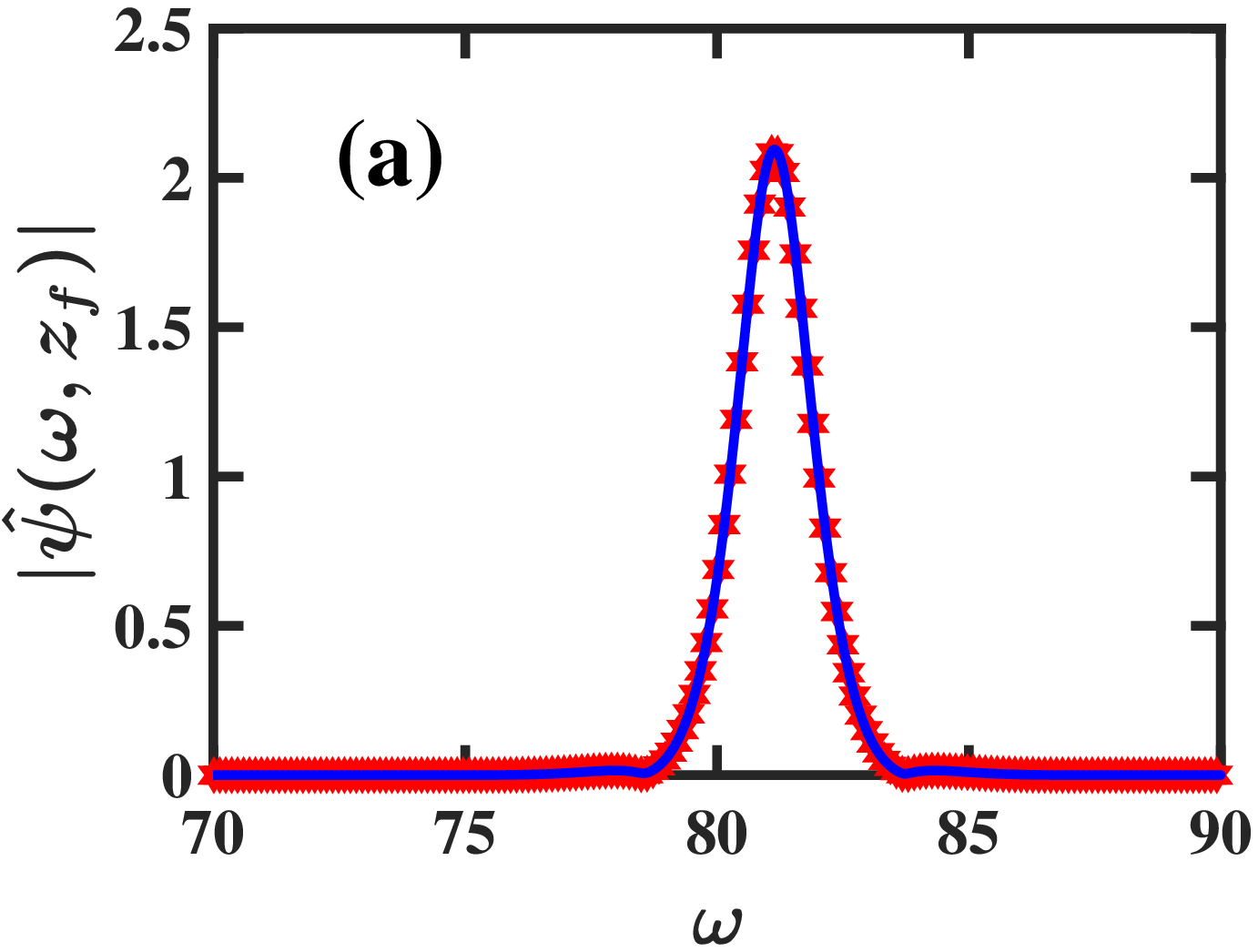} &
\epsfxsize=8.2cm  \epsffile{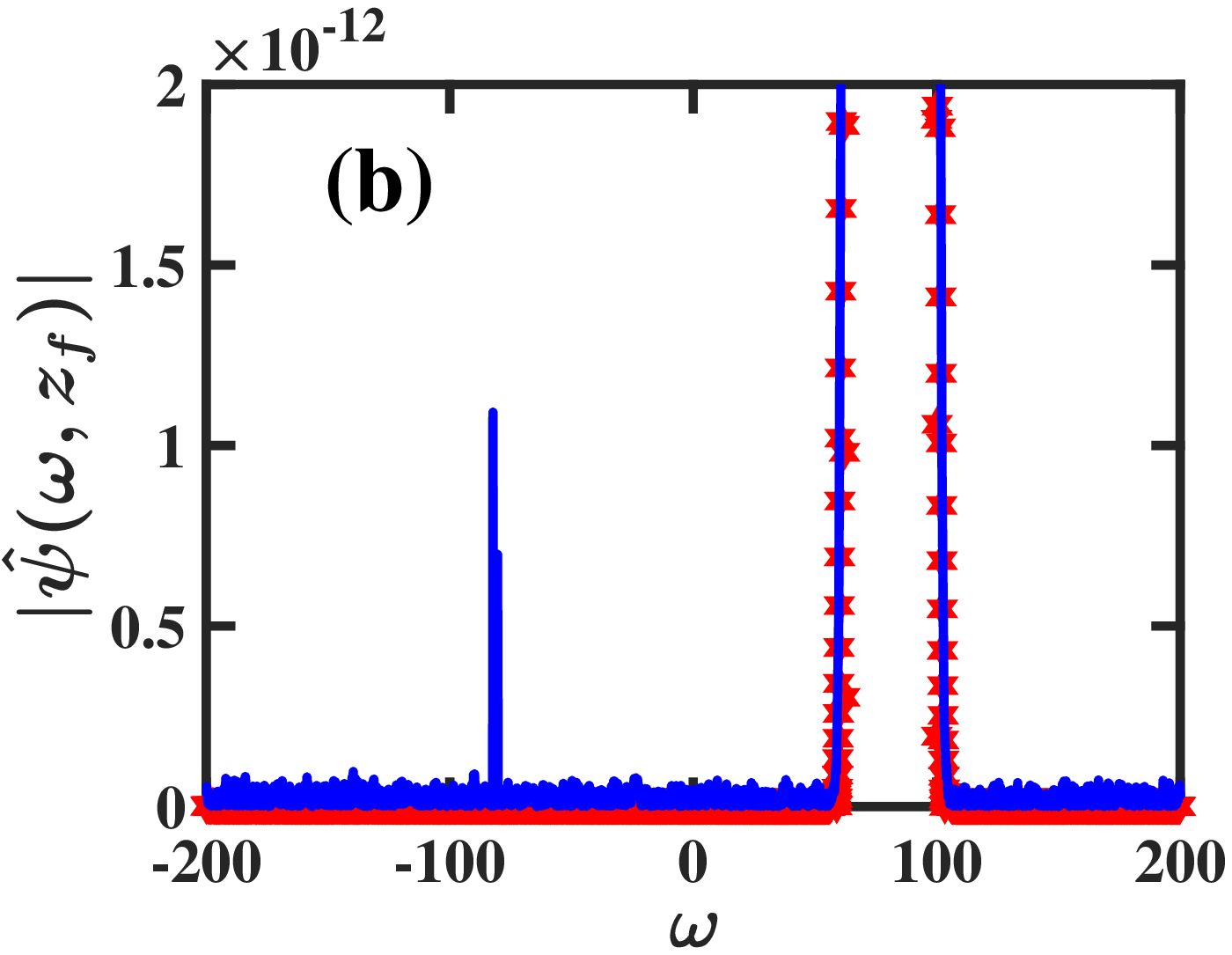}    
\end{tabular}
\end{center}
\caption{The shape of the Fourier spectrum $|\hat\psi(\omega,z_{f})|$,  
where $z_{f}=2000$, for propagation of a CQNLS soliton in a waveguide loop 
with weak frequency-dependent linear gain-loss, cubic loss, and delayed Raman response. 
The parameter values are the same as in Fig. \ref{fig9}. 
The solid blue curve represents the result obtained by numerical 
solution of Eqs. (\ref{cqnls31}) and (\ref{cqnls32}), and the 
red stars correspond to the perturbation theory prediction of 
Eqs. (\ref{Iz3}), (\ref{cqnls13}), and (\ref{cqnls38}).}                        
\label{fig11}
\end{figure}

Further insight into enhancement of transmission stability is gained 
by analyzing the Fourier spectrum of the pulse $|\hat\psi(\omega,z)|$. 
Figure \ref{fig11} shows a comparison between the numerically obtained 
Fourier spectrum $|\hat\psi(\omega,z)|$ at $z=z_{f}$ and the perturbation 
theory prediction of Eqs. (\ref{Iz3}), (\ref{cqnls13}), and (\ref{cqnls38}). 
The agreement between the theoretical and the 
numerical results is excellent. More specifically, the spectrum obtained 
in the simulation does not contain any radiative peaks, such at the peaks 
seen in Figs. \ref{fig3} and  \ref{fig7}. Furthermore, the graph of 
$|\hat\psi(\omega,z_{f})|$ vs $\omega$ does not contain any significant 
oscillations, such as the oscillations observed for the CNLS soliton in 
the absence of perturbation-induced frequency shifting in Ref. \cite{PC2018A}. 
These findings together with the findings in Sec. \ref{dynamics2} show that 
the interplay between perturbation-induced shifting of the soliton's frequency 
and frequency-dependent linear gain-loss indeed leads to enhanced 
transmission stability of the CQNLS soliton, and that it does so even for 
highly energetic CQNLS solitons near the flat-top soliton limit. Thus, 
the stabilization method that was developed in Ref. \cite{PC2018A} 
for solitons of the integrable CNLS equation also works for solitons 
of the nonintegrable CQNLS equation far from the CNLS limit. 
Transmission stabilization takes place in the following manner. 
The perturbation-induced frequency shift experienced by the soliton 
(due to the Raman effect, for example) leads to the separation of the 
soliton's spectrum from the radiation's spectrum, while the 
frequency-dependent linear gain-loss leads to efficient 
suppression of the emitted radiation.

\begin{figure}[ptb]
\begin{tabular}{cc}
\epsfxsize=8.2cm  \epsffile{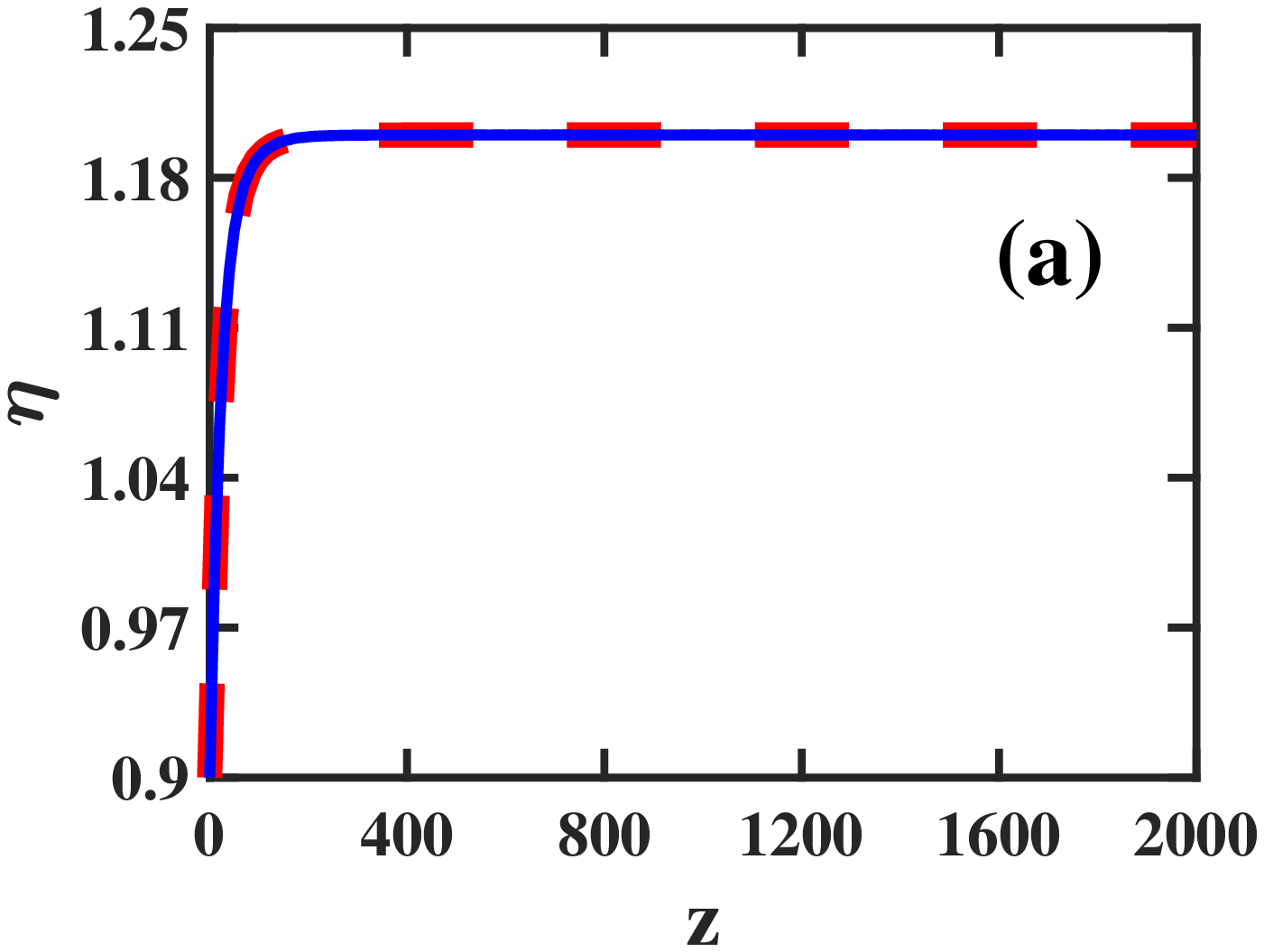} &
\epsfxsize=8.2cm  \epsffile{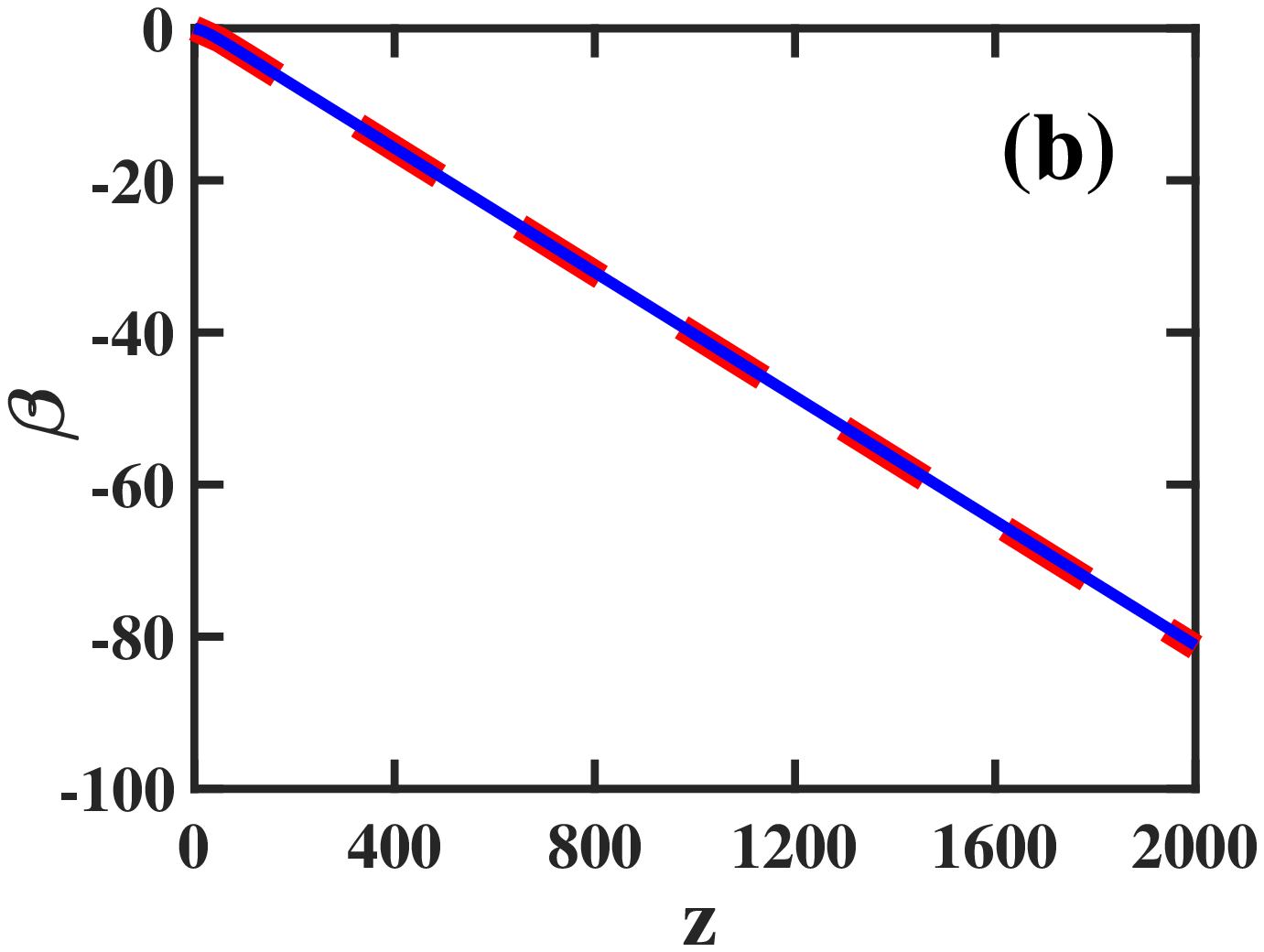}
\end{tabular}
\caption{The $z$ dependence of the CQNLS soliton's amplitude $\eta(z)$ (a) 
and frequency $\beta(z)$ (b) for the same waveguide setup as in 
Figs. \ref{fig9}-\ref{fig11}. The solid blue curves represent the results 
obtained by numerical solution of Eqs. (\ref{cqnls31}) and (\ref{cqnls32}). 
The dashed red curves correspond to the perturbation theory predictions 
of Eq. (\ref{cqnls38}) in (a), and of Eqs. (\ref{cqnls13}) 
and (\ref{cqnls38}) in (b).}
\label{fig12}
\end{figure}

The enhanced transmission stability in waveguides with frequency-dependent 
linear gain-loss and delayed Raman response has important consequences 
for the dynamics of the CQNLS soliton's amplitude and frequency. 
Figure \ref{fig12} shows the $z$ dependence of the soliton's amplitude 
and frequency obtained by numerical solution of Eqs. (\ref{cqnls31}) and (\ref{cqnls32})  
along with the perturbation theory predictions of Eqs. (\ref{cqnls38}) 
and (\ref{cqnls13}). The agreement between the perturbation theory predictions 
and the simulation's results is excellent. In particular, the numerically obtained 
amplitude value tends to the equilibrium value $\eta_{0}=1.2$ at short distances 
and continues to be close to this value throughout the propagation, in very good 
agreement with the theoretical prediction. In addition, the numerically obtained 
value of the soliton's frequency stays close to the $z$-dependent value predicted 
by Eqs. (\ref{cqnls13}) and (\ref{cqnls38}). Therefore, the efficient suppression 
of radiation emission and the enhanced stability of the CQNLS soliton in waveguides 
with weak frequency-dependent linear gain-loss, cubic loss, and delayed Raman response 
enable observation of stable amplitude and frequency dynamics along significantly 
larger distances compared with the distances obtained with the waveguide setups 
of section \ref{dynamics2}, which relied on weak frequency-independent linear gain.

{\it Comparison with transmission stability of the CNLS soliton}.   
Let us compare the results of the numerical simulation for transmission 
stabilization of the CQNLS soliton with the results of simulations 
for transmission stabilization of the CNLS soliton in the same optical 
waveguide setup. This comparison is very useful in determining whether 
the presence of defocusing quintic nonlinearity improves the transmission 
stabilization method that is based on the interplay between perturbation-induced 
frequency shifting and frequency-dependent linear gain-loss. The comparison is 
realized by performing numerical simulations with the following 
perturbed CNLS model \cite{PC2018A}:  
\begin{eqnarray} &&
i\partial_z\psi+\partial_t^2\psi+2|\psi|^2\psi=
i{\cal F}^{-1}(\hat g(\omega,z) \hat\psi)/2 - i\epsilon_{3}|\psi|^2\psi
+\epsilon_{R}\psi\partial_{t}|\psi|^2,  
\label{cqnls45}
\end{eqnarray} 
where $\hat g(\omega,z)$ is given by Eq. (\ref{cqnls32}). Note that 
Eq. (\ref{cqnls45}) is just Eq. (\ref{cqnls31}) without the quintic 
nonlinearity term. The initial condition for the simulation is in 
the form of the fundamental CNLS soliton (\ref{cqnls26}). 
In Ref. \cite{PC2018A}, we used the adiabatic perturbation theory 
for the CNLS soliton to show that the dynamics of its amplitude is 
described by 
\begin{eqnarray} &&
\frac{d\eta^{(c)}}{dz} = 
\left\lbrace g_{L}\left[\frac{\tanh(V^{(c)})}{\tanh(V^{(c)}_{0})} 
- 1 \right]
+\frac{4}{3}\epsilon_{3}\left[\eta_{0}^{(c)2}\frac{\tanh(V^{(c)})}{\tanh(V^{(c)}_{0})} 
- \eta^{(c)2} \right]\right\rbrace \eta^{(c)}, 
\label{cqnls46}
\end{eqnarray}  
where $V^{(c)}=\pi W/(4\eta^{(c)})$, $V^{(c)}_{0}=\pi W/(4\eta^{(c)}_{0})$, and 
\begin{equation}
g_{0} = -g_{L} 
+\frac{\left(g_{L} + 4\epsilon_{3}\eta_{0}^{(c)2}/3\right)}
{\tanh(V^{(c)}_{0})}.  
\label{cqnls47}
\end{equation}           
Additionally, the dynamics of the soliton's frequency is still given by 
Eq. (\ref{cqnls28}). It was shown in Ref. \cite{PC2018A} that $\eta^{(c)}=\eta^{(c)}_{0}$ 
is the only equilibrium point of Eq. (\ref{cqnls46}) with a nonnegative 
amplitude value and that it is a stable equilibrium point.

We solve Eqs. (\ref{cqnls45}) and (\ref{cqnls32}) numerically on a time domain 
$[t_{\mbox{min}},t_{\mbox{max}}]=[-400,400]$ with periodic boundary conditions,  
and with the initial condition (\ref{cqnls26}) with parameter values $\eta^{(c)}(0)$, 
$\beta^{(c)}(0)=0$, $y^{(c)}(0)=0$, and $\alpha^{(c)}(0)=0$. To enable a fair 
comparison with stabilization of the CQNLS soliton, we use the same parameter 
values that were used in the simulation with the perturbed CQNLS 
model (\ref{cqnls31}), i.e., $\epsilon_{3}=0.01$, $\epsilon_{R}=0.04$, 
$W=10$, $\rho=10$, and $g_{L}=0.5$. Additionally, we require that the initial 
power and the equilibrium power of the CNLS soliton would be equal to the 
initial power and the equilibrium power of the CQNLS soliton. Therefore, 
the values of $\eta_{0}^{(c)}$ and $\eta^{(c)}(0)$ used in the simulation 
with Eqs. (\ref{cqnls45}) and (\ref{cqnls32}) are $\eta_{0}^{(c)}=2.8086...$ 
and $\eta^{(c)}(0)=1.1503...$.
Due to the large frequency shift experienced by the CNLS soliton, 
we must employ a larger frequency domain, and smaller time and distance 
steps. As a result, the largest value of $z_{f}$ that can be used with 
a reasonable computation time is restricted to $z_{f}\simeq 200$ \cite{CNLS2}. 
Therefore, we run the simulation up to the final distance $z_{f}=200$.

\begin{figure}[ptb]
\begin{tabular}{cc}
\epsfxsize=8.1cm  \epsffile{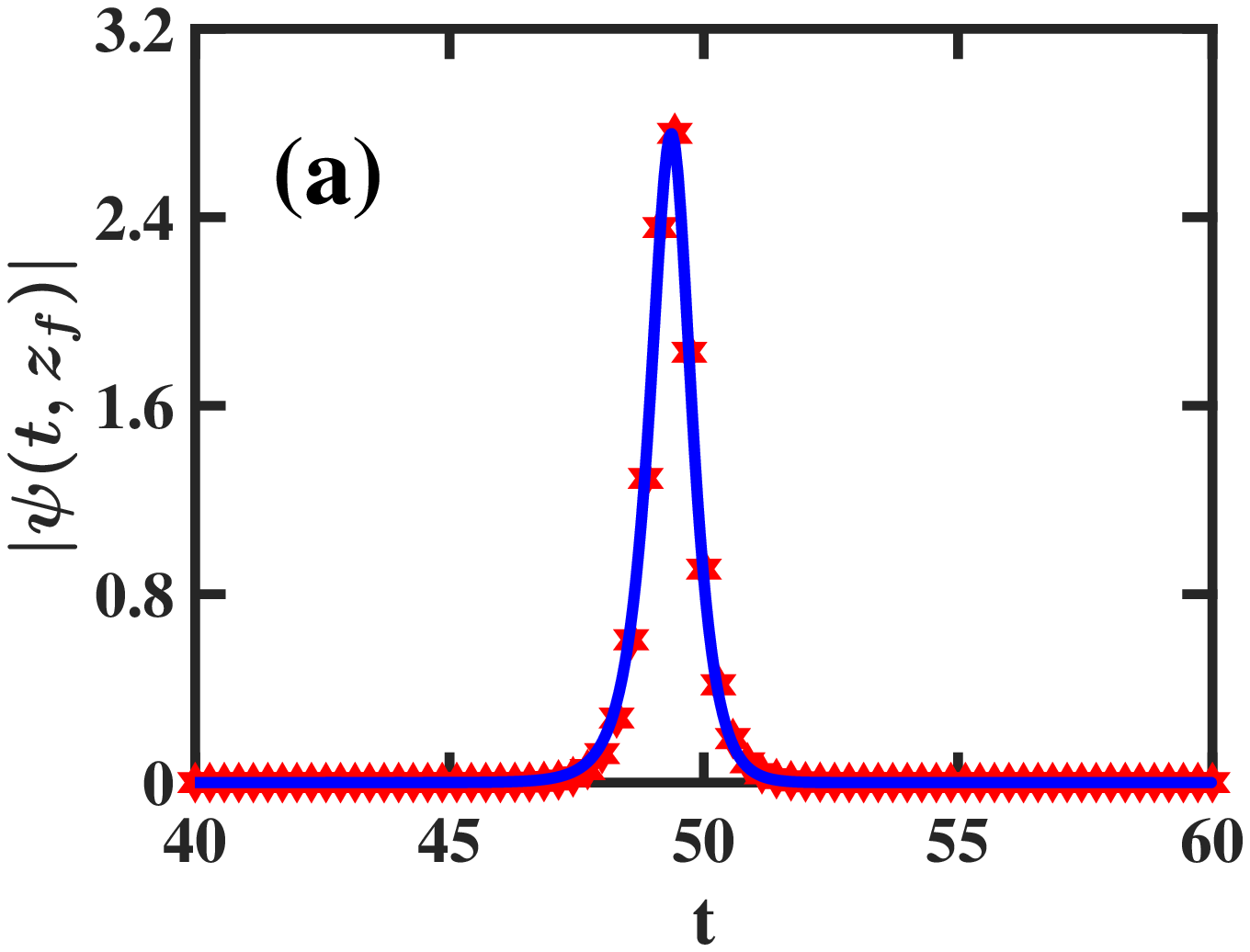} &
\epsfxsize=8.1cm  \epsffile{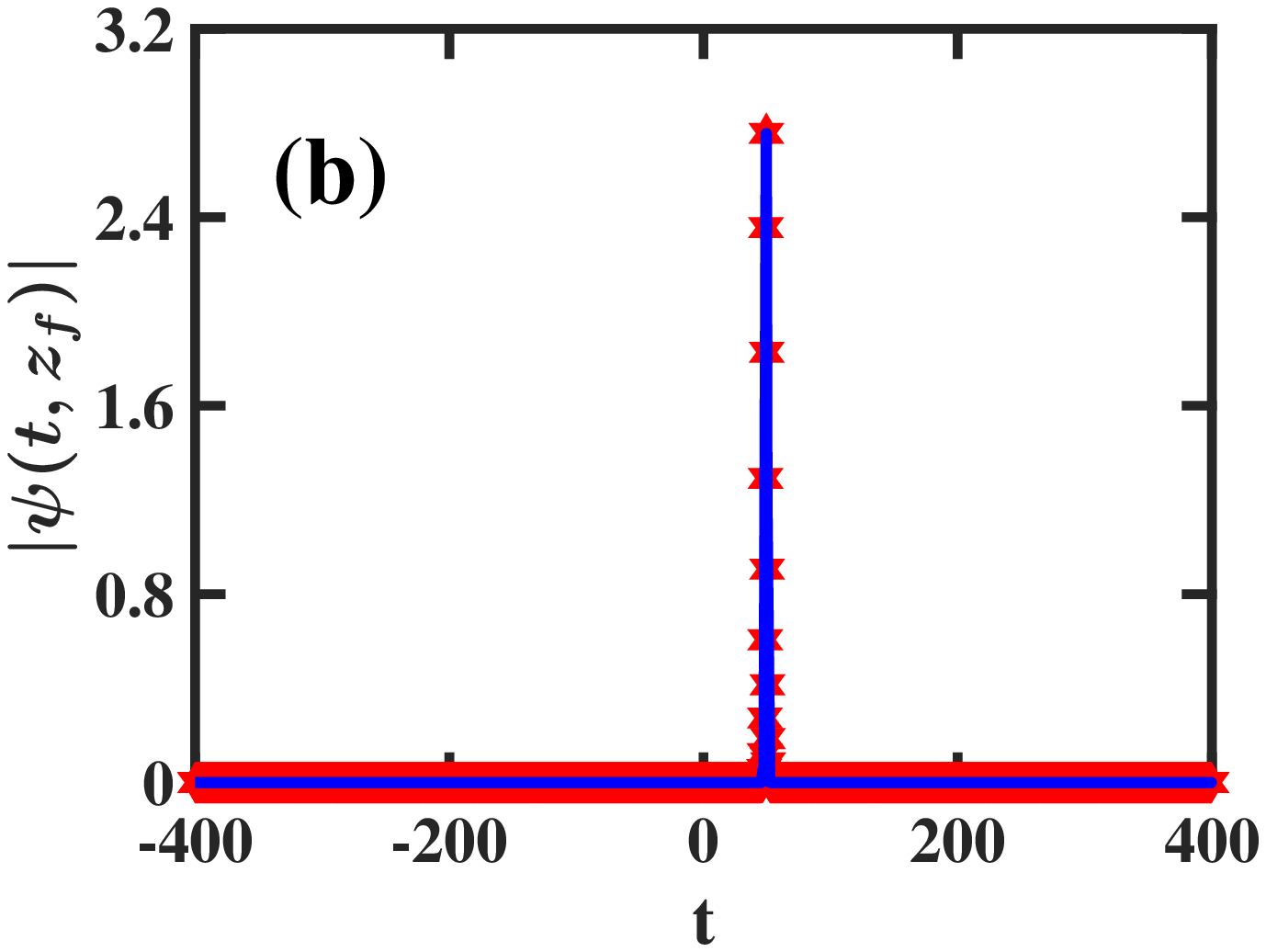} \\
\epsfxsize=8.1cm  \epsffile{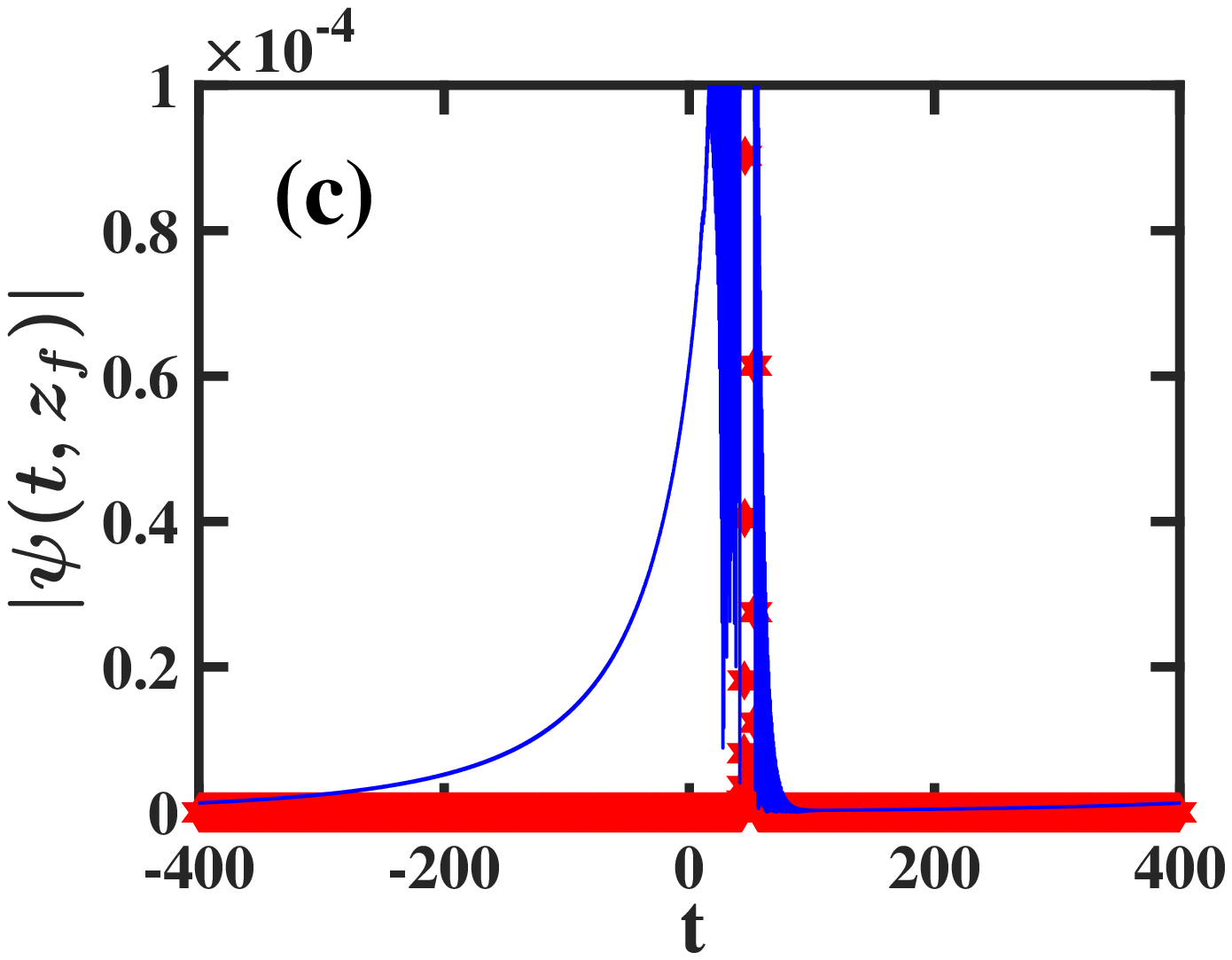} &
\epsfxsize=8.1cm  \epsffile{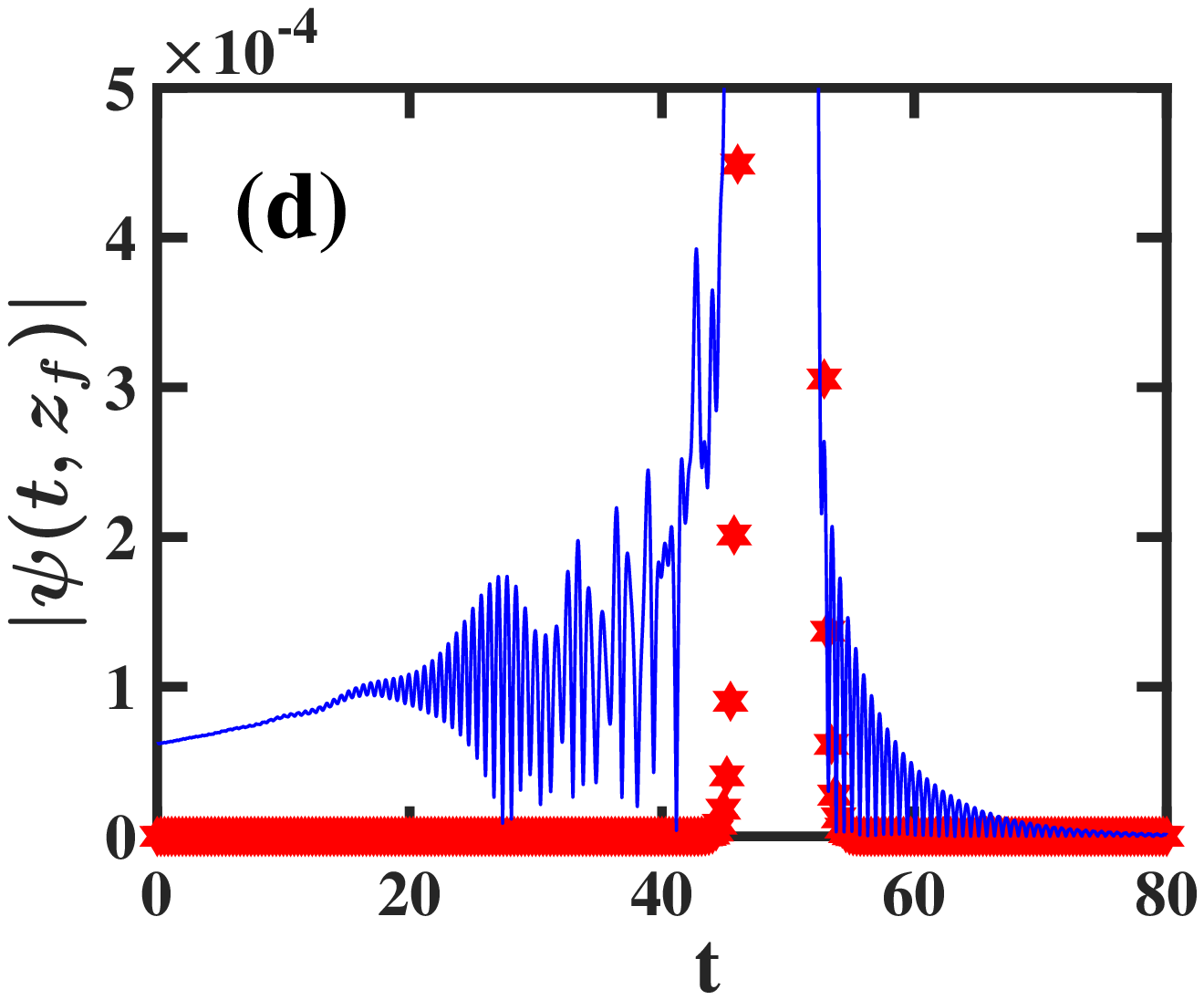} 
\end{tabular}
\caption{The pulse shape $|\psi(t,z_{f})|$, where $z_{f}=200$, for 
propagation of a CNLS soliton in a waveguide loop with weak 
frequency-dependent linear gain-loss, cubic loss, and delayed Raman response. 
The parameter values are $\epsilon_{q}=0.5$, $\epsilon_{3}=0.01$, 
$\epsilon_{R}=0.04$, $\eta_{0}^{(c)}=2.8086...$, $\eta^{(c)}(0)=1.1503...$,  
$W=10$, $\rho=10$, and $g_{L}=0.5$. The solid blue curve corresponds to the result 
obtained by the numerical simulation with Eqs. (\ref{cqnls45}) and (\ref{cqnls32}).  
The red stars represent the perturbation theory prediction of 
Eqs. (\ref{Iz5}) and (\ref{cqnls46}).}
\label{fig13}
\end{figure}

\begin{figure}[ptb]
\begin{tabular}{cc}
\epsfxsize=8.5cm  \epsffile{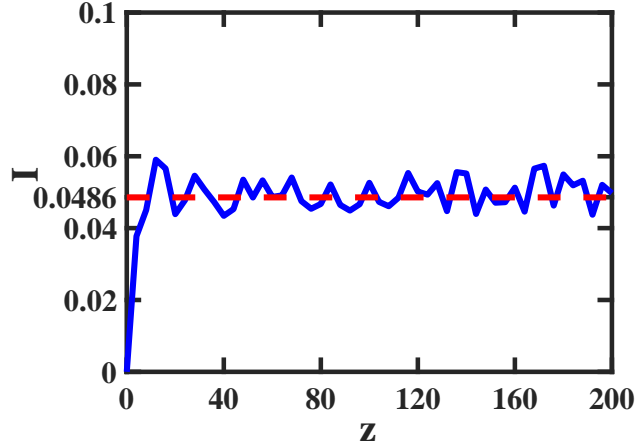} 
\end{tabular}
\caption{The $z$ dependence of the transmission quality integral $I(z)$ obtained 
by the numerical simulation with Eqs. (\ref{cqnls45}) and (\ref{cqnls32}) for the 
same waveguide setup as in Fig. \ref{fig13}. The solid blue curve represents $I(z)$ 
and the dashed red horizontal line represents $\langle I(z) \rangle$.}
\label{fig14}
\end{figure}

The numerical simulation's result for the pulse shape $|\psi(t,z)|$ at $z=z_{f}=200$ 
is shown in Fig. \ref{fig13} together with the theoretical prediction of 
Eqs. (\ref{Iz5}) and (\ref{cqnls46}). As seen in Figs. \ref{fig13}(a) and 
\ref{fig13}(b), the numerically obtained pulse shape at $z=z_{f}$ is 
close to the theoretical prediction and no significant radiation-induced 
pulse distortion is found. However, the comparison in Figs. \ref{fig13}(c) 
and \ref{fig13}(d) for small $|\psi(t,z_{f})|$ values reveals the presence 
of a small radiative tail and weak pulse distortion, which leads to an  
order $10^{-4}$ deviation of the numerical result from the theoretical 
prediction. This deviation is larger by 2 orders of magnitude compared 
with the deviation found for the CQNLS soliton at $z_{f}=2000$ in the 
same waveguide setup [see Fig. \ref{fig9}(c)]. On the other hand, the 
deviation is significantly smaller than the order 0.1 deviation observed 
for the CNLS soliton at $z_{f}=76$ in waveguides with frequency-independent 
linear gain [see Fig. \ref{fig5}(f)]. The latter comparison indicates 
that transmission of energetic CNLS solitons in waveguides with weak 
frequency-dependent linear gain-loss and delayed Raman response is stable. 
However, due to the relatively small value of $z_{f}$ in the current simulation, 
further numerical simulations with more powerful computers are required 
for verifying this conclusion. The conclusion about transmission 
stability of the CNLS soliton is also supported 
by the $I(z)$ curve for the numerical simulation, which is shown in 
Fig. \ref{fig14}. As seen in this figure, the value of $I(z)$ 
remains smaller than 0.059 throughout the propagation, and 
$\langle I(z) \rangle=0.04855$. Note that the value of $\langle I(z) \rangle$ 
is larger by a factor of 2.4 compared with the corresponding value for the 
CQNLS soliton, in accordance with the stronger pulse distortion experienced 
by the CNLS soliton.

\begin{figure}[ptb]
\begin{center}
\begin{tabular}{cc}
\epsfxsize=8.2cm  \epsffile{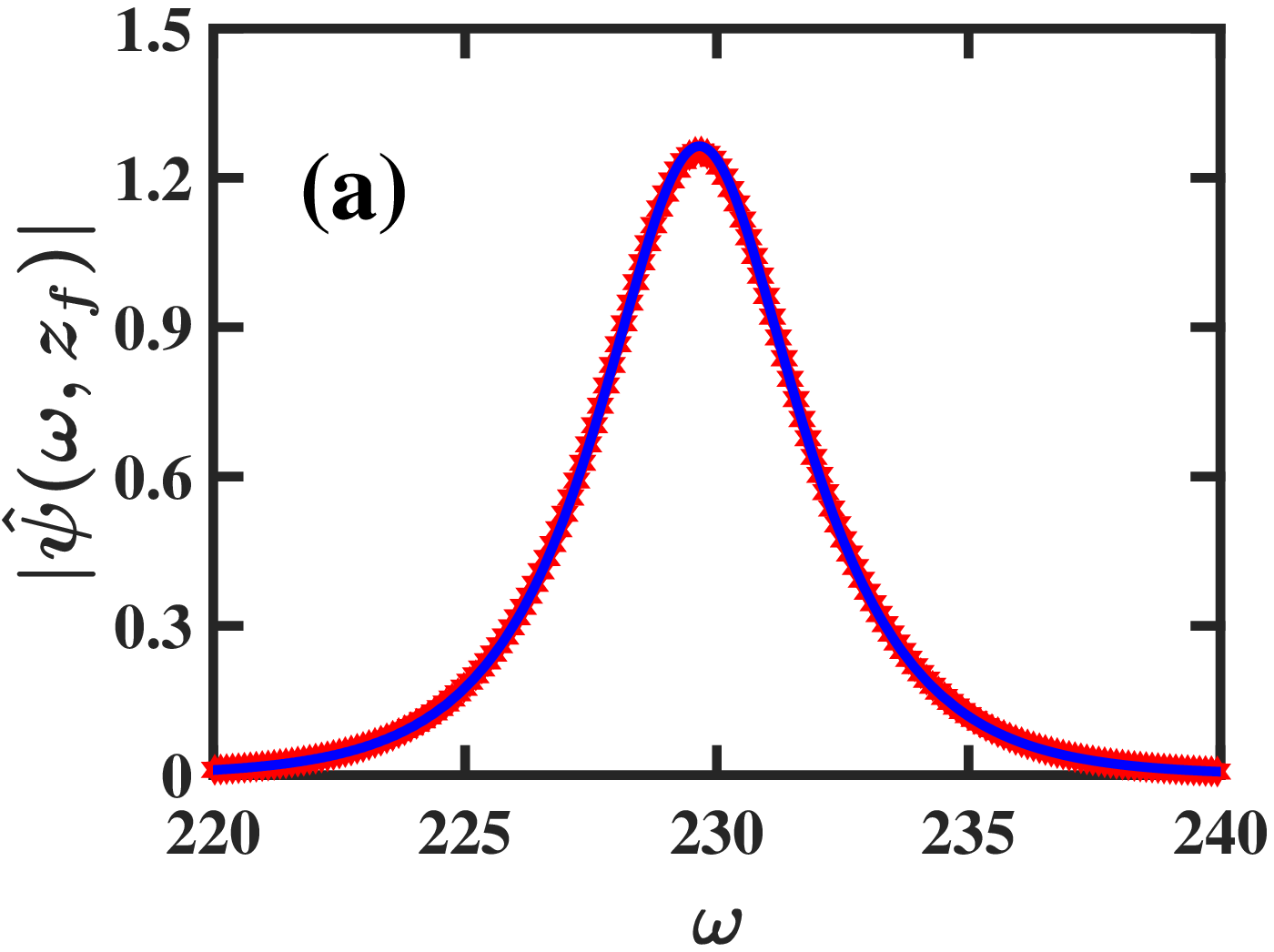} &
\epsfxsize=8.2cm  \epsffile{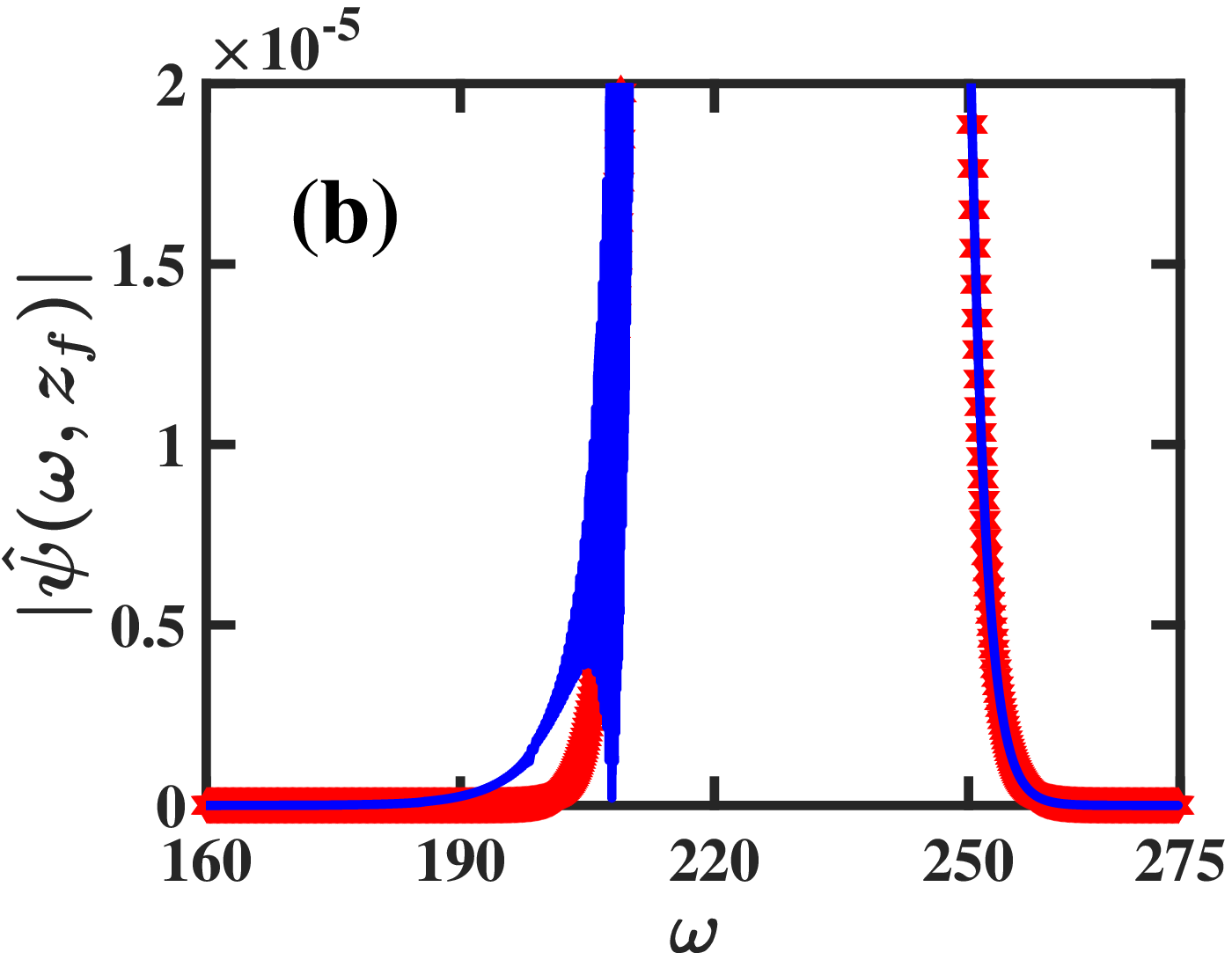} \\
\epsfxsize=8.2cm  \epsffile{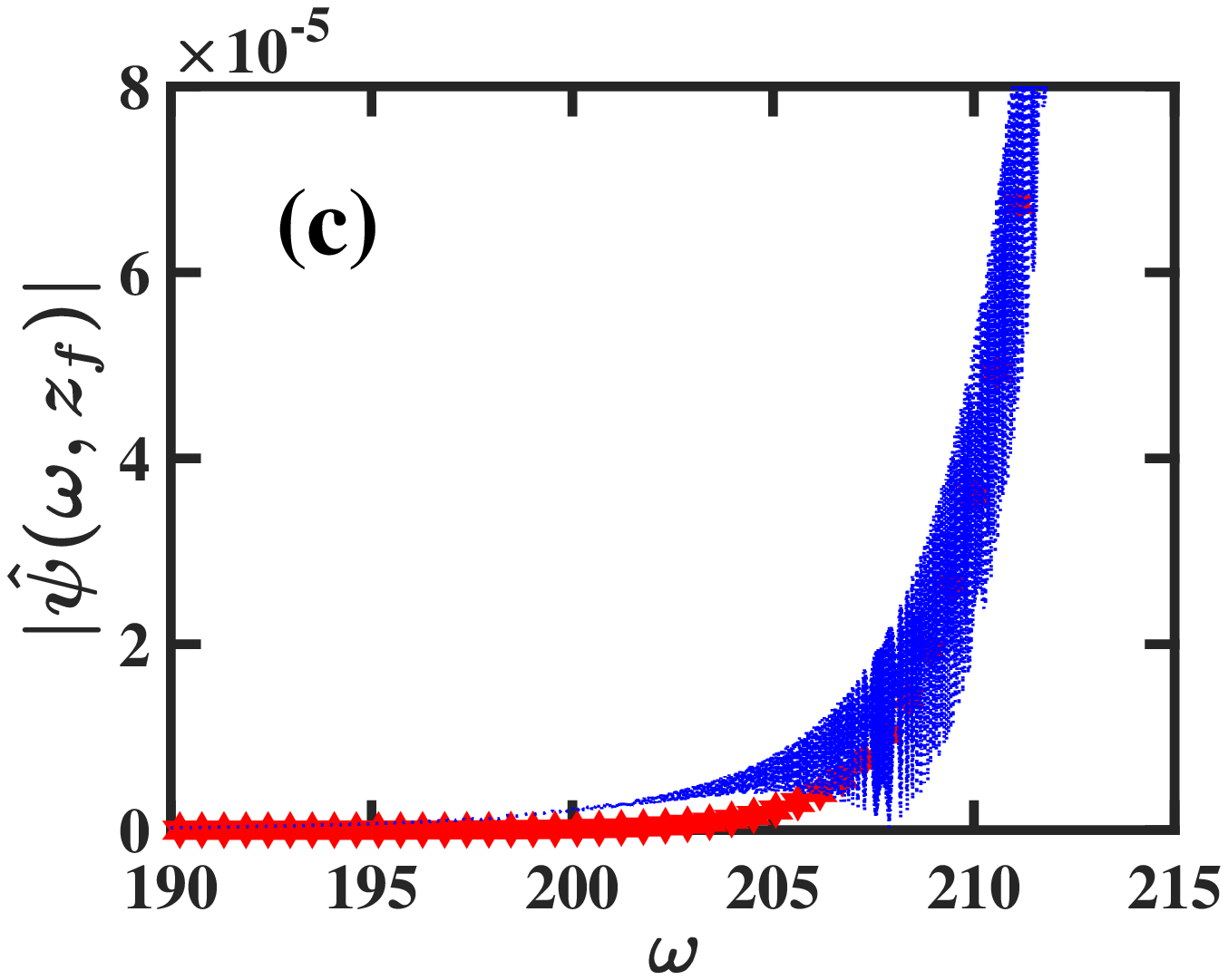}   
\end{tabular}
\end{center}
\caption{The shape of the Fourier spectrum $|\hat\psi(\omega,z_{f})|$,  
where $z_{f}=200$, for propagation of a CNLS soliton in a waveguide loop 
with weak frequency-dependent linear gain-loss, cubic loss, and delayed Raman response. 
The parameter values are the same as in Fig. \ref{fig13}. 
The solid blue curve represents the result obtained by numerical 
solution of Eqs. (\ref{cqnls45}) and (\ref{cqnls32}), and the red stars 
represent the perturbation theory prediction of Eqs. (\ref{Iz6}), 
(\ref{cqnls28}), and (\ref{cqnls46}).}                        
\label{fig15}
\end{figure}

Further understanding of transmission stabilization of the CNLS soliton can 
be gained by analyzing the Fourier spectrum of the pulse. A comparison 
between the simulation's result for the Fourier spectrum $|\hat\psi(\omega,z)|$ 
at $z=z_{f}=200$ and the perturbation theory prediction of Eqs. (\ref{Iz6}), 
(\ref{cqnls28}), and (\ref{cqnls46}) is shown in Fig. \ref{fig15}. As seen in 
Fig. \ref{fig15}(a), the agreement between the numerical and the theoretical 
results is very good and no significant radiative peaks are present. 
Additionally, the comparison in Figs. \ref{fig15}(b) and \ref{fig15}(c)  
reveals the existence of a weak oscillatory distortion at the low frequency 
tail of the soliton's spectrum. This distortion leads to an order $10^{-5}$ 
deviation of the numerically obtained spectrum from the theoretical prediction.
However, the latter deviation is significantly smaller than the order 1 deviation 
that is found for the CNLS soliton at $z_{f}=76$ in waveguides with weak  
frequency-independent linear gain [see Fig. \ref{fig7}(d)]. Therefore, the results 
shown in Fig. \ref{fig15} support our conclusion from Figs. \ref{fig13} and \ref{fig14} 
about transmission stability of the CNLS soliton in waveguides with weak 
frequency-dependent linear gain-loss, cubic loss, and delayed Raman response.

\begin{figure}[ptb]
\begin{tabular}{cc}
\epsfxsize=8.1cm  \epsffile{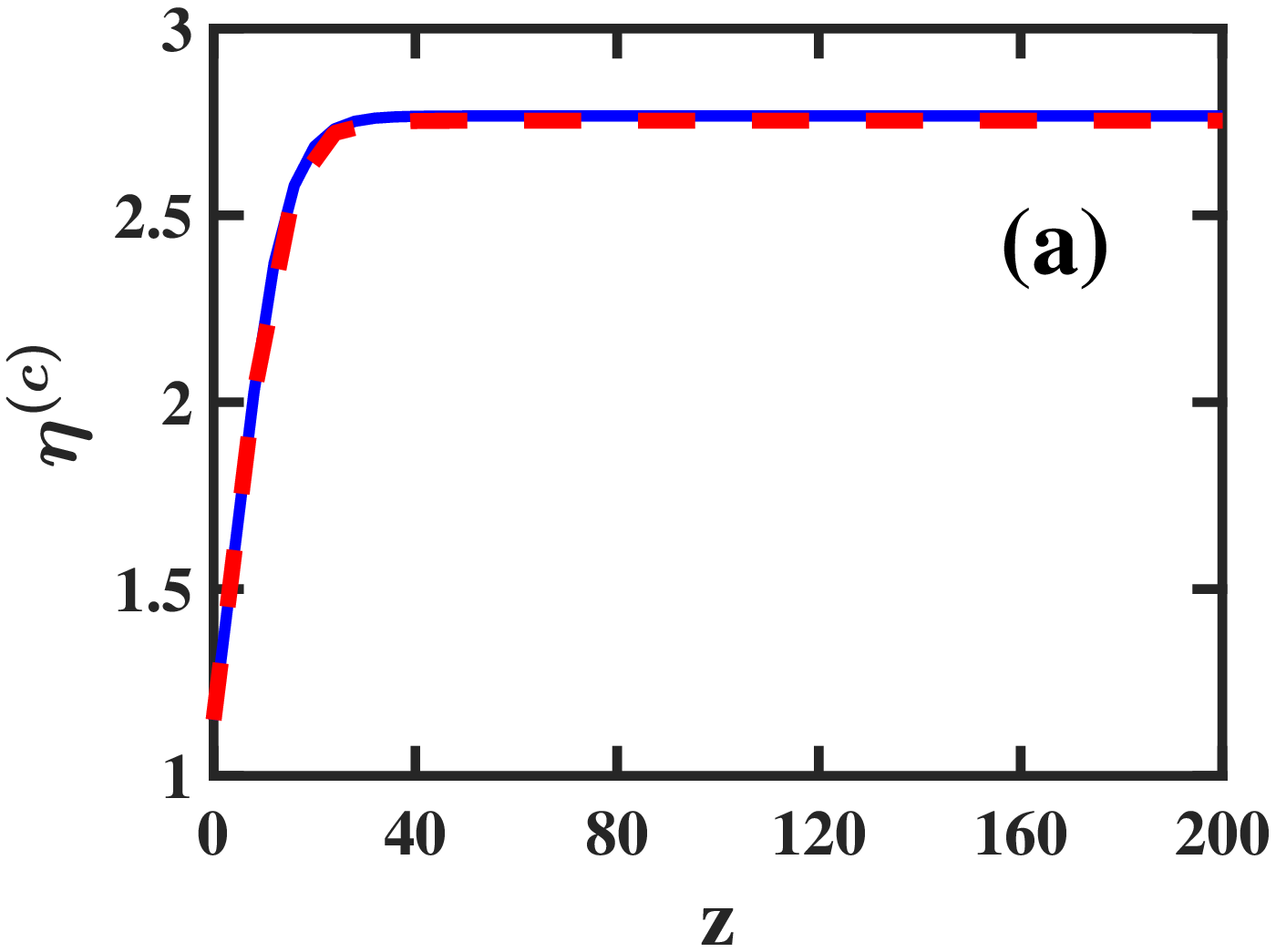} &
\epsfxsize=8.1cm  \epsffile{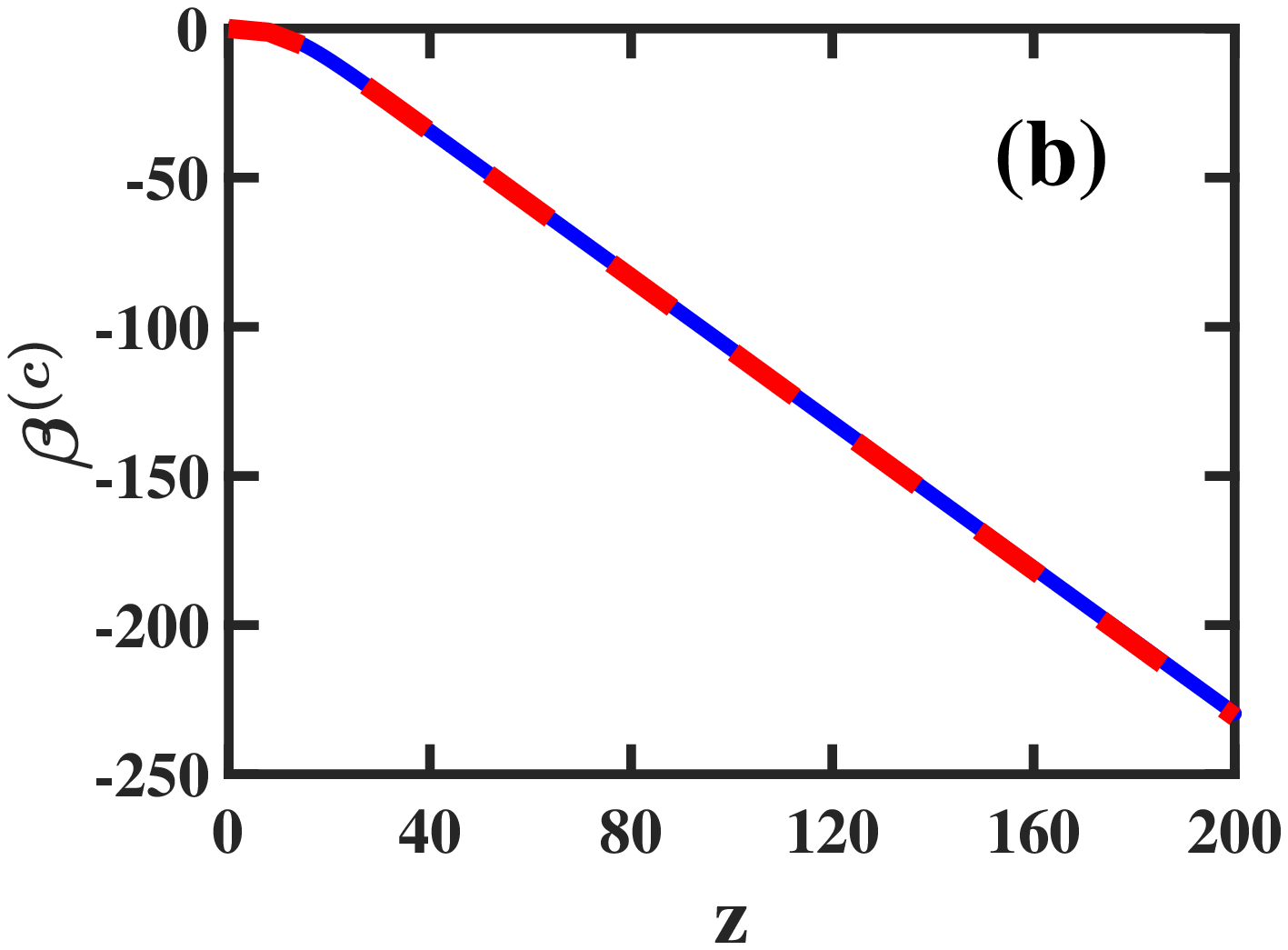}
\end{tabular}
\caption{The $z$ dependence of the CNLS soliton's amplitude $\eta^{(c)}(z)$ (a) 
and frequency $\beta^{(c)}(z)$ (b) for the waveguide setups considered in 
Figs. \ref{fig13}-\ref{fig15}. The solid blue curves represent the results 
obtained by numerical solution of Eqs. (\ref{cqnls45}) and (\ref{cqnls32}). 
The dashed red curves correspond to the perturbation theory predictions of 
Eq. (\ref{cqnls46}) in (a), and of Eqs. (\ref{cqnls28}) and (\ref{cqnls46}) in (b).}
\label{fig16}
\end{figure}

The suppression of pulse distortion for the CNLS soliton that is 
observed in Figs. \ref{fig13}-\ref{fig15} has important implications 
for the dynamics of the soliton's amplitude and frequency. The $z$ 
dependence of the soliton's amplitude and frequency obtained by 
numerical solution of Eqs. (\ref{cqnls45}) and (\ref{cqnls32}) 
is shown in Fig. \ref{fig16} together with the perturbation theory 
predictions of Eqs. (\ref{cqnls46}) and (\ref{cqnls28}).       
The agreement between the theoretical predictions and the simulation's 
results is very good. More specifically, the value of the soliton's 
amplitude obtained in the simulation tends to the equilibrium value 
$\eta_{0}^{(c)}=2.8086...$ at short distances and stays close to this value 
throughout the simulation, in full accordance with the theoretical prediction.
Furthermore, the numerically obtained value of the soliton's frequency remains 
close to the $z$-dependent value predicted by the perturbation theory. 
Therefore, the results shown in Fig. \ref{fig16} support our conclusion 
that transmission of the CNLS soliton in waveguides with weak 
frequency-dependent linear gain-loss, cubic loss, 
and delayed Raman response is stable despite the relatively 
large value of $\eta_{0}^{(c)}$. Nevertheless, due to the relatively small $z_{f}$ 
value in the current numerical simulation, additional simulations are needed 
for verifying the conclusion.

{\it Summary.} The comparison between Figs. \ref{fig9}-\ref{fig12} and 
Figs. \ref{fig13}-\ref{fig16} shows that enhancement of transmission 
stability is stronger for the CQNLS soliton than for the CNLS soliton. 
In particular, the distortion of the pulse shape, the average value of 
$I(z)$, and the distortion of the Fourier spectrum are significantly 
smaller for the CQNLS soliton. Furthermore, the Raman-induced frequency 
shift is remarkably smaller for the CQNLS soliton [compare Figs. 
\ref{fig12}(b) and \ref{fig16}(b)]. Based on these findings, we conclude 
that transmission stabilization of highly energetic solitons by 
frequency-dependent linear gain-loss and perturbation-induced frequency 
shifting is much more robust in waveguides with focusing cubic nonlinearity 
and defocusing quintic nonlinearity than in waveguides with focusing 
cubic nonlinearity.

\section{Conclusions}
\label{conclusions}

We demonstrated stabilization of energetic solitons of the nonintegrable 
CQNLS equation by a method that is based on the interplay between 
perturbation-induced shifting of the soliton's frequency and 
frequency-dependent linear gain-loss. The method was developed 
and demonstrated in Ref. \cite{PC2018A} for solitons of the integrable 
CNLS equation. However, it is unclear if the method works for nonintegrable 
nonlinear wave models far from their integrable limit. Furthermore, it is 
unclear whether soliton stability is enhanced or reduced in the nonintegrable 
case compared with the integrable one. In the current work, we addressed 
these important questions by considering the propagation of a highly energetic 
CQNLS soliton in a nonlinear optical waveguide with focusing cubic nonlinearity, 
defocusing quintic nonlinearity, and three weak perturbations due to weak 
linear gain-loss, cubic loss, and delayed Raman response. 
In this system, the weak linear gain-loss counteracts the effects of 
weak cubic loss, such that the value of the soliton's amplitude tends 
to an equilibrium value. Additionally, the shifting of the soliton's 
frequency is caused by the Raman perturbation. 
We considered two different waveguide setups, one with frequency-independent 
linear gain, and another with frequency-dependent linear gain-loss. 
For each waveguide setup, we used perturbative calculations to derive 
equations for the dynamics of the CQNLS soliton's amplitude and frequency 
in the presence of the three perturbations. Furthermore, we carried out 
numerical simulations with the perturbed CQNLS equation and characterized the 
distortion of the pulse shape and the Fourier spectrum. We also compared 
the simulations results with results of numerical simulations for 
propagation of a highly energetic CNLS soliton in similar waveguide 
setups in the absence of quintic nonlinearity.


In the case of waveguides with frequency-independent linear gain, our 
numerical simulations showed that transmission of the CQNLS soliton 
becomes unstable due to radiation emission effects. The radiation 
emitted by the soliton forms a highly oscillatory hump, which is 
spread over the entire computational domain. Additionally, the soliton's 
and radiation's Fourier spectra become separated due to the Raman-induced 
frequency shift of the CQNLS soliton. However, the presence of 
frequency-independent linear gain leads to continued growth of the radiation.  
As a result, the radiative hump continues to grow with increasing 
distance and the transmission becomes unstable. 
An important consequence of this radiative instability is that the numerically obtained 
values of the CQNLS soliton's amplitude and frequency 
deviate from the perturbation theory predictions at intermediate distances, 
i.e., the dynamics of the soliton's parameters is also unstable. 
A comparison with results of numerical simulations with the perturbed CNLS 
equation for soliton propagation in a similar waveguide setup showed that 
transmission instability is stronger for the CNLS soliton. We also found 
that the Raman-induced frequency shift is significantly larger for the 
CNLS soliton. Thus, our results indicated that transmission of highly 
energetic CQNLS solitons in the presence of weak nonlinear dissipative 
perturbations is advantageous compared with transmission of highly 
energetic CNLS solitons in the presence of the same dissipative perturbations.

Remarkable enhancement of transmission stability of the CQNLS soliton  
was realized in waveguides with frequency-dependent linear gain-loss. 
In this case, our numerical simulations showed that the pulse distortion 
experienced by the soliton is extremely small (of order $10^{-6}$ or 
smaller). Surprisingly, this pulse distortion is of the same order of 
magnitude as the pulse distortion observed in Ref. \cite{PC2018A} for 
CNLS solitons with a significantly smaller equilibrium power value. 
The enhanced transmission stability of the CQNLS soliton was also 
manifested in the dynamics of its amplitude and frequency. Indeed, 
the numerically obtained amplitude and frequency exhibited stable 
dynamics throughout the long-haul propagation (over a distance $z_{f}=2000$), 
in excellent agreement with the perturbation theory predictions. 
Further numerical simulations for propagation of a CNLS soliton 
in waveguides with frequency-dependent linear gain-loss indicated 
that the transmission is stable in this case as well. 
However, due to the large Raman-induced frequency shift 
experienced by the CNLS soliton, the CNLS simulations were run up to 
the relatively small final distance $z_{f}=200$. The pulse distortion 
of the CNLS soliton was significantly larger than the pulse distortion 
of the CQNLS soliton in the same waveguide setup, but considerably 
smaller than the distortion of the CNLS soliton in waveguides with 
frequency-independent linear gain. Based on these results, we concluded  
that transmission stabilization of highly energetic solitons by 
frequency-dependent linear gain-loss and perturbation-induced frequency 
shifting is much more robust in waveguides with focusing cubic nonlinearity 
and defocusing quintic nonlinearity than in waveguides with focusing 
cubic nonlinearity.

In summary, we demonstrated that the transmission stabilization method 
that was developed in Ref. \cite{PC2018A} for solitons of the CNLS 
equation also works for highly energetic solitons of the CQNLS equation 
far from the CNLS limit. The method is based on the interplay between 
perturbation-induced shifting of the soliton's frequency and frequency-dependent 
linear gain-loss. Our numerical simulations and perturbation analysis 
showed that the frequency shifting experienced by the CQNLS soliton 
(due to delayed Raman response in our study) indeed leads 
to a separation of its Fourier spectrum from the radiation's Fourier 
spectrum, while the frequency-dependent linear gain-loss leads to 
efficient suppression of radiation emission. As a result, we observed  
stable long-distance propagation of the highly energetic CQNLS soliton with 
almost no pulse distortion, and stable dynamics of its amplitude and frequency. 
Thus, our study provided the first demonstration of the stabilization method 
for solitons of a nonintegrable nonlinear wave model, far from the integrable limit.

\appendix
\section{Derivation of the equations for amplitude and frequency dynamics} 
\label{appendA}

In this appendix, we derive Eqs. (\ref{cqnls12}), (\ref{cqnls13}), and 
(\ref{cqnls35}), for the dynamics of the CQNLS soliton's amplitude and 
frequency in the presence of weak linear gain-loss, cubic loss, and 
delayed Raman response. The derivation is based on energy and momentum 
balance calculations that use the expressions for the power ${\cal M}$ 
and the momentum ${\cal P}$ in Eqs. (\ref{cqnls4}) and (\ref{cqnls5}).

We start by considering the dynamics of the amplitude and frequency 
parameters of the CQNLS soliton in waveguides with weak frequency-independent 
linear gain, cubic loss, and delayed Raman response. The propagation is 
described by the perturbed CQNLS equation (\ref{cqnls11}). We derive the equations 
for amplitude and frequency dynamics by employing a perturbation theory, 
which is similar to the adiabatic perturbation theory for the CNLS soliton 
(see, for example, Refs. \cite{Hasegawa95,Iannone98,PC2018A,Kaup90,PC2020} for a 
description of the latter theory). We calculate the dynamics in first-order 
in the two small parameters $\epsilon_{3}$ and $\epsilon_{R}$.   
In our perturbation theory, we write the solution to Eq. (\ref{cqnls11}) 
in the form    
\begin{eqnarray}
\!\!\!\!\!\!\!\!\! 
\psi(t,z)=\psi_{s}(t,z)+\psi_{rad}(t,z), 
\label{appendA1}
\end{eqnarray}
where $\psi_{s}(t,z)$ is the soliton solution (\ref{cqnls2}) of the 
unperturbed CQNLS equation with slowly varying parameters, and     
$\psi_{rad}(t,z)$ is the radiation part, which consists of terms 
that are proportional to $\epsilon_{3}$ and $\epsilon_{R}$. 
Thus, $\psi_{s}(t,z)$ in Eq. (\ref{appendA1}) is given by:  
\begin{eqnarray}
\!\!\!\!\!\!\!\!\! 
\psi_{s}(t,z)=\frac{\sqrt{2}\eta(z)\exp[i\chi(t,z)]}
{\left\{[1-\eta^{2}(z)/\eta_{m}^{2}]^{1/2}\cosh(2x)+1\right\}^{1/2}},
\label{appendA2}
\end{eqnarray}     
where $x=\eta(z)\left[t-y(z)\right]$, $\chi(t,z)=\alpha(z)-\beta(z)\left[t-y(z)\right]$, 
$y(z)=y(0)-2\int_{0}^{z} dz' \beta(z')$, 
and $\alpha(z)=\alpha(0)+\int_{0}^{z} dz' \left[\eta^{2}(z')+\beta^{2}(z')\right]$. 
Employing an energy balance calculation and the ansatz (\ref{appendA1}), 
we find: 
\begin{eqnarray}&&
\!\!\!\!\!\!\!\!\!\!\!\!
\partial_{z}\int_{-\infty}^{\infty} \!\!\! dt|\psi_{s}(t,z)|^{2}\!=\!
g_{0}\int_{-\infty}^{\infty} \!\!\! dt|\psi_{s}(t,z)|^{2}
-2\epsilon_{3}\int_{-\infty}^{\infty} \!\!\! dt|\psi_{s}(t,z)|^{4}.
\label{appendA3}
\end{eqnarray}       
We now use the following relations for the two integrals on the right 
hand side of Eq. (\ref{appendA3}) \cite{Ryzhik2007}: 
\begin{equation}
\int_{-\infty}^{\infty} \!\!\! dt|\psi_{s}(t,z)|^{2}= 
2\eta_{m}\mbox{arctanh}(\eta/\eta_{m}) , 
\label{additional1}
\end{equation}       
and 
\begin{equation}
\int_{-\infty}^{\infty} \!\!\! dt|\psi_{s}(t,z)|^{4} = 
4\eta_{m}^{3}\left[\mbox{arctanh}\left(\eta/\eta_{m}\right)
-\eta/\eta_{m} \right] . 
\label{additional2}
\end{equation} 
We obtain:  
\begin{eqnarray} &&
\!\!\!\!\!\!\!\!
\frac{d \eta}{dz}=
\frac{\left(\eta_{m}^{2}-\eta^{2}\right)}{\eta_{m}}
\left\{g_{0}\mbox{arctanh}\left(\frac{\eta}{\eta_{m}}\right)
-4\epsilon_{3}\eta_{m}^{2}\left[\mbox{arctanh}\left(\frac{\eta}{\eta_{m}}\right)- 
\frac{\eta}{\eta_{m}} \right]\right\},     
 \label{appendA4}
\end{eqnarray}   
which is Eq. (\ref{cqnls12}).

We now turn to derive Eq. (\ref{cqnls13}) for $d \beta/dz$. The derivation is valid 
for both perturbed CQNLS equations (\ref{cqnls11}) and (\ref{cqnls31}). Within 
the framework of the first-order perturbation theory, the rate of change of 
the soliton's momentum is given by 
\begin{eqnarray}&&
\!\!\!\!\!\!\!\!\!\!
\frac{d {\cal P}}{dz} \simeq 
i\partial_{z}\int_{-\infty}^{\infty} \!\! dt \, 
\left[\psi_{s}(t,z) \partial_{t}\psi_{s}^{*}(t,z) 
- \psi_{s}^{*}(t,z) \partial_{t}\psi_{s}(t,z)\right].  
\label{appendA5}
\end{eqnarray}
To obtain the equation for $d \beta/dz$, we derive two different expressions 
for $d {\cal P} /dz$ and equate them. The first expression is obtained by 
using Eq. (\ref{appendA2}) for $\psi_{s}(t,z)$. The second expression is 
obtained by using Eq. (\ref{cqnls11}) or Eq. (\ref{cqnls31}) to calculate 
the right hand side of Eq. (\ref{appendA5}).

Using Eq. (\ref{appendA2}) for $\psi_{s}(t,z)$ along with the definition 
of $\chi(t,z)$, we find: $\partial_{t}\psi_{s}=-i\beta\psi_{s}
+ \exp(i\chi)\partial_{t}|\psi_{s}|$. Substituting this relation and its 
complex conjugate into Eq. (\ref{appendA5}), we arrive at: 
\begin{eqnarray}&&
\!\!\!\!\!\!\!\!\!\!
\frac{d {\cal P}}{dz} \simeq 
-2\frac{d \beta}{dz} \int_{-\infty}^{\infty} \!\!\! dt|\psi_{s}(t,z)|^{2}  
-2\beta \partial_{z} \int_{-\infty}^{\infty} \!\!\! dt|\psi_{s}(t,z)|^{2}.
\label{appendA6}
\end{eqnarray}
On the other hand, we can express the right hand side of Eq. (\ref{appendA5}) 
in a different manner by performing the differentiation with respect to $z$ 
inside the integral, and by eliminating the mixed second-order partial 
derivatives with the help of integration by parts. This calculation yields: 
\begin{eqnarray}&&
\!\!\!\!\!\!\!\!\!\!
\frac{d {\cal P}}{dz} \simeq 
2i\int_{-\infty}^{\infty} \!\! dt \, 
\left[\partial_{t}\psi_{s}^{*}(t,z) \partial_{z}\psi_{s}(t,z) 
- \partial_{t}\psi_{s}(t,z) \partial_{z}\psi_{s}^{*}(t,z) \right].  
\label{appendA7}
\end{eqnarray}  
We now equate the right hand sides of Eqs. (\ref{appendA6}) and (\ref{appendA7}), 
and obtain 
\begin{eqnarray}&&
\!\!\!\!\!\!\!\!\!\!
-\frac{d \beta}{dz} \int_{-\infty}^{\infty} \!\!\! dt|\psi_{s}(t,z)|^{2}  
-\beta \partial_{z} \int_{-\infty}^{\infty} \!\!\! dt|\psi_{s}(t,z)|^{2} 
\nonumber \\&&
=i\int_{-\infty}^{\infty} \!\! dt \, 
\left[\partial_{t}\psi_{s}^{*}(t,z) \partial_{z}\psi_{s}(t,z) 
- \partial_{t}\psi_{s}(t,z) \partial_{z}\psi_{s}^{*}(t,z) \right].  
\label{appendA8}
\end{eqnarray}

In the next step, we use Eq. (\ref{cqnls11}) or Eq. (\ref{cqnls31}) 
to calculate the integral on the right hand side of Eq. (\ref{appendA8}). 
This calculation yields 
\begin{eqnarray}&&
\!\!\!\!\!\!\!\!\!\!
\!\!\!\!\!\!\!\!\!\!
i\int_{-\infty}^{\infty} \!\!\!\! dt \, 
\left[\partial_{t}\psi_{s}^{*}(t,z) \partial_{z}\psi_{s}(t,z) 
- \partial_{t}\psi_{s}(t,z) \partial_{z}\psi_{s}^{*}(t,z) \right] 
= \epsilon_{R}\int_{-\infty}^{\infty} \!\!\!\! dt \, 
\left[\partial_{t}|\psi_{s}(t,z)|^{2}\right]^{2} 
+ \dots \; . 
\label{appendA9}
\end{eqnarray}    
The first term on the right hand side of Eq. (\ref{appendA9}) is contributed 
by the delayed Raman response term in Eq. (\ref{cqnls11}) or Eq. (\ref{cqnls31}). 
The dots represent contributions coming from the gain and loss terms in 
Eq. (\ref{cqnls11}) or Eq. (\ref{cqnls31}). However, upon substitution of 
Eq. (\ref{appendA9}) into Eq. (\ref{appendA8}), the contributions from 
the gain and loss terms cancel out with the second term on the left 
hand side of Eq. (\ref{appendA8}). Therefore, we do not write these contributions here explicitly.         
We now substitute relation (\ref{appendA9}) into Eq. (\ref{appendA8}), take into account 
the aforementioned cancellation, and also use relation (\ref{additional1}). We obtain: 
\begin{eqnarray}&&
2\eta_{m}\mbox{arctanh}\left(\frac{\eta}{\eta_{m}}\right)
\frac{d \beta}{dz}
= -\epsilon_{R}\int_{-\infty}^{\infty} \!\!\!\! dt \, 
\left[\partial_{t}|\psi_{s}(t,z)|^{2}\right]^{2} . 
\label{appendA10}
\end{eqnarray}    
Using Eq. (\ref{appendA2}) for $\psi_{s}(t,z)$, we can show that 
\begin{eqnarray}&&
\!\!\!\!\!\!\!\!\!\!
\!\!\!\!\!\!\!\!\!\! 
\int_{-\infty}^{\infty} \!\!\!\! dt \, 
\left[\partial_{t}|\psi_{s}(t,z)|^{2}\right]^{2}= 
16\eta^{5}\left(1-\frac{\eta^{2}}{\eta_{m}^{2}}\right) 
\int_{-\infty}^{\infty} \!\!\!\! dx \, \frac{\sinh^{2}(2x)}
{\left[(1-\eta^{2}/\eta_{m}^{2})^{1/2}\cosh(2x)+1\right]^{4}} .
\label{appendA11}
\end{eqnarray}    
The integral on the right hand side of Eq. (\ref{appendA11}) can 
be calculated in a closed form \cite{Ryzhik2007}. This calculation 
yields:  
\begin{eqnarray}&&
\!\!\!\!\!\!\!\!\!\!
\!\!\!\!\!\!\!\!\!\! 
\int_{-\infty}^{\infty} \!\!\!\! dt \, 
\left[\partial_{t}|\psi_{s}(t,z)|^{2}\right]^{2}=
\frac{8}{3}\eta_{m}^{5}
\left[\frac{3\eta}{\eta_{m}} - \frac{2\eta^{3}}{\eta_{m}^{3}}
-3\left(1-\frac{\eta^{2}}{\eta_{m}^{2}}\right)
\mbox{arctanh}\left(\frac{\eta}{\eta_{m}}\right) \right].   
\label{appendA12}
\end{eqnarray}          
Substitution of Eq. (\ref{appendA12}) into Eq. (\ref{appendA10}) 
yields the following equation for the dynamics of $\beta$: 
\begin{eqnarray} &&
\!\!\!\!\!\!\!\!
\frac{d \beta}{dz}=
\frac{-4\epsilon_{R}\eta_{m}^{4}}{3\mbox{arctanh}\left(\eta/\eta_{m}\right)}
\left[\frac{3\eta}{\eta_{m}} - \frac{2\eta^{3}}{\eta_{m}^{3}}
-3\left(1-\frac{\eta^{2}}{\eta_{m}^{2}}\right)
\mbox{arctanh}\left(\frac{\eta}{\eta_{m}}\right) \right] .       
\label{appendA13}
\end{eqnarray}    
Equation (\ref{appendA13}) is Eq. (\ref{cqnls13}) of Sec. \ref{dynamics2}.

Let us derive Eq. (\ref{cqnls35}) for the dynamics of the CQNLS 
soliton's amplitude in waveguides with weak frequency-dependent 
linear gain-loss, cubic loss, and delayed Raman response. In this 
case, the propagation is described by Eq. (\ref{cqnls31}). Applying 
an energy balance calculation and ansatz (\ref{appendA1}), we obtain: 
\begin{eqnarray}&&
\!\!\!\!\!\!\!\!\!\!\!\!
\partial_{z}\int_{-\infty}^{\infty} \!\!\! dt|\psi_{s}(t,z)|^{2}\!=\!
K_{1}+K_{1}^{*}
-2\epsilon_{3}\int_{-\infty}^{\infty} \!\!\! dt|\psi_{s}(t,z)|^{4}, 
\label{appendA14}
\end{eqnarray}      
where 
\begin{eqnarray}&&
\!\!\!\!\!\!\!\!\!\!\!\!
K_{1}=\frac{1}{2}\int_{-\infty}^{\infty} \!\!\! dt \psi_{s}^{*}(t,z)
{\cal F}^{-1}(\hat g(\omega,z) \hat\psi_{s}(\omega,z))  \; . 
\label{appendA15}
\end{eqnarray}  
Using the convolution theorem and Eq. (\ref{appendA2}) for $\psi_{s}(t,z)$, 
we find 
\begin{eqnarray}&&
\!\!\!\!\!\!\!\!\!\!\!\!
K_{1}+K_{1}^{*} \!=\! \frac{\eta(z)}{(2\pi)^{1/2}}
\int_{-\infty}^{\infty} \frac{dx}
{\left\{[1-\eta^{2}(z)/\eta_{m}^{2}]^{1/2}\cosh(2x)+1\right\}^{1/2}}
\nonumber \\&&
\!\!\!\!\!\!\!\!
\int_{-\infty}^{\infty} \!\!\! ds \frac{\left\{g(s,z)\exp[i\beta(z)s] 
+ g^{*}(s,z)\exp[-i\beta(z)s]\right\}}
{\left\{[1-\eta^{2}(z)/\eta_{m}^{2}]^{1/2}\cosh[2x-2\eta(z) s]+1\right\}^{1/2}} .  
\label{appendA16}
\end{eqnarray}      
We assume that $\hat g(\omega,z)$ can be approximated by 
Eq. (\ref{cqnls33}). The inverse Fourier transform of the approximate 
form (\ref{cqnls33}) of $\hat g(\omega,z)$ is 
\begin{eqnarray}&&
g(t,z)=-(2\pi)^{1/2}g_{L}\delta(t) 
+\left(\frac{2}{\pi}\right)^{1/2} \!\!\!\!
(g_{0}+g_{L})\exp[-i\beta(z)t]\sin(Wt/2)/t,
\label{appendA17}
\end{eqnarray}         
where $\delta(t)$ is the Dirac delta function. Substitution of 
Eq. (\ref{appendA17}) into Eq. (\ref{appendA16}) yields: 
\begin{eqnarray}&&
\!\!\!\!\!\!\!\!\!\!\!\!
K_{1}+K_{1}^{*} \!=\!
-2g_{L}\eta_{m}\mbox{arctanh}\left(\frac{\eta}{\eta_{m}}\right) 
+\frac{2(g_{0}+g_{L})\eta}{\pi}J_{1}(\eta), 
\label{appendA18}
\end{eqnarray}   
where $J_{1}(\eta)$ is given by Eq. (\ref{cqnls36}). We now substitute 
Eq. (\ref{appendA18}) into Eq. (\ref{appendA14}) and also use relation 
(\ref{additional2}) to calculate the third term on the right hand side 
of Eq. (\ref{appendA14}). We obtain: 
\begin{eqnarray} &&
\!\!\!\!\!\!\!\!
\frac{d \eta}{dz}=
\frac{\left(\eta_{m}^{2}-\eta^{2}\right)}{\eta_{m}}
\left\{-g_{L}\mbox{arctanh}\left(\frac{\eta}{\eta_{m}}\right) 
+ \frac{(g_{0}+g_{L})\eta}{\pi\eta_{m}} J_{1}(\eta) 
\right.
\nonumber \\&&
\!\!\!\!\!\!\!\!\!
\left.
-4\epsilon_{3}\eta_{m}^{2}\left[\mbox{arctanh}
\left(\frac{\eta}{\eta_{m}}\right)- 
\frac{\eta}{\eta_{m}} \right]\right\},     
\label{appendA19}
\end{eqnarray}   
which is Eq. (\ref{cqnls35}) of Sec. \ref{dynamics3}.

\section{The transmission quality integral $I(z)$} 
\label{appendB}
       
In this appendix, we present the theoretical predictions for 
the shape and the Fourier spectrum of the CQNLS and the CNLS 
solitons, which were used in the analysis of transmission quality 
and stability. Furthermore, we present the method for calculating the 
transmission quality integral $I(z)$ and the transmission quality distance $z_{q}$ 
from the numerical simulations results.

The theoretical prediction for the shape of the CQNLS soliton is obtained 
by following the perturbation approach of appendix \ref{appendA}. 
In this approach, we use Eq. (\ref{appendA1}) to express the solution 
of the perturbed CQNLS equation $\psi(t,z)$ as a sum of the soliton part 
$\psi_{s}$ and the radiation part $\psi_{rad}$, where $\psi_{s}(t,z)$ 
is defined in Eq. (\ref{appendA2}). We therefore take $\psi_{s}(t,z)$ as the theoretical 
prediction for the soliton part: $\psi^{(th)}(t,z) \equiv \psi_{s}(t,z)$. 
As a result, the theoretical prediction for the shape of the CQNLS soliton 
is given by: 
\begin{eqnarray} 
|\psi^{(th)}(t,z)|=
\sqrt{2}\eta(z)
\left[(1-\eta(z)^{2}/\eta_{m}^{2})^{1/2}\cosh(2x)+1\right]^{-1/2} , 
\label{Iz1}
\end{eqnarray}     
where $\eta(z)$ is calculated by solving the perturbation theory's 
equation for $d\eta/dz$, and $y(z)$ is measured from the 
simulations \cite{y_z}. The Fourier transform of $\psi_{s}(t,z)$ 
with respect to $t$ is 
\begin{eqnarray} &&
\hat\psi_{s}(\omega,z)=
\left(\frac{\eta_{m}}{\pi^{2}\eta(z)}\right)^{1/2}
\Gamma \left(\frac{1}{2} - \frac{i(\omega+\beta(z))}{2\eta(z)} \right)
Q^{\frac{i(\omega+\beta(z))}{2\eta(z)}}_{-\frac{1}{2}}
\nonumber \\&&
\times
\exp\left\{i[\alpha(z)-\omega y(z)] 
+ \frac{\pi(\omega+\beta(z))}{2\eta(z)}\right\}, 
\label{Iz2}
\end{eqnarray} 
where $\Gamma(x)$ is the Gamma function and $Q_{\nu}^{\mu}$ is the 
associated Legendre function of the second kind \cite{Stegun1972}. 
Therefore, the theoretical prediction for the Fourier spectrum of 
the CQNLS soliton is 
\begin{eqnarray} 
\!\!\!\!\!\!\!
|\hat\psi^{(th)}(\omega,z)|=
\left(\frac{\eta_{m}}{\pi^{2}\eta(z)}\right)^{1/2}
\left|
\Gamma \left(\frac{1}{2} - \frac{i(\omega+\beta(z))}{2\eta(z)} \right)
Q^{\frac{i(\omega+\beta(z))}{2\eta(z)}}_{-\frac{1}{2}}
\right|
\exp\left[\frac{\pi(\omega+\beta(z))}{2\eta(z)}\right], 
\label{Iz3}
\end{eqnarray}   
where $\eta(z)$ and $\beta(z)$ are calculated by solving the perturbation 
theory's equations for $d\eta/dz$ and $d\beta/dz$.

The transmission quality integral measures the deviation of the pulse shape  
obtained in the simulations $|\psi^{(num)}(t,z)|$ from the soliton's shape 
predicted by the perturbation theory $|\psi^{(th)}(t,z)|$.  
We define $I(z)$ in the same manner as we did in previous studies 
of soliton propagation in nonlinear optical waveguide systems 
\cite{PNT2016,PC2018A}: 
\begin{eqnarray} &&
I(z)=   
\left[ \int_{t_{min}}^{t_{max}} dt\,
\left| \psi^{(th)}(t,z) \right|^2 \right]^{-1/2}
\left\{\int_{t_{min}}^{t_{max}} \!\!\!\!\!\!\!\!\! dt\,
\left[\;\left|\psi^{(th)}(t,z) \right| - 
\left|\psi^{(num)}(t,z) \right| \; \right]^2 \right\}^{1/2},  
\label{Iz4}
\end{eqnarray}  
where the integration is carried out on the entire time domain 
used in the simulations $[t_{\mbox{min}},t_{\mbox{max}}]$.    
Note that $I(z)$ measures both distortion in the pulse shape due to 
radiation emission and deviations of the numerically obtained values of 
the soliton's parameters from the values predicted by the perturbation theory. 
Additionally, we define the transmission quality distance $z_{q}$ as the distance at which the 
value of $I(z)$ first exceeds a constant value $K$. In the current paper we use $K=0.075$. 
We emphasize, however, that the values of $z_{q}$ obtained by using this definition 
are not very sensitive to the value of the constant $K$. In other words, small changes 
in the value of $K$ lead to small changes in the measured values of $z_{q}$.

The definitions of the theoretical predictions for the shape and the 
Fourier spectrum of the CNLS soliton are the same as the ones used in 
Ref. \cite{PC2018A}. That is, we use: 
\begin{eqnarray} 
|\psi^{(c)(th)}(t,z)|=
\eta^{(c)}(z)\mbox{sech}\left[\eta^{(c)}(z)\left(t-y^{(c)}(z)\right)\right],
\label{Iz5}
\end{eqnarray}     
and 
\begin{eqnarray} 
\!\!\!\!\!\!\!
|\hat\psi^{(c)(th)}(\omega,z)|=
\left(\frac{\pi}{2}\right)^{1/2}
\mbox{sech}\left[\pi\left(\omega+\beta^{(c)}(z)\right)/\left(2\eta^{(c)}(z)\right)\right], 
\label{Iz6}
\end{eqnarray}   
where $\eta^{(c)}(z)$ and $\beta^{(c)}(z)$ are calculated by the adiabatic perturbation 
theory for the CNLS soliton, and $y^{(c)}(z)$ is measured from the numerical 
simulations. The transmission quality integral for the CNLS soliton is 
calculated by employing Eq. (\ref{Iz4}) with $|\psi^{(th)}(t,z)|$ replaced 
by $|\psi^{(c)(th)}(t,z)|$.


\begin{thebibliography}{}
\bibitem{dimensions1} The dimensionless distance $z$ in Eq. (\ref{cqnls1}) 
is $z=X/(2L_{D})$, where $X$ is the dimensional distance, 
$L_{D}=\tau_{0}^{2}/|\tilde\beta_{2}|$ is the dispersion length,
$\tau_{0}$ is the soliton temporal width, and $\tilde\beta_{2}$ is 
the second-order dispersion coefficient. The dimensionless time is
$t=\tau/\tau_{0}$, where $\tau$ is time. 
$\psi=(\gamma_{3} \tau_{0}^{2}/|\tilde\beta_{2}|)^{1/2}E$, where $E$ 
is the electric field and $\gamma_{3}$ is the cubic nonlinearity coefficient.
The coefficient $\epsilon_{q}$ is related to the dimensional quintic 
nonlinearity coefficient $\gamma_{5}$ via 
$\epsilon_{q}=2\gamma_{5}|\tilde\beta_{2}|/(\gamma_{3}^{2}\tau_{0}^{2})$.
\bibitem{thresholds} Note that in Ref. \cite{PC2018A}, we used the 
same threshold values of 0.075 and 0.655 for determining the values 
of $z_{q}$ and $z_{f}$ from the simulations with the perturbed 
CNLS equation.
\bibitem{CNLS2} The duration of the simulation with $z_{f}=200$ on a workstation 
equipped with a 3.90 GHz processor was about 36 hours. The long duration 
was mainly due to the large value of $\beta(z_{f})$, $\beta(z_{f}) \simeq -230$, 
which made it necessary to employ a relatively large frequency domain $[-275,275]$, 
and small time and distance steps $\Delta t = 0.014$ and $\Delta z = 0.00004$.   
For a simulation with $z_{f}=400$, for example, the value of $\beta(z_{f})$ predicted 
by Eq. (\ref{cqnls30}) is $\beta(z_{f}) \simeq -476$. In this case, we would have 
to use a frequency domain $[-520,520]$ or larger, and $\Delta t$ and 
$\Delta z$ values of $\Delta t = 0.007$ and $\Delta z = 0.00001$. As a result, 
the simulation duration on the same workstation would be 48 days or longer.
\end{thebibliography}
\end{document}